\documentclass[aps,longbibliography,preprint,amsmath,amssymb,showpacs,showkeys,superscriptaddress]{revtex4-2}
\usepackage{amsmath,latexsym}
\usepackage{graphicx}
\usepackage{amssymb}
\usepackage{amsfonts}
\usepackage{amsbsy}
\usepackage{natbib}
\usepackage{comment}
\usepackage{color}
\usepackage{float}

\usepackage[colorlinks=true, urlcolor=blue, citecolor=blue, linkcolor=blue]{hyperref}

\usepackage{silence}

\renewcommand{\vec}[1]{\mbox{\boldmath$\mathrm{#1}$}}
\let\sb=_ \catcode`\_=\active \def_#1{\ensuremath \sb{\rm#1}}

\renewcommand{\vec}[1]{\mbox{\boldmath$\mathrm{#1}$}}
\newcommand{\be}{\begin{equation}}
\newcommand{\ee}{\end{equation}}
\newcommand{\ben}{\begin{eqnarray}}
\newcommand{\een}{\end{eqnarray}}

\begin{document}

\title{Pseudo-Hermitian Magnon Dynamics}

\author{Xi-guang Wang}
	\email{wangxiguang@csu.edu.cn}
	\affiliation{School of Physics, Central South University, Changsha 410083, China}
\author{Jamal Berakdar}
	\email{jamal.berakdar@physik.uni-halle.de}
	\affiliation{Institut f\"ur Physik, Martin-Luther Universit\"at Halle-Wittenberg, 06099 Halle/Saale, Germany}


\date{\today}

\begin{abstract}
A defining quantity of a physical system is its energy which is represented by the Hamiltonian. In closed quantum mechanical or/and coherent  wave-based systems the Hamiltonian is introduced as a Hermitian operator which ensures real energy spectrum and secures the decomposition  of any state over  a complete basis set spanning the space where the states live. Pseudo-Hermitian, or $PT$ symmetric, systems are a special class of non-Hermitian ones. They describe open systems but may still have real energy spectrum. The eigenmodes are however not orthogonal in general.  This qualitative difference to Hermitian physics has a range of consequences for the physical behaviour of the
 system in the steady state or when it is subjected to external perturbations. This overview reviews the recent progress in the field of pseudo-Hermitian physics as it unfolds when applied to low-energy excitations of magnetically ordered materials. The focus is mainly on long wave length spin excitations (spin waves) with magnons being the  energy quanta of these excitations.  Various setups including ferromagnetic, antiferromagnetic, magnonic crystals,  and hybride structures with different types of coupling to the environments as well as spatio-temporally engineered systems  will be discussed with a focus on the particular aspects that are brought about by the pseudo-Hermiticity such as mode amplifications, non-reciprocal propagation, magnon cloaking, non-Hermitian skin effect, PT-symmetric assisted Floquet engineering,  topological energy transfer, and field-induced enhanced sensitivity. 
\end{abstract}

\maketitle
\newpage
\section{Introduction to non-Hermitian and pseudo-Hermitian systems}
In conventional quantum systems, the Hamiltonian is usually  a Hermitian operator ensuring  real energy spectrum and the existence of a basis set of orthogonal eigenstates. The decomposition principle, meaning that an arbitrary state of the system can be expressed as a linear superposition of projections onto the subspaces spanned by  the basis,  is thus a consequence of Hermiticity. The multitude of phenomena rooted in this principle are well documented and conventionally  present first encounters  with principles of  quantum mechanics. 
 In reality, many  systems are  open  or undergo  energy gain or loss in which case the Hamiltonian typically fails to be Hermitian.
  Obviously, this is a situation  at the verge between the classical world, in which damping effects are ubiquitous, and  pure quantum energy-conserving  settings that are detached from environment.\\
 From a mathematical point of view,   non-Hermitian operators may have complex eigenvalues. The  eigenstates  are not necessarily orthogonal and may even non-normalizable. Thus, the decomposition principle is generally not operational. On the hand, we can expect static and dynamic behavior of such a system that differ qualitatively from a Hermitian counterpart.\\
  Early studies on non-Hermitian systems  focused mainly on  stability analysis, spectral changes due to environmental couplings, and the influence of dissipation on dynamical behavior.\cite{Gamow1928001,PhysRev.96.448,Lindblad1976} One of  the major milestones in  non-Hermitian  research is the identification of PT-symmetric systems, which are classified as a pseudo-Hermitian.\cite{PhysRevLett.80.5243,Bender2007repphysics} Pseudo-Hermitian systems form a distinct intermediate category between Hermitian and non-Hermitian problems. For pseudo-Hermitian setting, the Hamiltonian remains adjoint under a certain metric, fulfilling the pseudo-Hermiticity condition $ H  =  {\eta}  H ^{\dagger}  {\eta}^{-1} $.\cite{Das2011jphys} A pseudo-Hermitian Hamiltonian can still have  real energy spectrum but the associated system exhibits  features which are qualitatively different from Hermitian physics. In contrast to the complicated energy spectra of non-Hermitian systems, the eigenvalues imaginary parts of the pseudo-Hermitian  Hamiltonian  generally follow  a specific symmetry that can be externally tuned, as we will discussed in the course of this article.  Importantly,  pseudo-Hermitian systems may undergo critical transitions at specific parameters, referred to as non-Hermitian phase transitions, leading to significant changes in the energy spectrum and phenomena akin to phase transitions such as enhanced or diverging susceptibilities. The parameters controlling the phase transition  point  could be, for instance, the magnitude of gain or loss that are  introduced into the system or the strength of coupling between  constituent elements.\cite{bender47895492013}  These features have been shown to result  in new physical  effects, including spontaneous symmetry breaking, exceptional points (EPs, i.e. non-Hermitian spectral degeneracies where eigenvalues and eigenmodes coalesce), non-reciprocal or asymmetric wave propagation, and power oscillations that defy intuition by circulating energy between gain and loss regions.\cite{PhysRevX.4.031042,zhaohannwy0112018,Fleury2015naturecommun, Feng2013naturemater,ElGanainy2018naturephys} The accumulated findings so far 
 underline  the substantial  potential  of PT symmetric setups for applications in optics, condensed matter physics, and quantum information.

Pseudo-Hermitian concepts have been successfully emulated in a variety of classical systems – including optical, mechanical, electrical, and metamaterial platforms – where one can deliberately introduce gain and loss.\cite{Ruter2010nphys1515,PhysRevX.4.031042,zhaohannwy0112018,PhysRevA.84.040101} Optics, for example, provides an ideal ground because complex refractive index landscapes can be engineered with regions of optical amplification and absorption.\cite{Ruter2010nphys1515}  Similarly, mechanical and acoustic resonators can be coupled to active elements (for gain) and dampers (for loss) to form PT-symmetric oscillators.\cite{Ruter2010nphys1515,PhysRevX.4.031042} Electrical circuits with amplifiers and resistors, and electromagnetic/metamaterial structures with active inclusions, have further broadened the pseudo-Hermitian physics.\cite{PhysRevA.84.040101, Schindler2012jphysa} These efforts are motivated both by fundamental interest and emerging applications. For instance, exceptional-point physics has led to ultra-sensitive sensors and single-mode lasers, and PT-symmetric structures have enabled novel devices like invisible sensors and unidirectional waveguides that manipulate wave flow in unconventional ways.\cite{Chen2017nature,Hodaei2017nature, Kononchuk2022nature,PhysRevLett.112.203901,Fengscience2014, 10.1126science.aar7709}


\subsection{Mathematical formulation of pseudo-Hermiticiy}

For an overview on the different kind of dynamics we start with an energy-quantized system $S_1$ and consider the  dynamic in a small energy range (smaller than the energy level spacing)  around an  energy  level with the  real frequency  $a$. We assume $S_1$ is governed by a Schrödinger-type equation of motion, with $ H $ corresponding to  the "energy" operator (we call it henceforth Hamiltonian), and $t$  plays the role of   "time" in the Schrödinger-type wave equation.
If we have two identical systems $S_1$ and $S_2$ and they are  well separated, then  
$ H $ in the energy-range  of interest reads
$
	\begin{small}
		\begin{aligned} 
			\displaystyle  H  = \left( \begin{matrix} a  & 0  \\ 0  & a   \end{matrix} \right)
			\label{ham000}
		\end{aligned} 
	\end{small}
$, and we have two independent (orthonormal) states $ \left( \begin{matrix} 1    \\   0   \end{matrix} \right)$ and $ \left( \begin{matrix} 0    \\   1   \end{matrix} \right)$ at the frequency  $a$ and the dynamic of any given state $\psi(t_0)$ at time $t_0$ is completely quantified at time $t$  by $\psi(t)=  {U}(t,t_0)\psi(t_0)$, where $
\begin{small}
	\begin{aligned} 
		\displaystyle   U  = \left( \begin{matrix}  e^{-i a (t-t_0)}  & 0  \\ 0  & e^{-i a (t-t_0)}   \end{matrix} \right),
	\end{aligned} 
\end{small}
$ 
meaning that $S_1$ and $S_2$ propagate completely independent. If $S_1$ and $S_2$ are brought together such that we allow for  exchange of  information via some mechanism and name the amplitude  of information transfer  $b$ (which is proportional  to matrix elements for transitions between $S_1$ and $S_2$), then the coupled $ H _c$ attains  off-diagonal element $b$. The two-fold degenerate frequency $a$ level splits into  
$\omega^\pm= a \pm \frac{1}{2} \sqrt{4bb^*}=a\pm b$ and $  U _c$ becomes
$
\begin{small}
	\begin{aligned} 
		\displaystyle   U  _c= \left( \begin{matrix}  e^{-i \omega^+ (t-t_0)}  & 0  \\ 0  & e^{-i \omega^- (t-t_0)}   \end{matrix} \right).
	\end{aligned} 
\end{small}
$ 
The eigenmodes (often called acoustic ($+$) or optic $-$) of $ H _c$ are then 
$ \left( \begin{matrix} 0    \\   1   \end{matrix} \right)_\pm=\frac{1}{\sqrt{2}}\left[ \left( \begin{matrix} 1    \\   0   \end{matrix} \right) \pm  \left( \begin{matrix} 0    \\   1   \end{matrix} \right)\right].
$  If we prepare the total system $S$  at $t=t_0$ in $S_2$  (and $S_1$ is not populated) then at a later stage
 the system will evolve to $\psi(t)=  {U}_c(t,t_0)\psi(t_0)$ and the population of $S_2$ is obtained from $|\langle\psi(t)|\psi(t_0)\rangle|^2= \frac{1}{2}\left[1 + \cos(2|b| t)\right]$.  The  mode  population or mode power beats  between $S_1$ and $S_2$ and the  $t$ scale for  beating  is set by $b$. If the coupling  matrix element between $S_1$ and $S_2$ is harmonic, i.e., if  $b=b_0\cos(\omega_0 t) $ then we may also obtain analytical solutions if $b_0\gg a$ by diagonalizing the $t$ dependent  $ H _{c}$ via $R^{-1} H _c R$ with
 $R=\frac{1}{\sqrt{2}} \left( \begin{matrix}  1  & 1  \\ 1  & -1 \end{matrix} \right)$ 
 and the evolution operator reads
 $$\begin{small}
 	\begin{aligned} 
 		\displaystyle   U  _c(t,t=0)= \left( \begin{matrix}  e^{-i b_0 \int_0^t cos(\omega_0 t') dt'}  & 0  \\ 0  & e^{i b_0 \int_0^t cos(\omega_0 t') dt'}     \end{matrix} \right).
 	\end{aligned} 
 \end{small}
 $$ 
With the formula $$e^{-i\frac{b_0}{\omega_0}  \sin(\omega_0 t) }= \sum_{n=-\infty}^\infty J_n(b_0/\omega_0) e^{-i n\omega_0  t},$$ where $J_n$ is the Bessel function of the first kind with  the order $n$, we see that the dynamic is quasi-periodic, meaning the evolution will contain several frequencies depending on how much quantas of $n\omega_0$ are exchanged between the system and the driving coupling field but the dynamics is not dissipative, meaning there is no loss/gain in total  power. How much  $n\omega_0$ are exchanged we can infer  from the relation $$\lim_{|n|\gg 1} J_n(b_0/\omega_0)\to \frac{1}{2n\pi } ( \frac{b_0\, e}{2n\omega_0})^n$$ which means that if $\frac{b_0}{\omega_0}<1$ (and $a/b_0<1$) then only few additional frequencies $n\omega_0$ (with respect to the field-free case) are involved in the dynamic. All  three above cases ($b_0=0, \, b_0\neq 0 \wedge \omega_0=0,\, b_0\neq 0 \wedge \omega_0\neq 0)$, which  we will encounter in this article, present unitary evolution in $t$, the dynamic is $t$ reversal symmetric without loss of the input information. The situation is different in the presence of  environment with  coupling to $S_1$ and/or $S_2$ in which case information is not conserved within the system.  In our case 
the effect of the coupling  mechanism can be captured    in the most simple way  as 
\begin{equation}
\begin{small}
\begin{aligned} 
\displaystyle  H  = \left( \begin{matrix} a - i \gamma_1 & b \\ b^* & a + i \gamma_2 \end{matrix} \right).
\label{ham001}
\end{aligned} 
\end{small}
\end{equation}
$\gamma_{1,2}/\hbar\in \mathbb{R}$ describes energy dissipation or amplification rates of the decoupled $S_1$ and $ S_2$ (independent of whether   $b=0$ or generally complex).
With  $\gamma_{1,2}\neq 0 $, the Hamiltonian becomes non-Hermitian  $H \ne H^{\dagger}$.  The eigenvalues 
$$\lambda_{\pm} = a + [i(\gamma_2 - \gamma_1) \pm \sqrt{4 |b|^2 - (\gamma_1 + \gamma_2)^2}]/2$$  are in general complex and the eigenvectors $$ \vec{\psi}_{\pm} = ( (-i(\gamma_2 + \gamma_1) \pm \sqrt{4 |b|^2 - (\gamma_2 + \gamma_1)^2})/(2 |b|),1)^T$$  are no longer orthogonal.

In the particular  case of a balanced  gain and loss ($\gamma_1 = \gamma_2 = \gamma$), meaning the total system $S_1$ and $S_2$ do not gain or lose energy through environmental coupling even though the individual subsystems do,   the energy eigenvalues $\lambda_{\pm}$ may turn real in a certain range and  the Hamiltonian $ H $ is called pseudo Hermitian or  PT symmetric. 
In our particular case,   both eigenvalues $\lambda_{\pm}$ are real when $ |\gamma| < |b| $. \\
Mathematically,   a  pseudo-Hermitian $H$  is a non-selfadjoint operator  which is related to $H^\dagger$  via  a similarity transformation, i.e. $H^\dagger=\eta H\eta^{-1}$ with $\eta$ being a Hermitian operator \cite{nonad,schmuedgen}. 
The spectrum of $H$ is  real,   if $\eta $ is    positive-definite. This is seen by introducing  the $\eta$-inner product $ \langle \phi,H\psi\rangle_\eta$ with  $\eta$ playing the role of the metric  for an inner product on an arbitrary non-Euclidean space. Because  $H^\dagger\eta=\eta H$ we have 
$ \langle \phi,H\psi\rangle_\eta=\langle H \phi,\psi\rangle_\eta$ and it follows from the  spectral theorem  that $H$ has real  eigenvalues. Conventional Hermitian operators are a subset of pseudo-Hermitian one  with $\eta=\openone$.\\
 Pseudo-Hermitian operators are  also called parity(P)-time(T)-symmetric   operators,
because with  $T$ being antilinear operator,    $ (PT) H  
-  H  (PT)=0$ implies $ P \; T HT^{-1} =  P H^\dagger  =  H  P$ or $  H^\dagger  = P^{-1} H  P=P H  P=P H  P^{-1}$ (note $P^2=\openone$).  \\
 For the example discussed above, parity operation   often  physically means an exchange of $S_1$ by $S_2$. If we work with the Pauli matrices $\sigma_{i},\, i=x,y,z,$  then parity amounts to    $ \sigma_{x} $. 
From the above discussion of  Hamiltonian (\ref{ham001})  and its eigenfunctions we  conclude parity symmetry is given if  $ |\gamma| < |b| $,
in which case  $\eta$ amounts to $ \sigma_x$.  
  In contrast, when  $ |\gamma| > |b| $, the eigenvalues $\lambda_{\pm}$ are  complex  and  PT symmetry is broken.    Amplitude of modes with positive (negative) imaginary part  is amplified  (damped)  as $t$ increases. 
At the transition point  $ |\gamma| = |b| $, also  called  exception point (EP),   the  two eigenvalues and eigenvectors merge.  The order of this no-Hermitian degeneracy  at EP is set by the number of eigenvalues and eigenvectors that simultaneously coalesce.
Thus,  in the above example  we have EP of a second-order.  The phase transition is also reflected in the change of eigenvectors. Below the EP, the eigenvectors can be expressed as $$ \vec{\psi}_{\pm} = (\pm \exp(\mp i \theta), 1)^T,$$
 where $$\sin \theta = \gamma / b$$ and $$\cos \theta = \sqrt{b^2 - \gamma^2}/b.$$ 
The two eigenmodes are distributed in both $S_1$ and $S_2$. Still, they differ qualitatively in the behavior from the Hermitian case,  the two eigenvectors are namely 
 non-orthogonal for $\gamma \ne 0$, even though phase coherence is observable.
  At the EP, the two eigenvectors are parallel. Above the EP, the dynamic is dissipative. The non-orthogonal eigenvectors admit the form
  $$\vec{\psi}_{\pm} = ( i\exp(\pm \theta), 1)^T$$ with $$\cosh \theta = \gamma / b.$$
   In this regime, the symmetry of the eigenmodes with regard to $S_1$ and $S_2$ is broken. The mode 	amplitude or power is  mostly in one subsystem. This behavior is a special case of the general situation where not only $H\neq H^\dagger$ but also
 we have a non-normal $H$, meaning $\left[H,H^\dagger\right]\neq 0$, in which case one  distinguishes  right eigenstates $|R_\alpha\rangle$ from the
 left eigenstates  $\langle L_\alpha|$  where $\alpha$ is a collective set of quantum numbers defining the states. Mathematically, 
  problems involving non-normal eigenvalues  and   pseudospectra are well studied \cite{Trefethen,Bauer1960}.
   Here, we seek a quantification of the  effect of the
  non-normal  behavior,  $\left[H,H^\dagger\right]\neq 0$,  which is achieved by the  Petermann factor \cite{1070064,PhysRevA.39.1253,PhysRevLett.79.4357,PhysRevA.79.061801,PhysRevResearch.5.033042}  
  ($\langle| .|\rangle   \equiv \langle| .|\rangle_{\eta=\openone}$ )
  \begin{equation}
  	K_\alpha=\frac{\langle R_\alpha| R_\alpha\rangle \langle L_\alpha| L_\alpha\rangle}{|\langle L_\alpha| R_\alpha\rangle|^2}. \label{eq:petermann}
  \end{equation}
  The phase rigidity \cite{PhysRevX.6.021007,PhysRevE.55.R1,Jaiswal2023}  is indicated by
  \begin{equation}
  r_\alpha=\frac{|\langle L_\alpha| R_\alpha\rangle|}{\sqrt{\langle R_\alpha| R_\alpha\rangle \langle L_\alpha| L_\alpha\rangle}} \label{eq:phaserigid}.
  \end{equation}
  From Cauchy-Schwarz inequality  ${|\langle L_\alpha| R_\alpha\rangle|^2}\leq  \langle R_\alpha| R_\alpha\rangle \langle L_\alpha| L_\alpha\rangle$ we infer that 
  $K_\alpha =1$ when left and right states are equivalent but otherwise $K_\alpha> 1$ or  may even diverge.  The phase rigidity $  r_\alpha = 1/\sqrt{K_\alpha}$ 
  is bounded to $0\leq r_\alpha\leq 1$. 
  
   Above we discussed an example of two-coupled 
     PT symmetric systems   with balanced gain and loss as representative of  a specific class of pseudo Hermitian systems.
      There are other  forms of pseudo Hermiticity, defined by different forms of the operation $\eta$ leading to diverse spectral properties.\cite{S0219887810004816,SCHOLTZ199274}.


\section{Physical properties of Pseudo-Hermitian systems}

\begin{figure}[htbp]
	\includegraphics[width=0.9\textwidth]{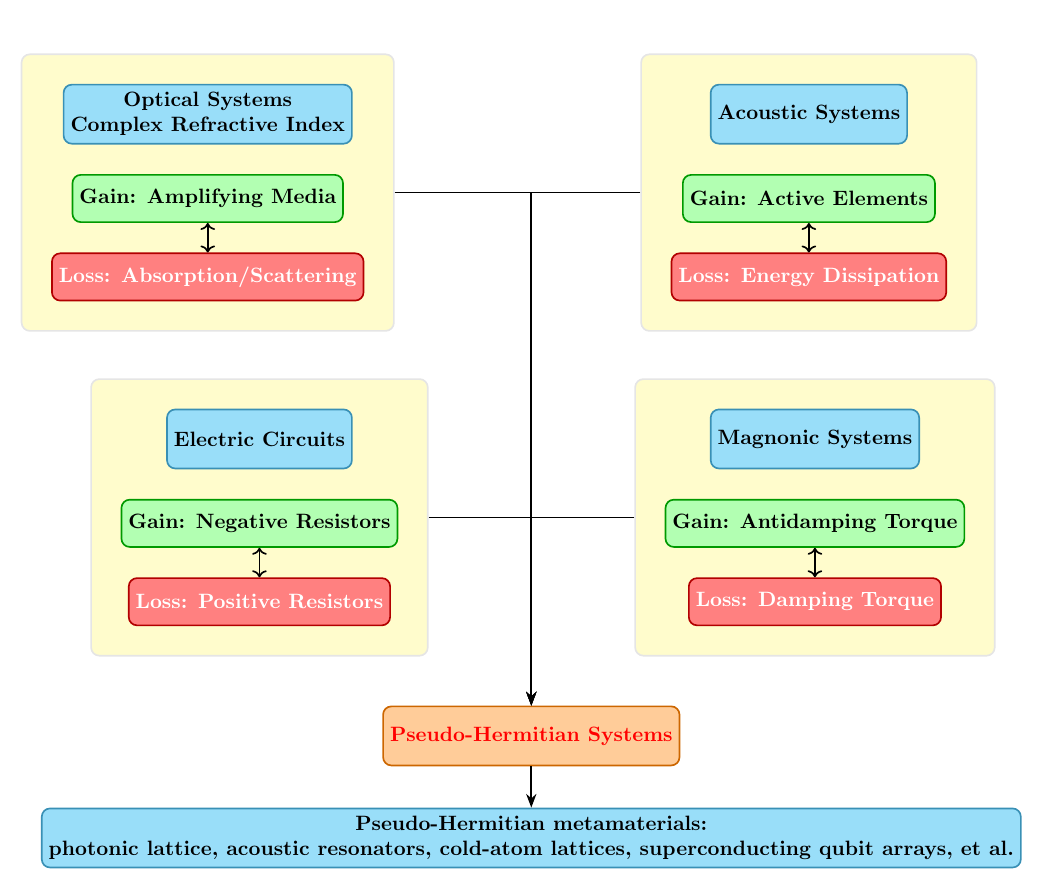}
	\caption{\label{models} Pseudo-Hermitian systems across different physical domains. }
\end{figure}

This chapter aims at a brief  overview on pseudo-Hermitian systems across optical, mechanical, electrical, and metamaterial platforms, and illustrate the realization mechanisms of gain and loss as  implemented in each platform. For an in-depth exposition we refer to the topic-specific literature. 

Generally, the PT symmetry condition is accomplished via  spatially distributed gain and loss with symmetric profiles. For example, an optical structure with refractive index $ n(r) = n^*(-r) $ with real parts and odd imaginary parts can formal PT symmetry.\cite{Ruter2010nphys1515} This means one region provides optical gain while the mirror region provides an equal amount of loss. Similarly, in mechanical or electrical oscillators, one element may supply energy (gain) and the other dissipates energy (loss) in equal measure.\cite{PhysRevX.4.031042,PhysRevA.84.040101} Under these balanced conditions, one can observes novel physics. In the unbroken phase, energy can oscillate between gain and loss components without net loss, leading to oscillatory dynamics. PT-symmetric structures can exhibit non-reciprocal wave propagation, for instance, unidirectional invisibility, where waves incident from one side are completely transmitted with no reflection, but from the opposite side there is reflective.\cite{Feng2013naturemater} Such effect occurs near the EP.\cite{PhysRevLett.106.213901,Feng2013naturemater} Besides,  EP can greatly amplify the response to small perturbations, where frequency splitting scales as the $n$th root of perturbation, and $n$ is the EP order.\cite{Chen2017nature,Zhang2024Microsyst} Also, selective PT-symmetry breaking has been used to enhance single-mode operation in lasers, by amplifying one mode while suppressing others.\cite{Hosseinscience2014}


\subsection{Pseudo-Hermitian optic systems}
Optics has been a suitable ground for realizing and manipulating the properties of pseudo Hermitian systems, particularly those exhibiting PT-symmetry \cite{Feng2017NaturePhotonics, PhysRevLett.103.093902}. 

In the PT-symmetry optics, engineering balanced optical  power gain and loss can be realized in coupled waveguides or microcavities, by modulating the  complex refractive index. The refractive index reads $n(x) = n_{\rm R}(r) + i n_{\rm I}(r)$, where $n_{\rm R}$ and $n_{\rm I}$ are the real (phase coherent refraction) and imaginary (gain/loss) parts, respectively. PT symmetry demands $n_{\rm R}$ be $n_{\rm R}(r) = n_{\rm R}(-r)$ and $n_{\rm I}(r) = -n_{\rm I}(-r)$. Optical gain is achieved  by amplifying media (such as pumped doped fibers, semiconductor optical amplifiers, or laser-pumped crystals). Optical loss is due to absorption or scattering (by using  intrinsically  lossy materials or tuning to radiation leakage).  For example, Guo et al.  used two parallel dielectric waveguides, one of which was coated to introduce loss while the other was pumped for gain and observed the PT-symmetry-breaking transition as the gain/loss strength  was varied.\cite{PhysRevLett.103.093902}  C. Rüter et al. reported the first observation of PT symmetry in optics using a pair of coupled waveguides in Fe-doped $\text{LiNbO}_3$, one waveguide undergoing optical gain via photorefractive two-wave mixing, the other experiencing equivalent absorption.\cite{Ruter2010nphys1515}  Their experiments verified spontaneous PT-breaking. Beyond a critical gain/loss strength, the guided-mode eigenstates transitioned from a regime where the propagation constants are real  to another regime where they became complex. Even below the phase transition, the presence of balanced gain and loss induced an asymmetric energy flow between the waveguides.

Since those pioneering studies, optical implementations of pseudo-Hermiticity have been advanced greatly. A typical  platform has been coupled microresonators or microrings with one lossy and one amplifying resonator. For example, Hodaei et al. and Feng et al. have independently demonstrated PT-symmetric microring laser systems.\cite{Hosseinscience2014,Fengscience2014} By coupling an active microring (gain provided by an electrically pumped InGaAsP semiconductor) with an identical passive ring (introducing loss), at the EP the laser’s frequency selection is significantly simplified. Here, one supermode was amplified to lasing while the other mode was suppressed by loss, overcoming the usual multi-mode competition in lasers, paving the way for single-mode lasers without the need for conventional isolators or filters. 

Another optical pseudo-Hermitian setup  involved optical lattices and metamaterials which can be realized by periodic structures with alternating gain and loss or complex index modulation. Feng et al. demonstrated a unidirectional invisible optical metamaterial near an EP.\cite{Feng2013naturemater} By engineering a Silicon-on-insulator waveguide array with carefully patterned loss and gain sections, light incident from one end experiences balanced gain-loss and transmit completely with no reflection. From the opposite end the symmetry was effectively broken and significant reflection occurred, meaning unidirectional reflectionlessness.

Across these optical platforms, one may identify  several common  physical PT-symmetry-induced phenomena. By tuning the coupling or gain–loss contrast, one can bring two resonances together in frequency and line width at EP. This feature is important   for developing new optical sensors. For example, a small change in the refractive index of the surrounding medium result in a measurable shift in the spectrum of the coupled optical microcavities operating near an EP resulting in ultra-high sensitivity sensors.  \cite{PhysRevLett.112.203901} This sensitivity enhancement is a consequence of  the fractional power dependence of the eigenvalues on the perturbation strength near an EP. For an EP of order $R$, the splitting of eigenvalues scales as the $R$-th root of the external perturbation.\cite{Hodaei2017nature,PhysRevLett.117.107402,Chen2017nature} This scaling means that even a  minor variation  in external parameters may  cause  substantial change in the spectral properties, including resonant frequencies or transmission coefficients. Such EP-based sensors have been developed for detecting minute changes in temperature, mass, or even the presence of single molecules.

Near the EP, balancing gain with additional loss causes  nonreciprocal light propagation.  A. Guo et al. integrate optical gain medium and loss material in  waveguides to form complex pseudo Hermitian potentials, and the related phase transition leads to a loss induced optical transparency \cite{PhysRevLett.103.093902}. B. Peng et al. has demonstrated near-perfect unidirectional transmission by engineering gain and loss in coupled cavity brought to an EP.\cite{Peng2014naturephysics} In this case, light propagates from one direction   experiencing  amplification and strong transmission, while light in the opposite direction undergoes  loss and weak transmission. The non-reciprocity also shows up  in the reflection on scattering potential with PT symmetry, and the feature is adopted to achieve unidirectional invisibility.\cite{PhysRevA.87.012103, PhysRevLett.106.213901} Additionally, if one introduces nonlinearity or time-periodic modulation in a PT-symmetric optical system, true optical isolation and one-way propagation can be achieved.\cite{PhysRevA.82.043803,NireekshanReddy14,Zhou16optexpress} For example, saturable gain in a PT dimer can make the effective gain-loss balance intensity-dependent, leading to nonreciprocal transitions for different input beam directions.\cite{PhysRevA.82.043803} More recently, Floquet PT-symmetric photonics (with time-periodic driving) also has been explored to achieve asymmetric energy transfer between modes.\cite{PhysRevA.94.043828,PhysRevLett.119.093901,10.1126sciadv.adu4653}

The combination of pseudo-Hermitian physics and nonlinear optics results in new phenomena. Specifically, EP offers unique opportunities for enhancing nonlinear optical processes. Near an EP, the system exhibits a significantly increased density of states, leading to amplified light-matter interaction and nonlinear effects such as new soliton solutions \cite{Driben11optlett,Wimmer2015naturecommun}.

Merging pseudo-Hermiticity with topological band structures is a new active area. PT-symmetric photonic lattices can be engineered to be topologically non-trivial and can  exhibit for example topological  edge states which in the case PT-symmetry  are enhanced or protected by gain-loss arrangements.\cite{Regensburger2012nature,zhaohannwy0112018} The photonic Su–Schrieffer–Heeger (SSH) chain with alternating gain and loss on the sublattices showed that the topological interface state remained robust while the continuum modes were attenuated. This provides a new handle to selectively amplify topological states or even induce topological phase transitions. Several proposals for non-Hermitian topological lasers and one-way edge lasers combine both topology and PT symmetry.

\subsection{Pseudo-Hermitian mechanical systems}
In mechanical systems, realizing a pseudo-Hermitian setup typically requires an active element to provide energy gain and an equivalent dissipative element to cause loss. One example is a pair of coupled oscillators where one oscillator is attached to an actuator that feeds energy into the system, while the other is attached to a damper that dissipates energy.  If the gain and loss are of equal strength, the dynamic can be shown to follow  a PT-like symmetry behavior, exhibiting  EPs, non-reciprocal wave propagation, and unidirectional invisibility \cite{PhysRevX.4.031042, 10.1063/5.0224250, PhysRevApplied.16.057001}. 

Carl Bender et al. demonstrated this concept in a tabletop mechanical experiment using two coupled pendula.\cite{bender47895492013} One pendulum had a small motor that supplied a constant drive (gain), while the other had a friction brake to provide equivalent damping (loss). The coupling was achieved via a spring connecting the pendula. By adjusting the gain/loss level, they observed the characteristic PT phase transition and EP.

Zhu et al. introduced the concept of PT symmetry in acoustics, showing theoretically that a carefully designed acoustic medium with complex density and modulus can exhibit one-way transparency.\cite{PhysRevX.4.031042}  The challenge in acoustics is realizing an effective acoustic gain to complement acoustic loss, since typical materials only dissipate energy. This has been overcome by using active piezoelectricity elements. For example, an acoustic resonator can be outfitted with a feedback control circuit that amplifies sound, acting as a negative resistance (gain).   Fleury, Sounas et al. demonstrated an invisible acoustic sensor consisting of two coupled acoustic cavities, where one was loaded with an absorptive element and the other with an active electrical network providing gain.\cite{Fleury2015naturecommun} At a certain frequency, the system perfectly absorbed incoming sound without reflection.

\subsection{Pseudo-Hermitian electric circuits}

Electrical circuits are a straightforward  platform to simulate pseudo-Hermitian Hamiltonians. Active (negative resistance) and passive (positive resistance) components constitute the gain-loss elements, combined with geometrically symmetric coupling configurations (e.g., matched inductive (L) and capacitive (C) pairs) fulfills the criteria for pseudo-Hermitian systems.\cite{PhysRevA.84.040101, Schindler2012jphysa} Schindler et al. performed an experiment with two RLC resonant circuits coupled together, one including an active negative resistor (gain) and the other a positive resistor (loss).\cite{PhysRevA.84.040101} This electronic dimer was shown to exactly obey the PT symmetry condition showing   typical features of pseudo-Hermiticity.

The advantage of electrical systems is the ease of reconfigurability and measurement. The dynamic control of pseudo-Hermitian phase transitions can be realized through electrical parameter modulation using tunable circuit components, such as variable resistors and semiconductor diodes, enabling real-time manipulation of non-Hermitian system. Researchers have since used lumped-circuit networks to explore higher-order and more exotic non-Hermitian effects. By cascading multiple resonant LC cells with alternating gain and loss, higher-order EPs (EP3, EP4, etc.) have been emulated confirming for example  the predicted sensitive dependence on perturbations.\cite{Tagouegni2023} Also, circuits have been used to realize not only PT symmetry but also anti-PT symmetry. Recent work demonstrated an anti-PT symmetric pair of RLC resonators with capacitive coupling, showing an EP transition that conserved the difference of energies in the two resonators.\cite{Choi2019naturecommun}

\subsection{Pseudo-Hermitian metamaterials}
Metamaterials  are engineered structures  designed to exhibit optical features  which can be grossly different from the optical  properties  of the constituents materials. Typically, this is achieved by   periodic or non-periodic arrangements to tune the 
permittivity or the permeability. The tunability of metamaterials makes them an ideal platform for realizing and controlling non-Hermitian phenomena. Pseudo-Hermitian metamaterials can be achieved through various design strategies, including the incorporation of specific arrangements of gain and loss elements, the integration of active components, or the exploitation of symmetries within the metamaterial structure.  Pseudo-Hermitian metamaterials enable the realization of  wave manipulation functionalities across different physical domains, photonic systems, acoustic resonators, cold-atom lattices, plasmonic system, and superconducting qubit arrays.\cite{Charumath2024Plasmonics, Ni2023ChemRev, PhysRevA.89.033829} Examples of such functionalities include unidirectional wave propagation, enhanced sensing capabilities, and the design of unconventional laser sources.

The feasibility of implementing optical gain and loss, coupled with the precise control offered by metamaterial design, has led to significant progress in this area.  By designing metamaterials with non-reciprocal couplings induced by gain and loss gradients or asymmetric unit cells, researchers have observed the accumulation of light at the boundaries of the metamaterial structure.\cite{Wang24advoptphoton} Liang Feng et al. employed electron beam lithography with sequential Ge/Cr and Si evaporation to fabricate a passive PT-symmetric metamaterial. This non-Hermitian system demonstrates EP-enhanced unidirectional reflection suppression, exhibiting complete optical transparency in forward propagation while maintaining conventional reflection in the reverse direction.\cite{Feng2013naturemater} Ye Geng et al. designed artificial photonic metamaterial structures that can transform ideal Weyl points into non-Hermitian exceptional rings .\cite{Geng24optlett} Researchers have built PT-symmetric photonic crystals, gratings, and even metasurfaces to achieve effects like asymmetric diffraction, coherent perfect absorption and lasing modes, and topological edge states. 

In the electromagnetic realm, microwave metamaterials have been designed by introducing gain in resonant meta-atoms. A split-ring resonator loaded with a transistor were adopted to provide gain, paired with another ring with a resistor for loss,  which results in  a pair forming  a PT unit cell.\cite{Xu12optexpress} When arrayed, these cells produced a metamaterial with an extraordinary transmission band under one polarization and a reflection band under the opposite propagation direction.

Metamaterials also apply for acoustic and mechanical implementations. Pseudo-Hermiticity can be introduced through the integration of active components such as piezoelectric transducers and electronic feedback control systems. Yangyang Chen et al. designed active feedback loops to amplify or attenuate mechanical vibrations in a direction-dependent manner, effectively breaking time-reversal symmetry and resulting in a  pseudo-Hermitian effective Hamiltonian.\cite{Chen2021NatureCommun} Yanghao Fang et al. introduced new designs of parity-time-symmetric elasto-dynamic metamaterials, allowing for applications include hypersensitive sensing and elastic wave control.\cite{Fang2021NewJPhys} V. Achilleos et al. demonstrated enhanced absorption at EP in acoustic pseudo-Hermitian  metamaterials.\cite{PhysRevB.95.144303} Furthermore, Xinhua Wen et al. adopted space-time modulated Pseudo-Hermitian metamaterials for efficient frequency conversion and amplification for advanced acoustic devices.\cite{Wen2022CommunPhys} 

Regarding  the topological aspects,  non-Hermitian metamaterials, including pseudo-Hermitian ones, can exhibit topological phases beyond Bloch band theory, generating non-Hermitian skin effect. Di Zhou et al. demonstrated  such non-Hermitian skin effect in 1D and 2D lattice, with applications in various active materials and biological systems.\cite{PhysRevResearch2023173} Nonlinearity in PT-symmetric materials  results in further interesting phenomena  such as discrete breathers, as introduced by N. Lazarides et al., with potential applications in optical devices and energy gathering. \cite{PhysRevLett110053901}

\section{Magnonics}
The main purpose of this overview is to discuss PT-symmetry in connection with low-energy excitations or spin waves of spin-ordered materials.
Magnons are the energy quanta of spin waves which are  transverse,  collective small-amplitude  oscillations around the ground state ordering.
Magnons possess spin but no charge, and the transmission of magnons typically does not involve the transport of charge and hence  Joule heating is absent, making magnons ideal carriers for information  transmission, processing, and storage.\cite{Chumak2015naturephysics} Magnonic devices offer potential advantages such as low energy consumption, the operation at GHz-THz frequencies, and the possibility of integration in   nanoscale devices and spintronics.\cite{Flebus2024JPhys,Serga2010jphys} How to generate, manipulate, and detect magnon flow and use them to develop high-performance devices is a rapidly developing  interdisciplinary field, meanwhile known as   magnonics.\cite{Pirro2021naturerev,Kruglyak2010jphys,Demokritov,Rezende}

\subsection{Magnetic excitations}

Considering a localized spin $s_1$ and approaching it with an another spin $s_1$, then we find from von Neumann's equation of motion \cite{doi:10.1142/6094} that $s_1$ and will experience a torque with a strength depending on the interaction between $s_1$ and $s_2$, which is usually rooted in the exchange interaction. Physically, $s_1$ will responds with a precessional motion \cite{DanielDStancilspinwave,hillebrands2003spin,Rezende}.
For an ordered collection of spins, this precession sums up to a collective wave but in a finite-size system such as a stripe the exchange energy scale is a much higher (typically order of magnitude) than that set by  other interactions, most importantly  magnetostatic dipolar interactions. \cite{9706176,Barman2021,Krawczyk2014} The motion is locally precessional building up to a collective dispersive wave with energies in GHz in ferromagnets  to THz for antiferromagnets. 
In reality the precession is damped due to various spin-dependent scattering events.  The energy spectrum is dense with typically a large number of excitations even at quite small temperatures  in which case one may  use the corresponding coherent state which is in turn  
well described classically. In reality, these waves might  be quantized but in the case of the coherent state this quantization  is usually due to (classical) finite-size effects. Starting from the magnonic Fock states to construct the coherent state can be one way but it is more practical to start from the (semi)classical end in which one can encompass more easily the magnetostatic interactions, geometry and closure fields effects. 
Technically, one may start from an atomic lattice with classical spins localized on each site (this approach is called atomistic spin simulations), or coarse grain the (quantum or classical) magnetic moments and go over to continuum approach to define a magnetization vector field $\mathbf{m}(\mathbf{r},t)$, where $\mathbf{r}$ is the center of the coarse-grain cell. This method commonly dubbed micromagnetic simulations and  is reasonable if we study phenomena relevant  at wavelengths much larger than the atomic lattice constant. 
%
%
The temporal evolution and spatial propagation of the spin waves is   mathematically described by the Landau-Lifshitz-Gilbert equation,\cite{Gilbert1353448}
\begin{equation}
	\displaystyle \frac{\partial \vec{m}}{\partial t} = - \gamma \vec{m} \times \vec{H}_{\rm{eff}}+  \alpha \vec{m} \times \frac{\partial \vec{m}}{\partial t}.
	\label{LLG}
\end{equation}
The equation incorporates the material properties and geometry as well as external fields in the effective magnetic field $\vec{H}_{\rm{eff}}$. The various  damping mechanisms are collectively subsummed in the parameter $\alpha$, called Gilbert damping, $\gamma$ is the gyromagnetic ratio.
To determine $\vec{H}_{\rm{eff}}$, one starts from  the functional for the magnetic energy density $E_m$ 
\begin{equation}
	\displaystyle E_m[\mathbf{m}] = A_{ex} (\vec{\nabla} \vec{m})^2 + E_K -\mu_0 M_s (\vec{H}_a \cdot \vec{m}) - \frac{1}{2}\mu_0 M_s (\vec{H}_d \cdot \vec{m}),
	\label{energy}
\end{equation} 
which includes the exchange energy with exchange coefficient $A_{ex}$, the Zeeman energy due to an  external magnetic field $\vec{H_a}$. $E_K$ is the magnetic anisotropy energy. 
which is a functional of the  the magnetization vector field $\vec{m}$ which is for convenience usually   normalized to unity in magnitude. The magnetostatic energy is determined by the  demagnetizing field $H_d$, which arises from volume charge $\rho = - \mu_0 M_s (\nabla \cdot \vec{m})$ and surface charge $\sigma_s = \mu_0 M_s (\vec{m} \cdot \vec{n})$ (here $M_s$ is the saturation magnetization, $\mathbf{n}$ is the surface normal,  and $\mu_0$ is the permeability of the vacuum). The demagnetizing field is determined by\cite{hillebrands2003spin, DanielDStancilspinwave}
\begin{equation}
	\begin{aligned}
		\displaystyle \vec H_d(\vec{r}) = - \nabla\Psi_d(\vec{r}), \\
		\Psi_d(\vec{r}) = \frac{1}{4\pi\mu_0}[\int\frac{\rho(\vec{r'})}{|\vec{r}-\vec{r'}|}d^3\vec{r'} + \int\frac{\sigma_s(\vec{r'})}{|\vec{r}-\vec{r'}|}d^2\vec{r'}].
		\label{demag}
	\end{aligned}
\end{equation}

The equilibrium magnetization is obtained by minimizing the functional   $\delta E_m [\mathbf{m}]= 0$ which can be also  expressed as a torque equation $\vec{m} \times  \vec{H}_{\rm{eff}} = 0$.  The effective field reads 
\begin{equation}
	\displaystyle \vec{H}_{\rm{eff}} = -\frac{1}{\mu_0 M_s} \frac{\delta E_m}{\delta \vec{m}} = \frac{2 A_{ex}}{\mu_0 M_s}\nabla^2 \vec{m} + \vec{H}_a + \vec{H}_d -\frac{1}{\mu_0 M_s}\frac{\delta E_K}{\delta \vec{m}}.
	\label{Heff}
\end{equation}

For clarity we employ  uniform magnetic field $\vec{H_a} = H_0 \vec{y} $ applied along a direction which we define as the $+y$ direction. Contributions from demagnetizing field and anisotropy field are neglected for simplicity (they will be added and discussed in the following sections). Magnon or spin wave (SW) excitations are  described by small deviations $ \delta m_{x} = m_{x,s} e^{i(k r - \omega t)}$ and $ \delta m_{z} = m_{z,s} e^{i(k r - \omega t)}$ from the equilibrium state ($\vec{m}_0 = (0, 1, 0)$).  Inserting our ansatz in LLG equation (\ref{LLG}) and seeking  an equation of motion which is linear in  $\delta m_{x}, \delta m_{z} $, we obtain
\begin{equation}
	\begin{aligned}  
		\omega m_{x,s} + i \alpha (\omega_H + \omega_{ex} k^2) m_{x,s} - i (\omega_H + \omega_{ex} k^2) m_{z,s} &= 0, \\
		\omega m_{z,s} + i \alpha (\omega_H + \omega_{ex} k^2) m_{z,s} + i (\omega_H + \omega_{ex} k^2) m_{x,s} &= 0.
		\label{equationsw}
	\end{aligned}              
\end{equation}
The determining frequencies are  $ \omega_H = \frac{\gamma H_0}{1 + \alpha^2} $, 
and $ \omega_{ex} = \frac{2 \gamma A_{ex}}{(1+\alpha^2) \mu_0 M_s} $. Introducing $\psi = m_{x,s} + i m_{z,s}$, the equation becomes,
\begin{equation}
	\begin{aligned}  
		\omega \psi - (1-i\alpha)(\omega_H + \omega_{ex} k^2)  \psi  = 0.
		\label{equationswp}
	\end{aligned}              
\end{equation}
The intrinsic frequency of the magnetic excitation is obtained by $\omega = (1-i\alpha)(\omega_H + \omega_{ex} k^2)$.  the negative imaginary component describes the attenuation  of the magnetic excitation, which is determined by the $\alpha$ term. Its real component $\omega_H + \omega_{ex} k^2$ represents the dispersion relation above the cut-off frequency $ \omega_H $. Such parabolic dispersion relation is typical for   exchange-interaction dominated  SW and hence such magnons are  called exchange spin wave, or magnons.

For extended  ferromagnetic medium, the dipolar interaction can be captured  by the Herrings-Kittle formula,\cite{PhysRev.81.869}
\begin{equation}
	\begin{aligned}  
		\omega = \sqrt{(\omega_H + \omega_{ex} k^2)(\omega_H + \omega_{ex} k^2 + \omega_M \sin^2\theta_k)}.
		\label{equationsd}
	\end{aligned}              
\end{equation}
We introduced the definitions  $\omega_M = \gamma M_s$, and $\theta_k$ as the angle between the wave vector and the magnetization. In the limit of long wavelength, i.e. small $k$, the effect of exchange  is weak  and the excitation frequency $\omega = \sqrt{\omega_0(\omega_0 + \omega_M \sin^2\theta_k)}$  is dominated by the dipolar interactions and is anisotroic with propagation direction (an example is shown in Fig.\ref{dipolarmodel}).


\subsection{Spin waves in confined geometries}

\begin{figure}[htbp]
	\includegraphics[width=0.9\textwidth]{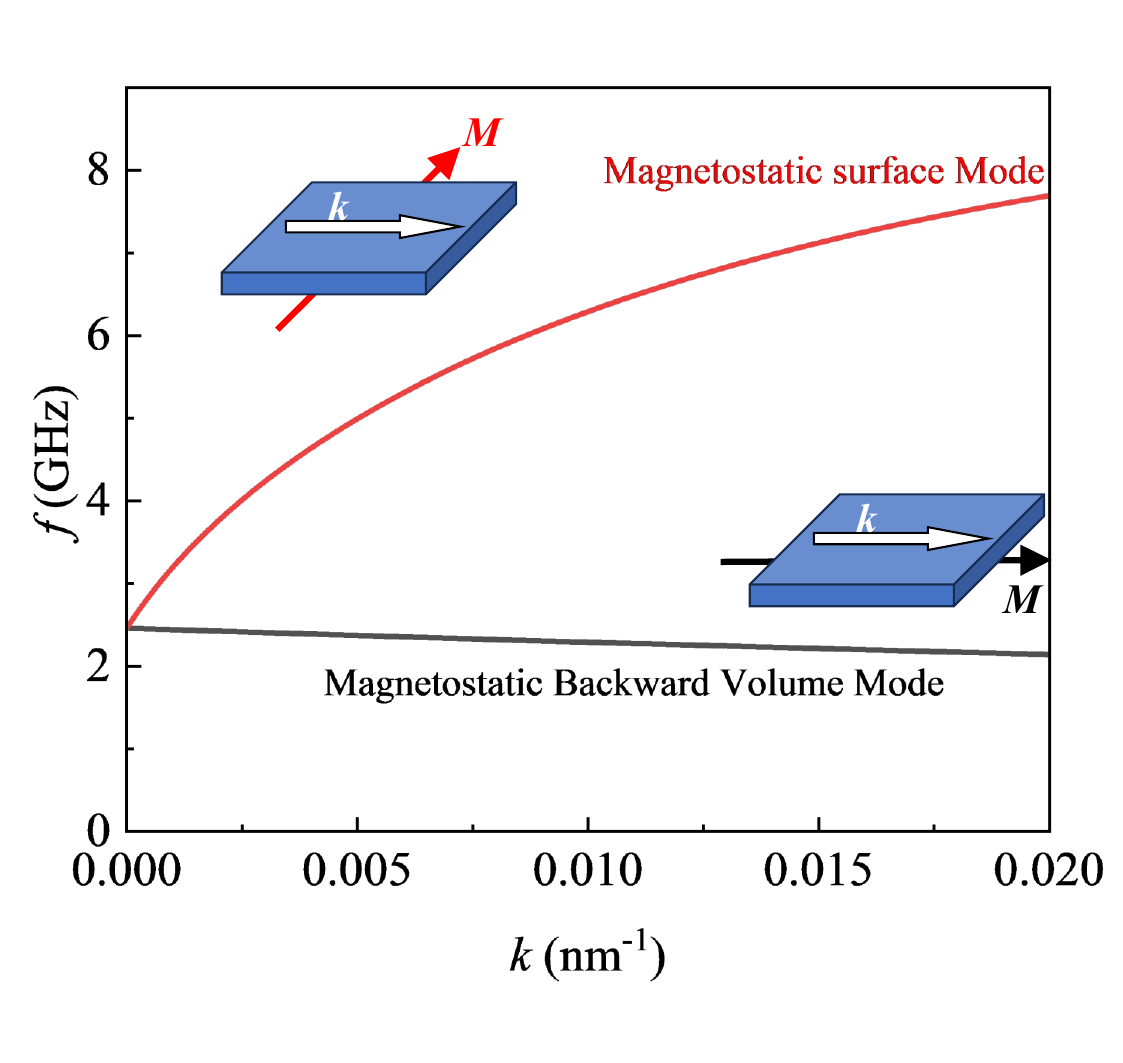}
	\caption{\label{dipolarmodel} Typical dispersion of   dipolar surface spin waves (wave vector $\mathbf{k}$ perpendicular  to magnetization $\mathbf{M}$), and backward volume spin waves (wave vector $\mathbf{k}$ parallel to magnetization $\mathbf{M}$) in a thin ferromagnetic stripe. }
\end{figure}

The properties of magnetostatic spin waves are strongly affected by the geometry and the dominant interaction inside the material. At first, we consider the properties of spin waves in ferromagnetic thin films confined only in  thickness, i.e. in the $y$ direction. Such a case is also encountered  for thin strip  with a thickness $d$ being much smaller than the width $w$. The magnetization is saturated in-plane along a magnetic  field $\vec{H_a} = H_0 \vec{z} $ that we apply along  the $ z $ direction. 
In this case, the film boundaries generate magnetic surface charges and  affect the dipolar field. The dipolar field can be approximated by \cite{BAKalinikos1986jpc}
\begin{equation}
	\begin{aligned}  
		\vec{H}_d = -[f(kd) \frac{\vec{k} \cdot \vec{m}}{k^2} \vec{k} + (1-f(kd)(\vec{n}\cdot \vec{m})\vec{n})].
		\label{dipolarfilm}
	\end{aligned}              
\end{equation}
Here, $f(kd) = 1 - (1 - e^{-kd})/kd$ with $k^2 = k_x^2 + k_z^2$.  Inserting  the film dipolar field into the linearized LLG equation yields  the spin wave dispersion relation 
\begin{equation}
	\begin{aligned}  
		\omega = \sqrt{(\omega_H + \omega_{ex} k^2)(\omega_H + \omega_{ex} k^2 + \omega_M F_m)},\\
		F_m = 1 - f(kd) \cos^2(\theta) + \frac{\omega_M f(kd) (1-f(kd)) \sin^2(\theta)}{\omega_H + \omega_{ex} k^2}.
		\label{dispersionfilm}
	\end{aligned}              
\end{equation}
Here, $\theta$ is the angle between the  static magnetization and the wavevector. In the exchange limit with large wavevectors, the dispersion relation is reduced to  $\omega = \omega_H + \omega_{ex} k^2$ for unrestricted system. For the dipolar limit with small $ k $ with the wavevector being  parallel to the equilibrium magnetization ($ \theta = 0 $), the dispersion relation $ \omega = \sqrt{\omega_H [\omega_H + \omega_M (1 - f(kd))]} $ corresponds to  a typical backward volume magnon mode (Fig. \ref{dipolarmodel}), where the spin wave frequency decreases with $k$.  The backward volume wave group velocity is antiparallel to the wavevector and phase velocity. On the other hand,  for $ \theta = \pi/2 $ we have  magnetostatic surface magnon with an amplitude decaying exponentially along the thickness direction, and its dispersion relation is $ \omega = \sqrt{\omega_H (\omega_H + \omega_M) + \omega_M^2 f(kd) (1 - f(kd))} $, see Fig. \ref{dipolarmodel}. The surface spin wave has both positive group velocity and phase velocity.\cite{Kalinikos1980ieee}. M. Mohseni et al. studied the propagation of surface magnon in  thin film with defects, and the dipolar field symmetry breaking induced backscattering-immune feature is identified, uncovering the robustness and chiral features of magnetostatic surface magnon.\cite{PhysRevLett.122.197201}

For a system confined in two directions (for example $y$ and $z$ directions), the demagnetization tensor describes shape effects induced by dipolar fields. The scenario is especially suitable for the confined nanostripe. Considering the shape function $D(\vec{r})$ (equal to 1 inside the magnet and 0 outside it), the magnetization field is expressed as $ \vec{M}(\vec{r}) = M_s D(\vec{r}) \vec{m}(\vec{r}) $. The Fourier transform of $ D(\vec{r}) $ can be written as,\cite{jmmmBELEGGIA2004270}
\begin{equation}
	\begin{aligned}  
		D(\vec{k}) = \int D(\vec{r}) e^{-i \vec{k} \cdot \vec{r}} d\vec{r}. 
		\label{shapek}
	\end{aligned}              
\end{equation}
By computing the inverse Fourier transform, the demagnetization field is obtained as,
\begin{equation}
	\begin{aligned}  
		\vec{H}_d = -\frac{M_s}{8 \pi^3} \int d^3 \vec{k} \frac{D(\vec{k})}{k^2}\vec{k}(\vec{m} \cdot \vec(k)) e^{i \vec{k}\cdot \vec{r}}. 
		\label{demagshape}
	\end{aligned}              
\end{equation}
Via a point-function demagnetization tensor $N_{ij}$, the demagnetization tensor field is defined by $\vec{H}_{d,i} = - N_{ij} M_j$ with,
\begin{equation}
	\begin{aligned}  
		N_{ij}(\vec{k}) = \frac{D(\vec{k})}{k^2}k_i k_j. 
		\label{dmgfactor}
	\end{aligned}              
\end{equation}
The real space factor $N_{ij}(\vec{r})$ can be obtained by  inverse Fourier transformation. By integrating over the $k_z$ along $z$ axis with constant height $d$, the tensor $ \vec{N}(\vec{k}) $ is obtained as,\cite{PhysRevB85014427, sciadv1701517, jmmmBELEGGIA2004270}
\begin{equation}
	\begin{small}
		\begin{aligned} 
			\displaystyle \vec{N}((\vec{k}) = \frac{|\sigma_k|^2}{w} \left( \begin{matrix} \frac{k_x^2}{k^2}f(kd) & \frac{k_x k_y}{k^2}f(kd) & 0 \\ \frac{k_x k_y}{k^2}f(kd) & \frac{k_y^2}{k^2}f(kd)  & 0 \\0 & 0 & 1-f(kd) \end{matrix} \right).
			\label{dmgfactork}
		\end{aligned} 
	\end{small}
\end{equation}
Here, $ \sigma_k = \int_{-w/2}^{w/2} e^{-ik_y y} dy$ is the Fourier transform of the magnet shape across the width $w$ of the system in the $ y $ direction. Then, the demagnetization tensor can be determined using the Fourier-space approach $\vec{N}(\vec{r}) = \frac{1}{2\pi}\int \vec{N}((\vec{k}) e^{i \vec{k} \cdot \vec{r}} d \vec{k}$ \cite{PhysRevB85014427, sciadv1701517}. Using the equilibrium magnetization along $+y$ direction as an example, and submitting the demagnetization field into the linearized LLG equation (\ref{LLG}) delivers the spin-wave dispersion relations 
\begin{equation}
	\begin{aligned}  
		\omega = \sqrt{(\omega_H + \omega_{ex} k^2 + \omega_M N_{xx})(\omega_H + \omega_{ex} k^2 + \omega_M N_{zz})}.
		\label{dispersionconfined}
	\end{aligned}              
\end{equation}

\subsection{Magnons in hybrid materials}

\begin{figure}[htbp]
	\includegraphics[width=0.9\textwidth]{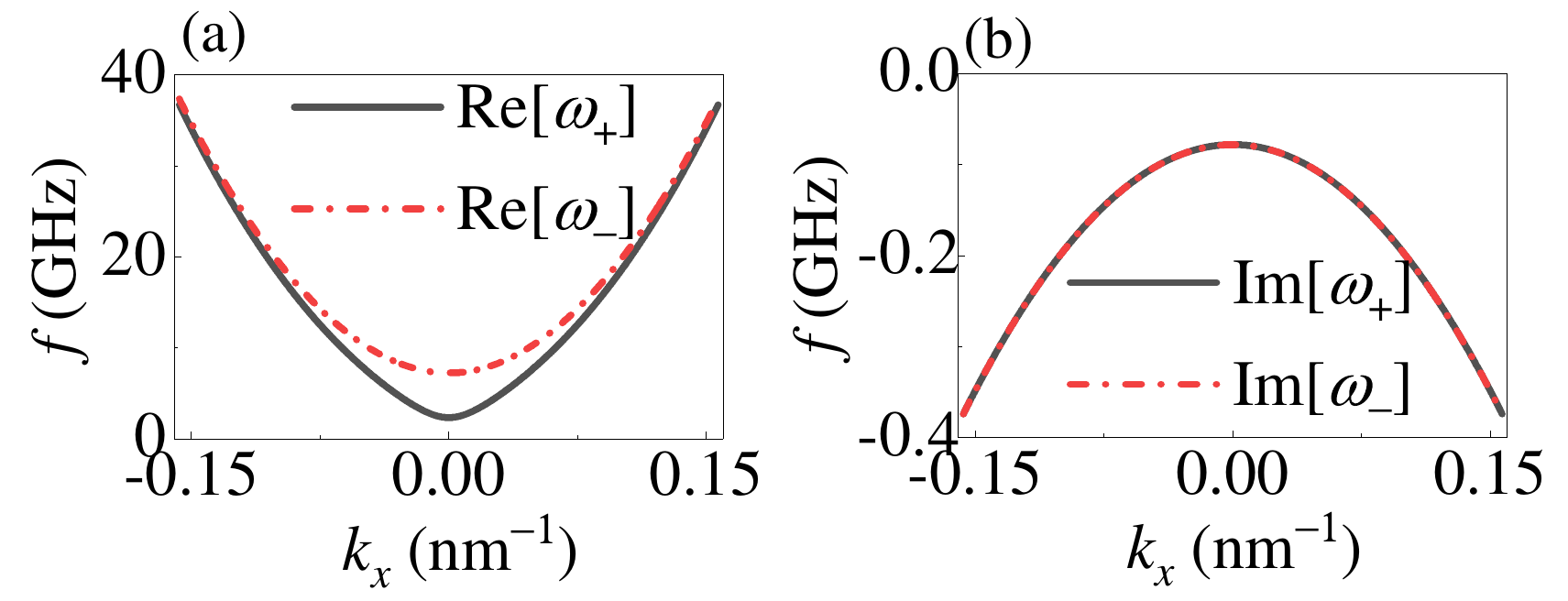}
	\caption{\label{SAFmagnon} In the synthetic antiferromagnet (c.f. model of Fig. \ref{pseudomodels}(c)), (a) real and (b) imaginary parts of the two magnon frequencies as functions of the wavevector $k_x$.\cite{apl50029523} }
\end{figure}

\begin{figure}[htbp]
	\includegraphics[width=0.9\textwidth]{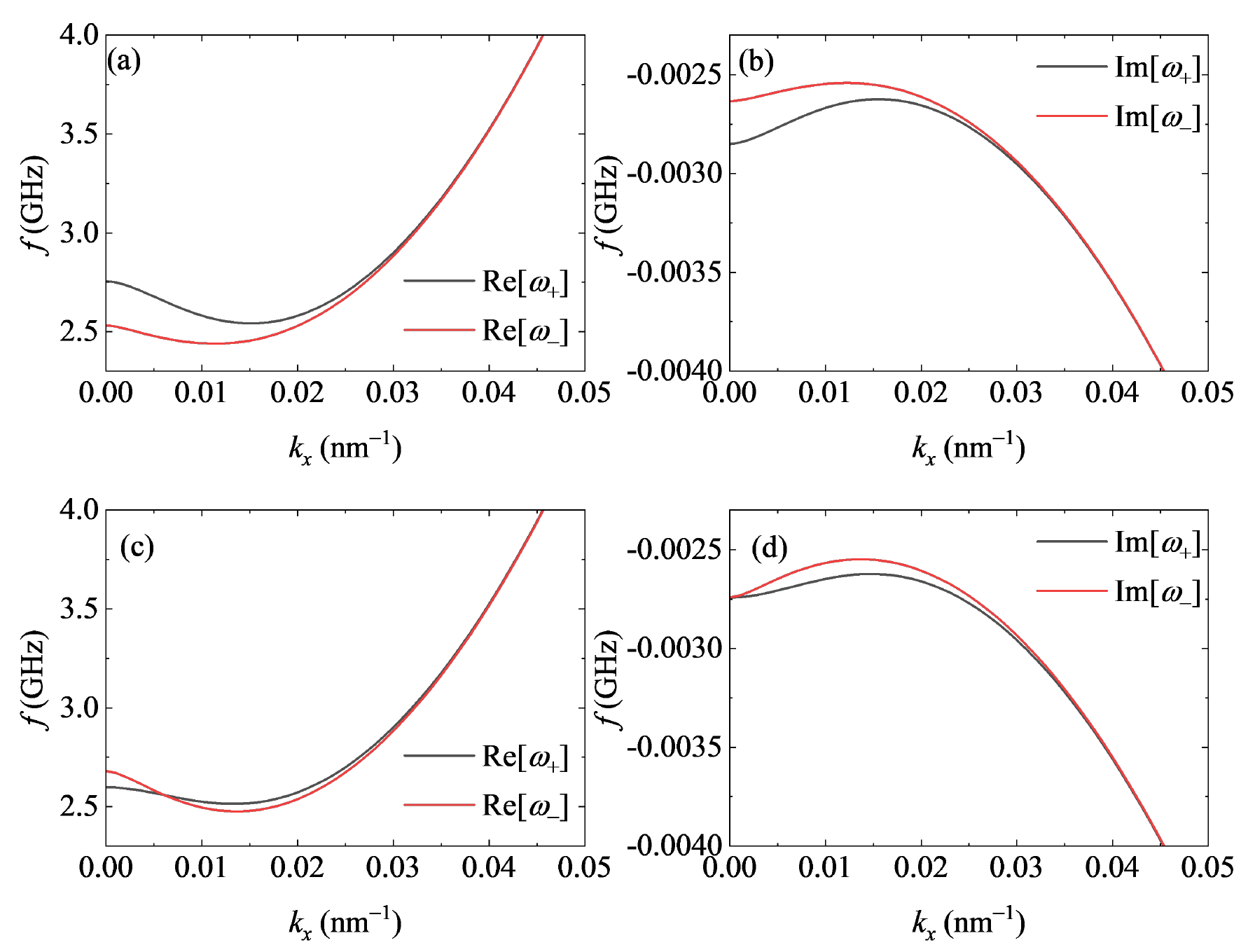}
	\caption{\label{dipolarcouplemagnon} For dipolarly coupled waveguides with antiparallel and parallel magnetization (c.f. model of Fig. \ref{pseudomodels}(d)), (a) real and (b) imaginary parts of the two magnon frequencies as functions of the wavevector $k_x$.\cite{PhysRevApplied.18.024080}}
\end{figure}
In the introduction we discussed the importance of the coupling between the two systems $S_1$ and $S_2$ with respect to the unfolding dynamics upon particular excitations. Considering $S_1$ and $S_2$ as magnonic subsystems we need to specify how this coupling comes about and how to incorporate it in the theory in a realistic way such that the predictions becomes experimentally relevant. \\
The coupling mechanism depends essentially on the setup. For example,  one may  consider two stripes, such as the ones shown in Fig.(\ref{dipolarmodel}),
to be grown in such a way that a nanometer-thick metallic layer is sandwiched between them.  In this case the coupling between the two layers is caused by
 the Ruderman-Kittel-Kasuya-Yosida (RKKY) interaction. The RKKY interaction  is rooted in the spin sensitivity of scattering of  the spacer’s itinerant electron  at the interface with the two magnetic layers, in particular on the direction of the magnetization of the layers. This  causes a dependence  the formed standing electron wave  on the magnetic states of the magnetic stripes  and eventually an effective  coupling between them.\cite{PhysRev9699, PhysRev106893,PBruno1999jphysccondens, PhysRevB447131} The interlayer RKKY coupling is much weaker than the intralayer exchange coupling of the magnet, but it has the advantage that it can be manipulated in a versatile way, which bears  obvious advantages for practical application. The coupled magnon modes can be tuned by engineering the interlayer coupling. The RKKY interaction affects the preferred relative orientation of the magnetic moments in coupled magnets under the ground state, e.g., parallel for ferromagnetic coupling or antiparallel for antiferromagnetic coupling.  The nature of the ground state directly determines the staring point for magnons. Besides, the additional RKKY coupling energy directly alters the magnon dispersion relation, and the contribution is usually isotropic.\\
 Another type of coupling is of  dipolar nature. The classical dipolar coupling stems  from the demagnetization field.  Such dipolar interaction is  present even across vacuum/insulators, long-range, highly anisotropic, and plays a  crucial role for shaping the long-wavelength and directional properties of magnons. Via material engineering and structuring it can be controlled to a certain extent.\\
Although both RKKY  and dipolar interactions can be present and contribute, the for flat stripes and nanometer thick spacer layer  the  RKKY interaction often dominates. Therefore, we  neglect at first he dipolar interaction and  analyze the influence of RKKY interaction. For two thin magnetic layers coupled via ferromagnetic RKKY coupling, the effective coupling field $$\vec{H}_{c,p} = \frac{J_{RKKY}}{\mu_0 M_s t_p} \vec{m}_{p'}$$ prefers  parallel magnetization directions ($\vec{m}_1 = \vec{m}_2 $), dependent on the thickness  of spacer and $t_p$,  the thickness of magnetic layer, $J_{RKKY} > 0$ is the interlayer exchange coupling strength, $p = 1,2$ enumerates two magnetic layers, and $p' \ne p  $. Under an uniform magnetic field $\vec{H_a} = H_0 \vec{y} $ applied along $+y$ direction, the equilibrium magnetization $\vec{m}_{0,p}= (0, 1, 0) $. Again we seek a wave solution for the small magnon deviations  
$$ \vec{m}_{s,p} = (\delta m_{x,p}, 0, \delta m_{z,p}) e^{i(k r - \omega t)}.$$ By linearzing the LLG Eq. (\ref{LLG}) and introducing  $ \psi_p = \delta m_{x,p} + i \delta m_{z,p} $, the magnon equation is expressed as
 \begin{equation}
	\begin{aligned}  
		\omega \psi_1 - (1 - i\alpha)(\omega_H + \omega_{ex} k^2) \psi_1 + (1 - i\alpha)q \psi_2  &= 0, \\
		\omega \psi_2 - (1 - i\alpha)(\omega_H + \omega_{ex} k^2) \psi_2 + (1 - i\alpha)q \psi_1  &= 0.
		\label{coupleequationfm}
	\end{aligned}              
\end{equation}
The coupling strength is given by $$q = \frac{\gamma J_{RKKY}}{(1+\alpha^2)\mu_0 M_s t_p}.$$
The above equation follow then from the eigenvalue problem $$ \omega \vec{\psi} =  H _m \vec{\psi} $$ with $ \vec{\psi} = (\psi_1, \psi_2). $ The $2 \times 2$ with the non-Hermitian  Hamiltonian having the form,
 \begin{equation}
	\begin{aligned}  
	 	  H _m = (1 - i\alpha) \left( \begin{matrix} \omega_H + \omega_{ex} k^2 & -q \\ -q & \omega_H + \omega_{ex} k^2 \end{matrix} \right).
		\label{hamfm}              
   \end{aligned}  
\end{equation}
The eigenvalues of magnon modes reads
 \begin{equation}
	\begin{aligned}  
		\omega_{\pm} = (1 - i\alpha)(\omega_H + \omega_{ex} k^2 \pm q), 
		\label{dispersionfm}              
	\end{aligned}  
\end{equation}
which also represents the magnon dispersion relations. Here, two magnon modes are optical and acoustic modes separated by the coupling $q$, along the lines discussed in the introduction.
 The separation is always $ \omega_+ - \omega_- = 2(1-i\alpha) q $ independent of wavevector.\cite{PhysRevLett.63.1645}

For an antiferromagnetic coupling $ J_{RKKY} < 0 $, neighboring magnetic layers have opposite magnetization, and such heterostructures is referred to as synthetic antiferromagnets (SAFS).\cite{PhysRevLett.125.017203,Duine2018naturephys,Jungwirth2018naturephys} Here, we consider two coupled magnetic layers with easy-axis along $y$ (with anisotropy constant $K_y$) and in-plane anisotropy in $x$-$y$ plane (with anisotropy constant $K_z$). The total effective field is $ \vec{H}_{\rm{eff},p} = \frac{2 A_{\mathrm{ex}}}{\mu_0 M_s} \nabla^2 \vec{m}_p - \frac{J_{\rm{AF}}}{\mu_0 M_s t_p} \vec{m}_{p'} + \frac{2 K_y}{\mu_0 M_s} m_{p,y} \vec{y} - \frac{2 K_z}{\mu_0 M_s} m_{p,z} \vec{z} $. In this case, the stable equilibrium magnetization is $\vec{m}_{0,1} = -\vec{m}_{0,2} = (0, 1, 0) $, and the magnon dispersion relation is,
 \begin{equation}
	\begin{aligned}  
		\omega_{\pm} = \sqrt{\omega_0^2 - (1+\alpha^2)(\omega_z \pm q)^2} - i \alpha \omega_0, 
		\label{dispersionafm}              
	\end{aligned}  
\end{equation}
In the above equation, we define $$ \omega_0 = \omega_{ex} k^2 + \omega_{Ky} + \omega_{Kz} $$ and $$\omega_{Ky,Kz} = \frac{\gamma K_{y,z}}{(1+\alpha^2)\mu_0 M_s}.$$
 Different from the above case with ferromagnetic coupling, the separation $ \omega_+ - \omega_- $ is not a constant anymore, and the imaginary parts of two modes are identical, as demonstrated by  Fig. \ref{SAFmagnon}.

Turning to  the dipolarly  coupled magnetic layers, which will be the planar waveguides for our magnons, we not the dipolar coupling strength is weaker and wave-vector dependent, for example in the case depicted in Fig.\ref{dipolarmodel}. For two nanostripes separated by the distance $d_s$, using the Fourier-space approach we obtain  $\vec{N}_d = \frac{1}{ 2 \pi}\int \vec{N}((\vec{k}) e^{i k_y d_s} d \vec{k}$, and the coupled spin-wave dispersion relations is,\cite{sciadv1701517,PhysRevApplied.18.024080,PhysRevApplied.12.034015}
\begin{equation}
	\begin{aligned}  
		\omega = \sqrt{(\omega_H + \omega_{ex} k^2 + \omega_M N_{xx} \pm \omega_M N_{d,xx})(\omega_H + \omega_{ex} k^2 + \omega_M N_{zz} \pm \omega_M N_{d,zz} )}.
		\label{dispersionconfineddipolar}
	\end{aligned}              
\end{equation}
Compared to the dipolar interaction inside each stripe, the coupling between two different stripes is smaller. Due to the nature of dipolar interaction, the coupling strength becomes weaker with the increase of wavevector and separation distance. As proved by the results in Fig. \ref{dipolarcouplemagnon}, the separation between two modes is larger in the lower wavevector range. Here, magnetization is in the same direction with magnon propagation, indicating the coupled backward volume magnon mode. The dispersion relations is dependent on the magnetization profile. Different from the case with antiparallel magnetization (Fig. \ref{dipolarcouplemagnon}(c)) in two coupled layers, for parallel magnetization a cross point appears of the  two dispersion curves at a low wavevector. 

Q. Wang et al. exploited the features of dipolarly coupled back volume magnons for the application in directional coupler.\cite{sciadv1701517,Wang2020natureelectron} The interference between two coupled magnon modes periodically transfer between two magnetic waveguides, and the transfer length is dependent on the separation between two dispersion curves, recalling the example of two coupled systems from the introduction we note that the propagation distance plays the role of $t$ (the time in the Schrödinger-type equation) a case which also applies to optical waveguide particularly in the paraxial limit.
 Applying external magnetic field, changing the distance between two waveguides, or switching between parallel and antiparallel magnetization, one  can reconfigure the magnon transfer enabling so as dynamically reconfigureable signal directional coupler. Also, the coupled magnetostatic  surface magnon modes can serve for  similar periodic transfer. Chiral features of magnetostatic surface magnon can be a source for  nonreciprocal dispersion relations and transfer length. By applying  electric field or varying equilibrium magnetization, it is possible to control the magnon propagation in the selected waeveguide.\cite{PhysRevApplied.12.034015}

\subsection{Current-controlled magnonic damping and antidamping}
Magnonic damping describe the decay rate of magnon amplitude in time or distance. Since magnons  excitations in our regime of interet is basically
precessional, this damping means a loss of angular momentum. Hence, immediate causes for this damping are external torques. This observation will have an important consequences for the discussion below. While environmental coupling often views as loss of energy and the system is coupling to an external bath of harmonic oscillators, for example, the torques as vectors have a direction and hence allow for a directional damping. \\
  In  magnonic devices generally, low damping is preferred  to achieve  stronger magnon excitation and longer magnon propagation distance. The description of magnon damping is often captured by the $ \alpha $ term in the LLG Eq. (\ref{LLG}).\cite{Gilbert1353448} In this framework, the damping constant $ \alpha $ phenomenologically describe  the magnetization relaxation  towards its equilibrium direction without specifying the underlying mechanisms, such as magnon-phonon coupling, multi-magnon scattering processes (e.g. three-magnon scattering), impurities et. al.\cite{chen10.10635.0232608,PhysRevB.76.104416,PhysRevB.61.9553,Bozhko10.106310.0000872,PhysRevB.103.L220403,PhysRevLett.102.257602,PhysRevLett.99.027204,Heinrich1042236} Materials like the ferromagnetic insulator Yttrium Iron Garnet (YIG) are highly sought for due to their intrinsically low Gilbert damping, which can be ranged from $10^{-5}$ to $ 10^{-2} $ depending on the sample quality.\cite{PhysRevMaterials.4.024416, Hauser2016scirep,DingJinjun9076316,Onbasli10.10631.4896936,Mitra2017SciRep,PhysRevB.98.224422,YAMADA2020167253,PhysRevB.106.054405,PhysRevLett.107.066604}

Recent developments uncover the ability to actively control magnonic damping directionally  using electrical currents, primarily through spin torques. This allows for tuning magnon lifetimes, compensating for loss (antidaming), or even driving auto-oscillations.\cite{Collet2016naturecommun,PhysRevLett.113.197203, PhysRevB.68.024404,Locatelli2014,Moriyama10.10631.4918990,PhysRevLett.106.036601,Fulara10.1126sciadv.aax8467,Demidov2010NatureMater}

In structures where a spin-polarized electrical current passes through or is injected into a magnetic layer, angular momentum can be transferred to the local magnetization, generating a spin transfer torque (STT).\cite{RALPH20081190,SLONCZEWSKI1996L1,PhysRevB.54.9353} This STT can act either as an effective damping (enhancing energy dissipation) or antidamping (reducing effective damping), depending on the relative orientation of the spin polarization and the magnetization.\cite{PhysRevB.68.024404,Urazhdin2014Nanotechnology} The STT usually requires conductive material, and can not apply on the insulator (for example, YIG). Aside from the STT, in bilayers comprising a heavy metal (HM) or topological insulator (TI) and a ferromagnet, the in-plane charge current in the HM or TI layer can generate a pure spin current via the spin Hall effect or Rashba-Edelstein effect, which flows into the adjacent ferromagnetic layer and exerts a torque on its magnetization.\cite{PhysRevLett.83.1834, Garello2013, EDELSTEIN1990233, Brataas2012naturemater,Hoffmann6516040, Garello2013,10.1126science.1218197,PhysRevB.87.174411,Snchez2013naturecommun}  The torque arises from spin-orbit interaction effects, and is called  spin orbit torque (SOT).  SOT generally acts in two ways, as a  damping-like torque and a field-like torque. The damping like SOT can directly modify the effective magnonic damping, providing a mechanism to electrically manipulate magnon density  loss or gain, and the effect is dependent on the electric current direction, geometry and magnetization. Different from the STT, the SOT can effectively act on  insulating magnets. By providing sufficient antidamping torque, SOT can compensate for the intrinsic damping, enabling sustained auto-oscillations in spin Hall nano-oscillators or efficiently switching magnetization.\cite{Chen7505988,Haidar2019naturecommun,Awad2017NaturePhys,Demidov2012naturemater,Divinskiyadma.201802837,10.1126science.1218197,Miron2011nature, PhysRevLett.109.096602,Garello2013} STT and SOT allow for dynamic manipulation of magnon properties by applying electric current, enabling functionalities like signal amplification, modulation, and non-reciprocal propagation in magnonic circuits.\cite{Demidov2016naturecommun,PhysRevLett.102.147202,Padron10.10631.3660586,Divinskiyadma.201802837,PhysRevB.104.134422,PhysRevLett.132.036701}

Our particular focus is on functionalizing SOT for pseudo-Hermitian magnon physics.   This is achieved by exploiting the directionality of the  damping like SOT. When the charge current $\vec{J}_c$ flows through the HM layer, the spin Hall effect deflects electrons with opposite spins in opposite direction transverse to  $\vec{J}_c$, generating a spin current $\vec{J}_s$ with spin polarization $\vec{\sigma}$ perpendicular to both $\vec{J}_c$ and the interface normal. This injected spin current acts on the local magnetization $\vec{m}$ in the attached magnetic layer, transferring spin angular momentum and exerting the SOT $$\vec{T} =  \tau (\vec{m} \times (\vec{\sigma} \times \vec{m}))$$ on it.\cite{Hoffmann6516040, Garello2013} Here, unit vector $ \vec{\sigma} $ represents the spin polarization of the injected spin current. For a current $\vec{J}_c$ along $ x $ in an HM layer attached to the magnetic layer (interface normal along $ z $), the spin current flows along $ z $ with polarization $ \vec{\sigma} $ typically along $ y $, i.e. $ \vec{\sigma} = \vec{z} \times \vec{j}_c $, where the unit vector of electric current $ \vec{j}_c = \vec{J}_c / |J_c| $.\cite{PhysRevLett.83.1834} The SOT amplitude $$ \tau = \frac {\theta_{SH} J_c \gamma \hbar }{2 e \mu_0 M_s t} $$ is related to the spin Hall angle $\theta_{SH}$. The SOT direction $ \vec{m} \times (\vec{\sigma} \times \vec{m}) $ is analogous to the Gilbert damping term, and thus acts as  an additional effective damping or antidamping. If $ \vec{T} $ counteract (enhance) the net damping, which is known as antidamping (damping) effect. By introducing the SOT with $ \vec{\sigma} = \vec{y} $ for the single magnetic layer, Eq. (\ref{equationswp}) becomes,
 \begin{equation}
	\begin{aligned}  
		\omega \psi - (1-i\alpha)(\omega_H + \omega_{ex} k^2 - i \omega_J)  \psi  = 0.
		\label{equationswpsot}
	\end{aligned}              
\end{equation}
The intrinsic frequency of the magnetic excitation becomes $$\omega = (1-i\alpha)(\omega_H + \omega_{ex} k^2 - i \omega_J)$$
 with $$ \omega_J = \frac{\tau}{1+\alpha^2}. $$
  The equation clearly evidences  that a positive $\omega_J$ increases the imaginary part of the magnon frequency, meaning  additional decrease in the magnon amplitude, and the $\alpha \omega_0$ term is the intrinsic loss due to  Gilbert damping. Reversing the direction of the electric current ($\omega_J < 0$), the SOT turns into an antidamping torque and plays the role of gain  enhancing the magnon amplitude. 


\section{Pseudo-Hermitian Magnonics}

\begin{figure}[htbp]
	\includegraphics[width=0.9\textwidth]{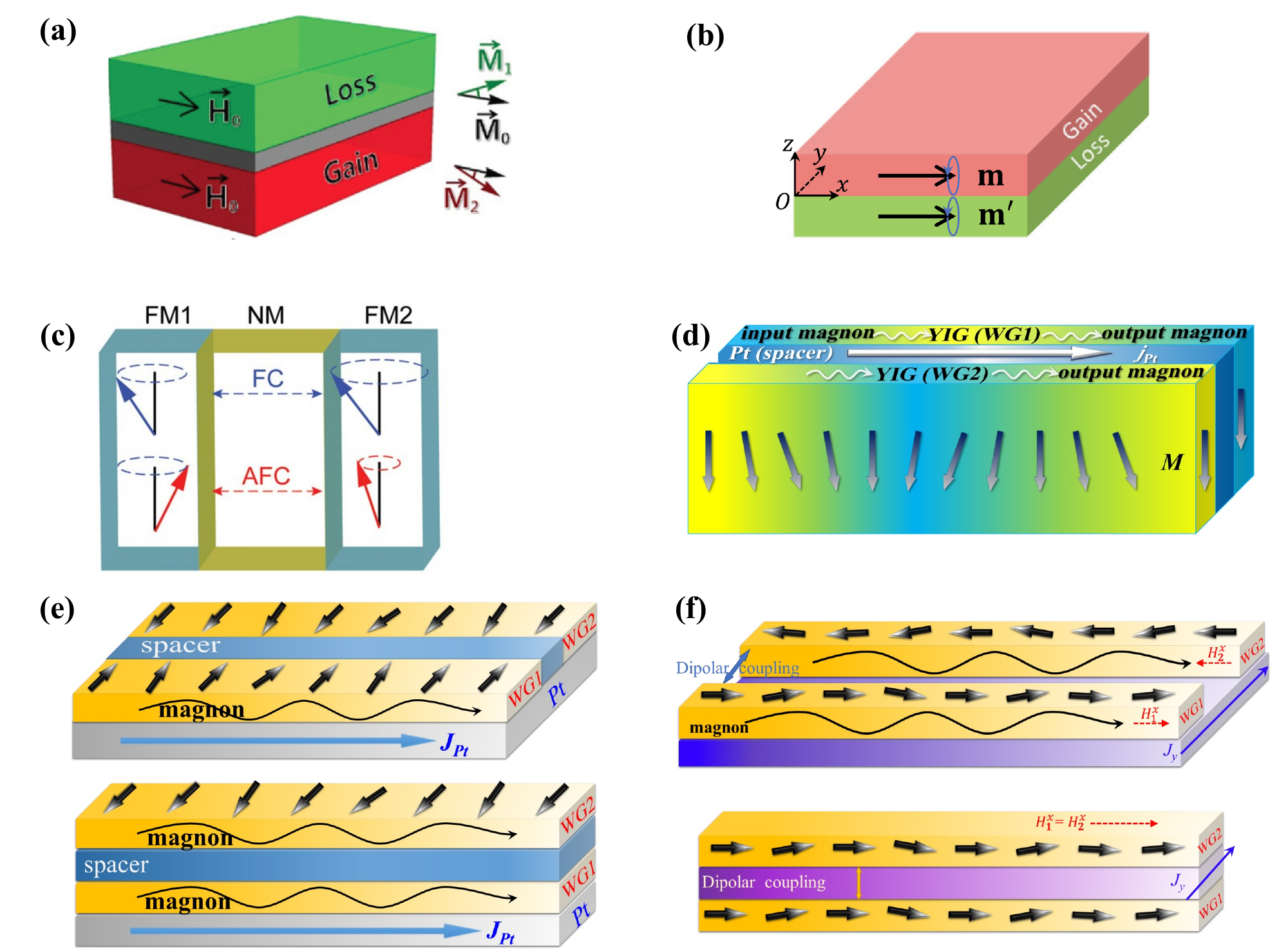}
	\caption{\label{pseudomodels} Several schematics for  model magnonic  pseudo-Hermitian setups. (a) Two coupled magnets with symmetric gain and loss in magnon amplitudes.\cite{PhysRevB.91.094416} (b)Coupled FM bilayers with balanced gain (red layer) and loss (green layer) with equilibrium magnetizations along the $x$ direction.\cite{PhysRevLett.121.197201} (c) Two coupled FM layers with different damping.\cite{Haoliangsciadveaax9144} (d) Two RKKY coupled magnonic waveguides. The charge current in the spacer layer results in spin-orbit torque(SOT)-induced  magnonic gain and loss in magnon density  in the waveguides.\cite{Wang2020nc} (e) Schematics for synthetic antiferromagnetic heterostructure with gain-loss, and loss-more loss.\cite{apl50029523} (f) Schematics of two dipolarly coupled magnonic waveguides WG1 and WG2 with SOT induced gain and loss.\cite{PhysRevApplied.18.024080} }
\end{figure}

\begin{figure}[htbp]
	\includegraphics[width=0.9\textwidth]{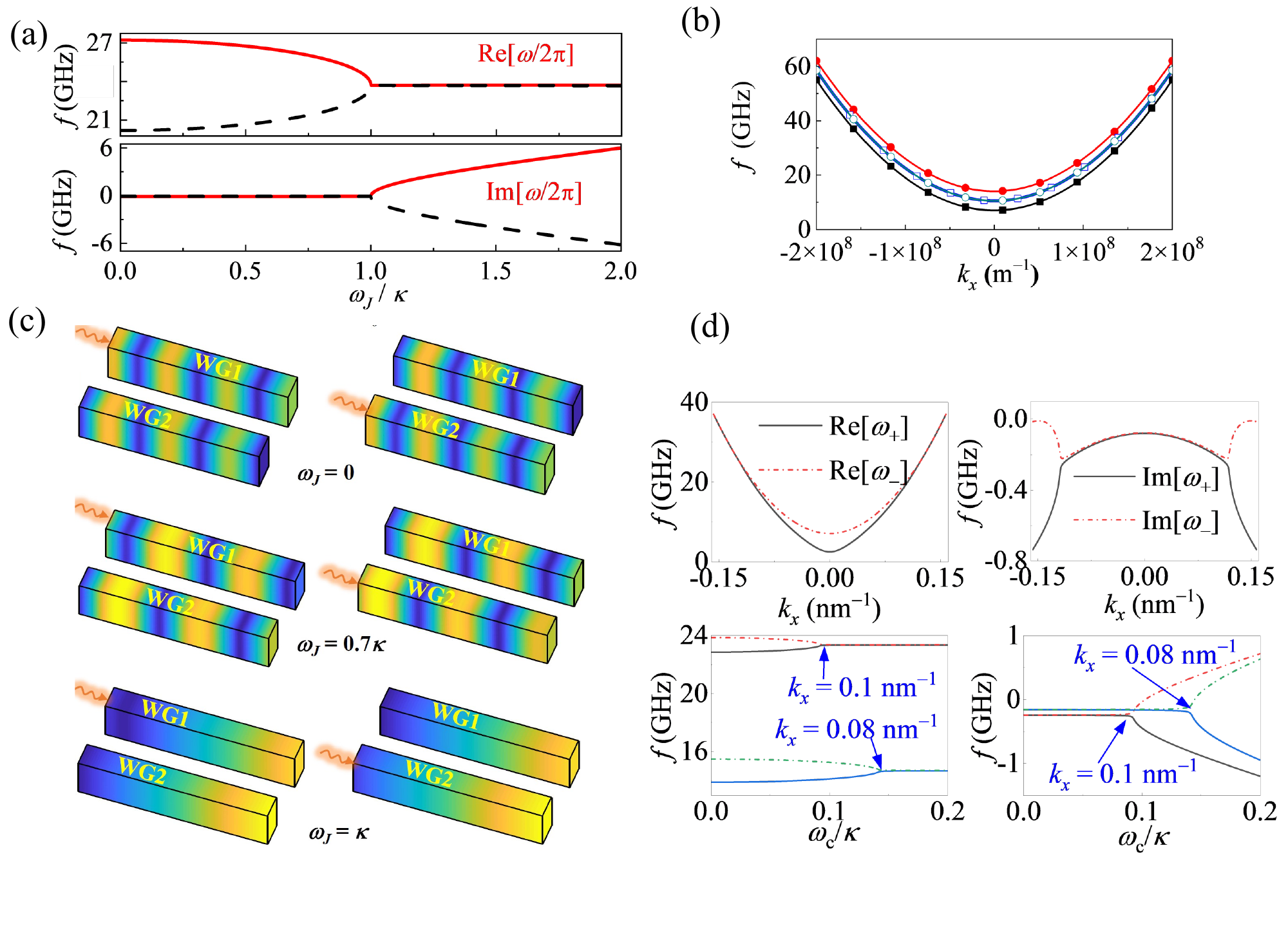}
		\includegraphics[width=0.8\textwidth]{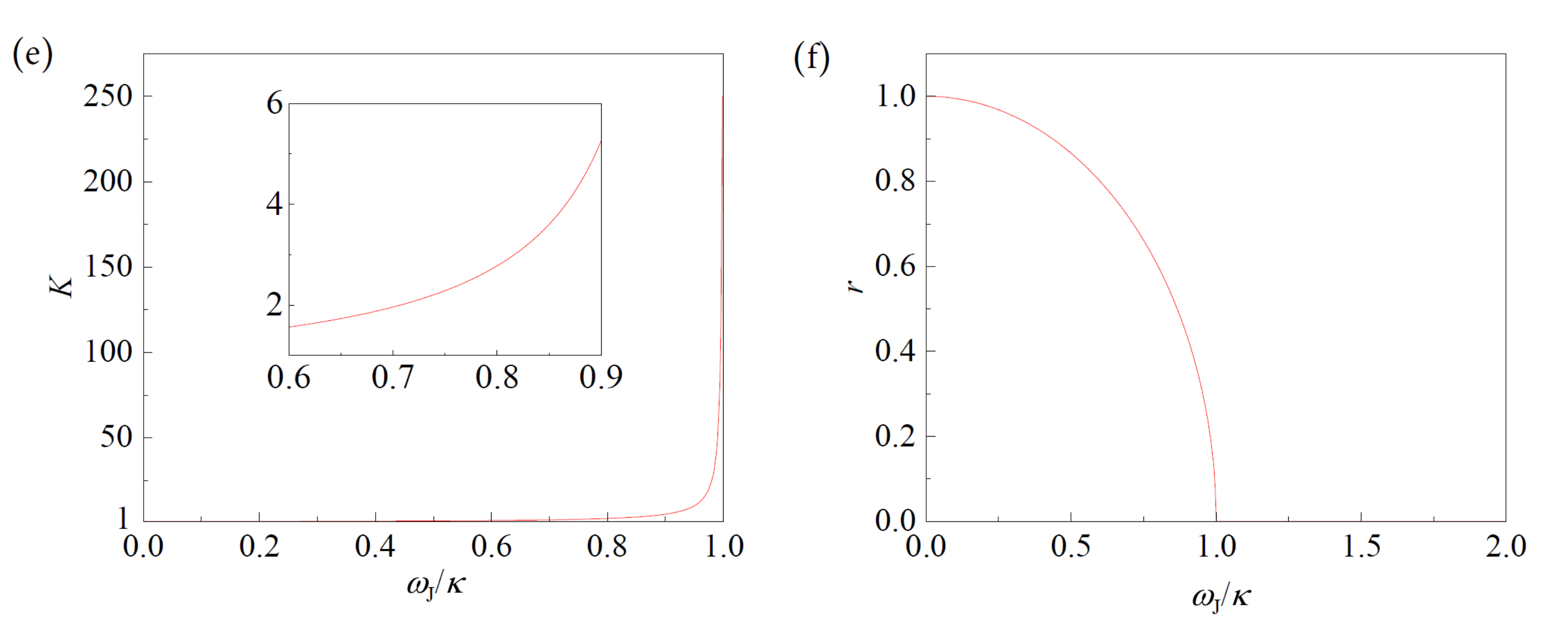}
	\caption{\label{pseudoep}  (a) For the system depicted in Fig. \ref{pseudomodels}(d), real and imaginary parts of the two magnon eigenfrequencies $f = \omega/(2\pi)$ as function of charge current density in the spacer $J_c\propto \omega$ relative to the interlayer coupling strenght $\kappa$.  (b) At $\omega_J = 0$ and EP, magnon dispersions of two magnon modes. (c) Spatial profiles of propagating magnon amplitudes for a different  strengths of the balanced magnonic loss/gain.\cite{Wang2020nc} (d) For antiferromagnetic heterostructure of  Fig. \ref{pseudomodels}(e), magnon dispersion relations under a finite $ \omega_J $, and real and imaginary parts of two magnon eigenfrequencies $f = \omega/(2\pi)$ as scanning $ \omega_J $.\cite{apl50029523}
	(e) Is the Petermann factor (Eq. (\ref{eq:petermann}) (inset shows the regime of small balanced magnonic gain/loss). (f) Is the phase rigidity (Eq. (\ref{eq:phaserigid})) corresponding to the modes in (a). } 
\end{figure}

\begin{figure}[htbp]
	\includegraphics[width=0.9\textwidth]{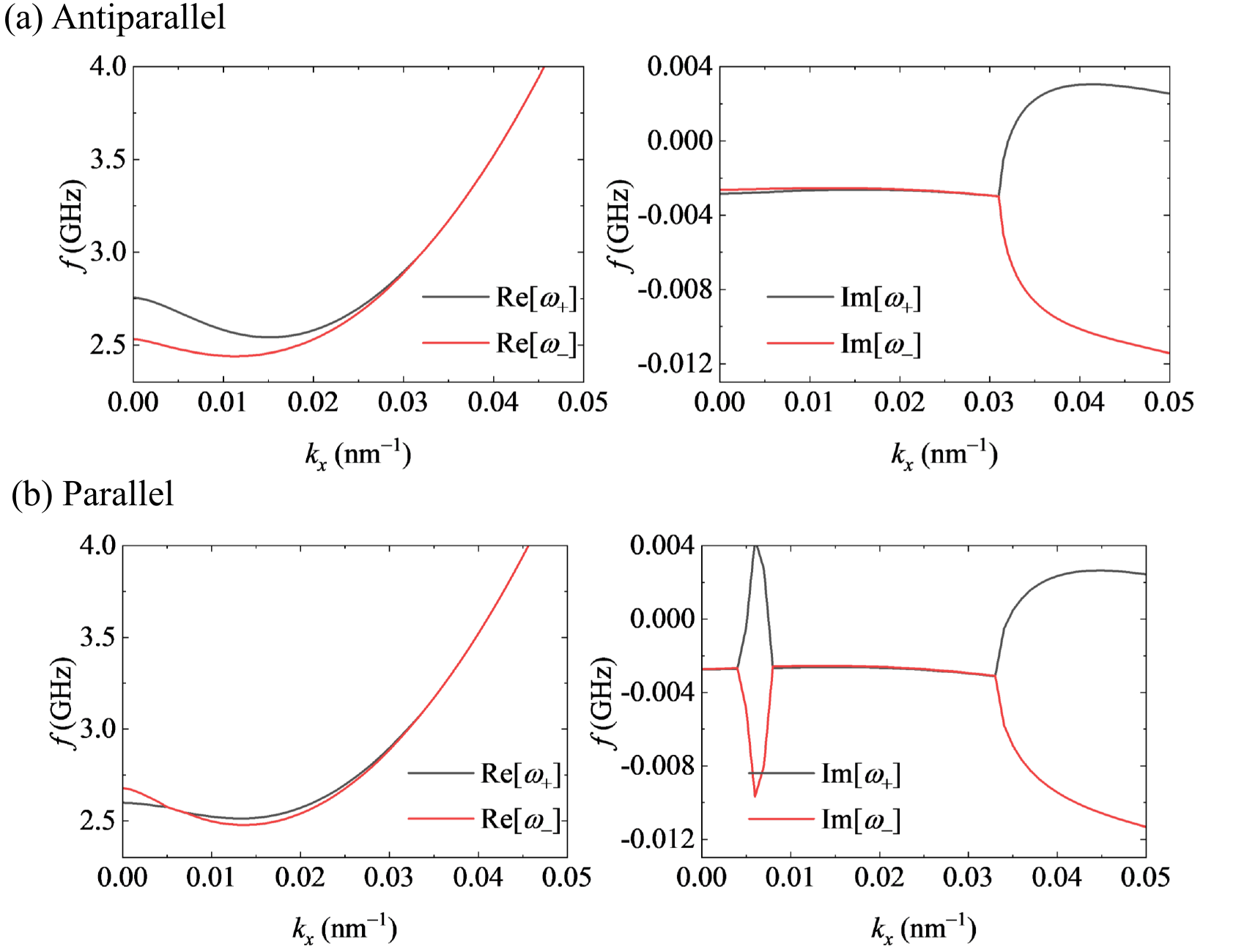}
	\caption{\label{dipolarep} For the setup shown  in Fig. \ref{pseudomodels}(f) with antiparallel and parallel magnetization in magnetic layers, the real (dispersion) and the imaginary parts of two magnon eigenfrequencies $f = \omega/(2 \pi)$.\cite{PhysRevApplied.18.024080}}
\end{figure}

Magnonic systems present a special platform for investigating pseudo-Hermitian physics: Magnetic materials are inherently coupled to intrinsic damping processes, such as Gilbert damping, which represent a natural loss mechanism. This inherent loss means that magnonic systems are fundamentally open and non-Hermitian systems. Furthermore, 
As clear from the equation of motion (\ref{LLG}), the dynamics can be highly nonlinear ($\mathbf{H}_{eff}$ depends on sought after solution $\mathbf{m}$). This, one can study nonlinear pseudo-Hermitian physics in a realistic systems. Importantly, the non-linearity can be tuned by external electric and magnetic fields as well as the degree of damping and antidamping can be controlled by the injecting charge current density. Additionally, as discussed in details below, the possibility to grow experimentally multilayers also with alternating magnetic order \cite{Gruenberg2016,Ustinov2017,doi:10.1142/2359,Duine2018naturephys} renders the possibility of realizing  higher order EPs and anti-PT symmetric systems. Another interesting aspect concerns the interplay between intrinsic non-reciprocity caused by particular types of anti-symmetric exchange interactions and that due to pseudo-Hermitian effects that we mentioned above.  \\
Experimentally, the  development of spintronics \cite{doi:https://doi.org/10.1002/9781119698968.ch8,doi:https://doi.org/10.1002/9781119698968.ch7,doi:https://doi.org/10.1002/9781119698968.ch6,doi:https://doi.org/10.1002/9781119698968.ch5,doi:https://doi.org/10.1002/9781119698968.ch4,doi:https://doi.org/10.1002/9781119698968.ch3} has provided powerful tools to actively manipulate magnon populations and integrate them in nanoscale circuits as well as to
fabricate the needed structures with  effective gain and loss. Mechanisms like STT and SOT allow for electrical control over magnon damping – either enhancing loss (via damping like torque) or compensating it to achieve gain (via anti-damping torque).  This combination of intrinsic loss and controllable gain makes magnonics exceptionally well-suited for engineering Hamiltonians with tailored non-Hermitian properties. Pioneering theoretical work in 2015 introduced PT symmetry into magnetism using two coupled ferromagnets – one with loss and one with an equal amount of gain – and predicted a spontaneous PT-breaking EP with coalescing magnon modes.\cite{PhysRevB.91.094416} Since then, the field is progressing   rapidly   both theoretically   and experimentally.\cite{10.1002aelm.202300674}

The advent of hybrid magnonic systems further expanded the possibilities. By coupling magnons to other degrees of freedom, such as microwave photons in cavities (cavity magnonics)\cite{ZARERAMESHTI20221,Harder10.10635.0046202}, mechanical vibrations (magnomechanics)\cite{PhysRevX.11.031053}, or superconducting qubits,\cite{Dany10.1126science.aaz9236,PhysRevLett.125.117701,CRPHYS20161777290} hybrid platforms offer enhanced tunability and control, facilitating the realization and exploration of pseudo-Hermitian magnonics phenomena as applied to other research fields. 

\subsection{Pseudo-Hermitian magnons in ferromagnets and synthetic antiferromagnets}
The  pseudo-Hermitian magnons based on waveguides or thin films are engineered  such as one element experiences intrinsic or enhanced damping (loss), while the other is subjected to gain. The straightforward realization of magnonic PT symmetry is in two directly coupled magnets with intentionally unbalanced damping.\cite{PhysRevB.107.094435} The concept introduced by Lee et al. in 2015 used two coupled macroscopic magnetic structures (Fig. \ref{pseudomodels}(a)).\cite{PhysRevB.91.094416} One magnet with a magnonic  loss is caused by intrinsic damping, while the other has an equal amount of gain realized by spin-transfer torque (STT) or parametric pumping of the magnet. The magnetization dynamics for the coupled macrospins ($\vec{m}_1$ and $ \vec{m}_2 $) with positive and negative dampings can be described the following equations:\cite{PhysRevB.91.094416}
 \begin{equation}
	\begin{aligned}  
	    \frac{\partial \vec{m}_1}{\partial t} = -\gamma \vec{m}_1 \times \vec{H}_1 - \gamma K \vec{m}_1 \times \vec{m}_2 + \alpha \vec{m}_1 \times \frac{\partial \vec{m}_1}{\partial t}, \\
		 \frac{\partial \vec{m}_2}{\partial t} = -\gamma \vec{m}_2 \times \vec{H}_2 - \gamma K \vec{m}_2 \times \vec{m}_1 - \alpha \vec{m}_2 \times \frac{\partial \vec{m}_2}{\partial t}.
		\label{macrospin1}
	\end{aligned}              
\end{equation}
Here, the first(second) equation with positive(negative) $\alpha$ is used for loss(gain) mechanism. With the symmetric gain and loss, the system in invariant under the combined parity and time reversal operations resulting in  pseudo-Hermitian magnons.

Later, A. Galda and V. M. Vinokur constructed a non-Hermitian Hamiltonian for coupled Gilbert damping induced loss and STT induced gain in macrospin model and chain model, leading to the appearance of PT symmetry and its broken phase.\cite{PhysRevB.94.020408, PhysRevB.97.201411}  H. Yang et al. theoretically maps ferromagnets with gain/loss to antiferromagnets with loss/gain (Fig. \ref{pseudomodels}(b)), finding that antiferromagnetism and AFM Skyrmions emerge in the gain layer when the system's PT symmetry is broken.\cite{PhysRevLett.121.197201} For the magnets with partial  gain and partial loss, the effective Hamiltonian is obtained as,\cite{PhysRevLett.121.197201}
\begin{equation}
	\begin{small}
		\begin{aligned} 
			\displaystyle  H  = c \left( \begin{matrix} \chi_1(\vec{k}) + \alpha[\chi^*_2(\vec{k})-2iK'] & \chi_2(\vec{k})+\alpha\chi_1(\vec{k}) &  \alpha \chi^*_0 & \chi_0 \\ -\alpha\chi_1(\vec{k}) + [\chi^*_2(\vec{k})-2iK'] & -\alpha \chi_2(\vec{k})+\chi_1(\vec{k}) & \chi^*_0 & \alpha \chi^*_0 \\ \alpha \chi_0 & \chi_0 & \chi_1(\vec{k}) - \alpha[\chi^*_2(\vec{k})-2iK'] &  \chi_2(\vec{k})-\alpha\chi_1(\vec{k}) \\ \chi^*_0 & \alpha \chi_0 & \alpha\chi_1(\vec{k}) + [\chi^*_2(\vec{k})-2iK'] & \alpha\chi_2(\vec{k})+\chi_1(\vec{k}) \end{matrix} \right).
			\label{afmhameffective}
		\end{aligned} 
	\end{small}
\end{equation}
Here, the authors define $c = \gamma/[(1+\alpha^2)\mu_0 M_s a^3]$, $\chi_0 = i \gamma J$, $\chi_1(\vec{k}) = 2 D \sin k_y a$, $ \chi_2(\vec{k}) = 2 i J(\cos k_x a + \cos k_y a) - i (4+\gamma) J - 2 i K'$ and $K' = K+ \mu_0 M_s^2 a^3/2$ for the easy-plane anisotropy and demagnetizing energy. The Hamiltonian allows for PT symmetry phase and its phase broken.

The experimental studies for pseudo-Hermitian magnons was realized by Liu et al.\cite{Haoliangsciadveaax9144} via a passive PT-symmetric magnet pair. They fabricated two ferromagnetic layers (NiFe with $ \vec{m}_1^0 $ and Co with $ \vec{m}_2^0 $) separated by a thin spacer, with different intrinsic damping rates, and coupled via interlayer exchange (RKKY) coupling, as demonstrated by Fig. \ref{pseudomodels}(c). This asymmetric-loss bilayer achieves PT symmetry in a passive sense, where no external gain is added, but one layer has simply a lower loss than the other.The magnons of the passive PT-symmetric system can be described by the linearized equations,\cite{Haoliangsciadveaax9144}
 \begin{equation}
	\begin{aligned}  
		-i \omega \vec{m}_1^0 &= (\omega_H + \omega_2^K - i \alpha_1\omega)\vec{z} \times \vec{m}_1^0 - \omega_1^K \vec{z}\times \vec{m}_2^0, \\
		-i \omega \vec{m}_2^0 &= (\omega_H + \omega_1^K - i \alpha_2\omega)\vec{z} \times \vec{m}_2^0 - \omega_2^K \vec{z}\times \vec{m}_1^0.
		\label{passiveequation}
	\end{aligned}              
\end{equation}
The passive PT symmetry can be manipulated by alternating the RKKY interaction between magnets via the spacer layer thickness.

For  waveguides coupled via the interlayer coupling (RKKY or dipolar coupling), by inserting a heavy metal spacer (for instance, Pt layer) between two waveguides, the SOT can simultaneously tune gain or loss in magnonic amplitudes  in both guides and control where  PT symmetric magnons appear.
 As demonstrated by Fig. \ref{pseudomodels}(d), the charge current flowing along $x$ axis generates the SOT $ \vec{T}_1 =  \tau (\vec{m}_1 \times (\vec{y} \times \vec{m}_1)) $ in the first waveguide WG1 enhancing the effective magnonic damping (loss), and the SOT $ \vec{T}_2 =  -\tau (\vec{m}_2 \times (\vec{y} \times \vec{m}_2)) $ on WG2 weakening the effective damping (gain). Supposing the interlayer coupling is dominated by the RKKY interaction, the magnon equation (\ref{coupleequationfm}) with SOT becomes,\cite{Wang2020nc}
 \begin{equation}
	\begin{aligned}  
		\omega \psi_1 - (1 - i\alpha)(\omega_H + \omega_{ex} k^2 - i \omega_J) \psi_1 + (1 - i\alpha)q \psi_2  &= 0, \\
		\omega \psi_2 - (1 - i\alpha)(\omega_H + \omega_{ex} k^2 + i \omega_J) \psi_2 + (1 - i\alpha)q \psi_1  &= 0.
		\label{coupleequationfmsot}
	\end{aligned}              
\end{equation}
If the intrinsic damping is very small ($ \alpha \rightarrow 0 $), the magnonic guided modes in  the WG1 are subject to the PT-symmetric confining complex potential  $V(z) = V_R(z) + i V_I (z) $ with $ V_R(z_0) = \omega_H + \omega_{ex} k^2 $ and $ V_I(z_0) = -\omega_J $.
In WG2 the potential is  $V_R(-z_0) = \omega_H + \omega_{ex} k^2 $ and $V_I(-z_0) = \omega_J $. In this case, the Hamiltonian operator is extracted as,
\begin{equation}
	\begin{small}
		\begin{aligned} 
			\displaystyle  H  = \left( \begin{matrix} \omega_H + \omega_{ex} k^2 - i \omega_J & -q \\ -q & \omega_H + \omega_{ex} k^2 + i \omega_J \end{matrix} \right).
			\label{magnonhamrkky}
		\end{aligned} 
	\end{small}
\end{equation}
Similar to the PT-symmetric Hamiltonian Eq. (\ref{ham001}), using the transformation $\eta = \sigma_x$ leads to pseudo-Hermitian condition $ \eta  H  =  H ^{\dagger} \eta $, indicating the pseudo-Hermiticity feature of the system. We note that, the small damping $ \alpha $ slightly affect the PT symmetry, and the system cannot precisely satisfy the pseudo-Hermitian condition. Depending on the material and quality of the waveguide, $ \alpha $ can be in the order of $10^{-5}$. For a small $ \alpha $ the influence is negligible, and the pseudo-Hermitian behavior still hold.\cite{Wang2020nc}

Fig. \ref{pseudomodels}(e) shows a different design for SOT driving gain and loss in synthetic antiferromagnets.\cite{apl50029523} The SAF is deposited on a heavy metal film (in $x$-$y$ plane). The two magnetic layers of the SAF are both in the $x$-$y$ plane and have opposite local magnetization directions $\vec{m}_{1,2}$. The charge current in the heavy metal layer apply the SOT $ \vec{T}_{1,2} \tau (\vec{m}_{1,2} \times (\vec{y} \times \vec{m}_{1,2})) $, and simultaneously drive coupled gain and loss as $\vec{m}_1 = - \vec{m}_2$. Such setting also generates the coupled gain-loss induced magnonic PT-symmetry. Besides, in the case without RKKY coupling, adopting the dipolar interaction also can couple two waveguides with symmetric SOT induced gain and loss and realize the PT-symmetry condition.

\begin{figure}[htbp]
	\includegraphics[width=1\textwidth]{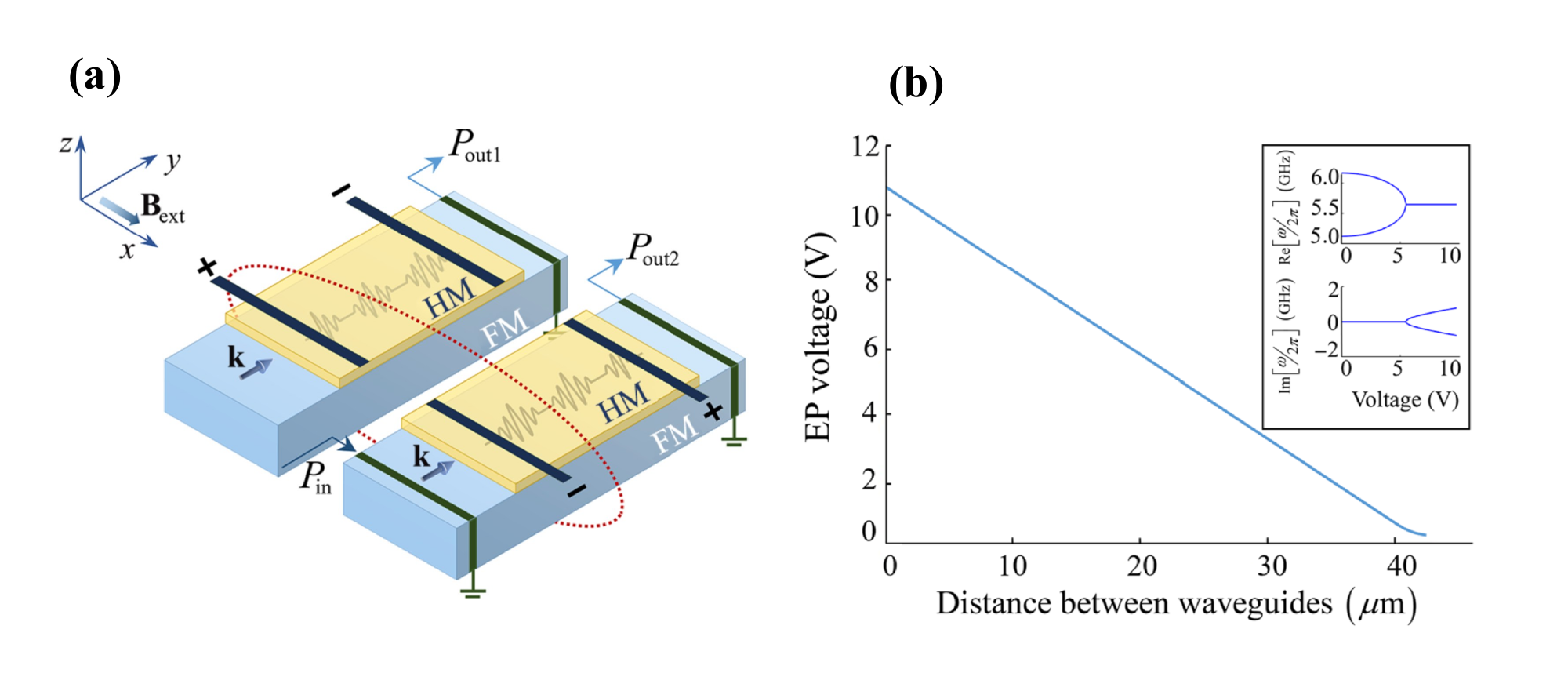}
	\caption{\label{planar} (a) Schematic of the PT-symmetric planar dipolarly coupled ferromagnetic waveguides with opposite spin-orbit torques. (b) The EP voltage as a function of the distance between two coupled waveguides, and the inset shows the variation of the normal nodes with applied voltage.\cite{PhysRevApplied.18.014003}}
\end{figure}

 O. O. Temnaya et al. also introduced a different design for PT-symmetric planar dipolarly coupled ferromagnetic waveguides, as demonstrated by Fig. \ref{planar}(a).\cite{PhysRevApplied.18.014003} Here, PT symmetry is achieved by  balancing  magnonic gain and loss via the attached metal layers with  strong spin-orbit interactions and opposite electric currents.  The corresponding magnon equation with opposite effective damping is written as,
 \begin{equation}
 	\begin{small}
 		\begin{aligned} 
 			\displaystyle \dot{c}_v = -i \Omega c_v - \Gamma_v c_v-i\Omega_c c_{v'}.
 			\label{planareq}
 		\end{aligned} 
 	\end{small}
 \end{equation}
 $v,v' = 1,2$, $v \ne v'$ $\dot{c} = a m_y + i b m_z$ is the complex amplitudes with $a = \sqrt{\Omega_{xx}/\Omega}$ and $b = \sqrt{\Omega_{zz}/\Omega}$. The effective damping coefficients for two waveguides read $ \Gamma_{1,2} = \alpha(\Omega_{xx}+\Omega_{zz})/2 \pm \Gamma_I$. $\Omega_{xx}$ and $\Omega_{zz}$ are related to dipolar interaction tensor. $\Gamma_I$ describe the SOT induced gain and loss in waveguides.

The above design is a suitable  ground for an experimental realization, sophisticated deposition techniques of multilayered structure can imprint the desired variation of the dielectric function. A magnonic realization offers a workaround without a permanent material modification; by just changing the voltage at the ends of heavy metal wire and tuning the current density to the specific ratio the EP vicinity is approached. This flexibility allows not only for an easy control of the PT symmetry-related features.

\begin{figure}[htbp]
	\includegraphics[width=0.6\textwidth]{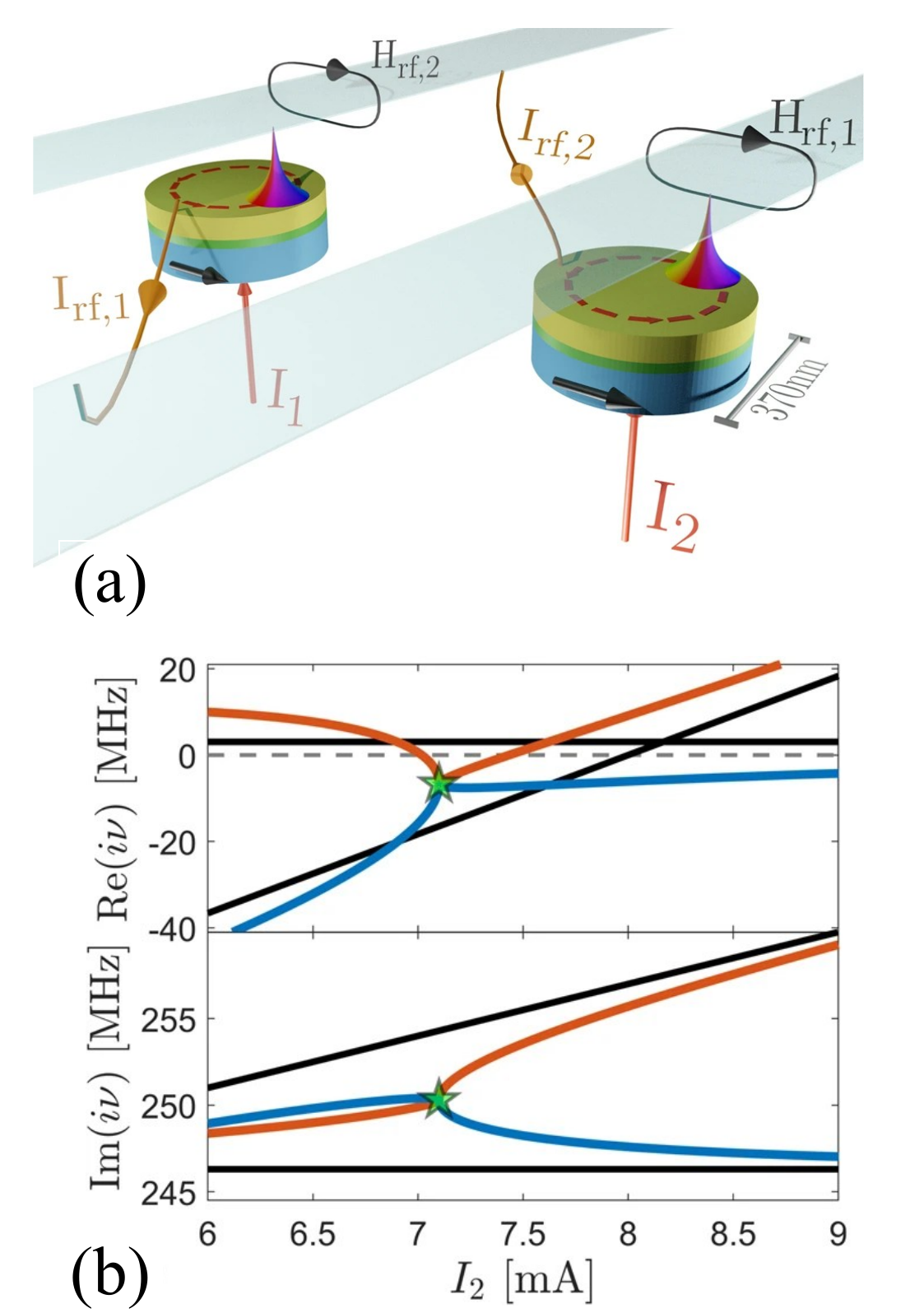}
	\caption{\label{oscillator} (a) Schematic of the pseudo-Hermitian coupled spin-torque nano-oscillators with vortexes. By adjusting individual drive currents, condition of PT symmetry and EP can be achieved.(b) Eigenfrequencies of two coupled magnetic oscillators as functions of $I_2$ under a fixed $I_1$ (red and blue lines). Black lines are calculated in the uncoupled case.\cite{Wittrock2024naturecommun}}
\end{figure}

Later, S. Wittorck et al. introduce  pseudo-Hermitian magnonics into  auto-oscillatory magnetic systems.\cite{Wittrock2024naturecommun} By inductively coupling two spin-torque nano-oscillators with magnetic vortex, and adjusting their individual drive currents, they achieved the balanced gain-loss condition characteristic of PT symmetry. The pseudo-Hermitian magnetization dynamics can be described by Thiele equation,
\begin{equation}
	\begin{small}
		\begin{aligned} 
			\displaystyle \frac{d}{dt}\left[ \begin{matrix} z_1\\ z_2 \end{matrix} \right] = i  {A} \cdot \left[ \begin{matrix} z_1\\ z_2 \end{matrix} \right],\\
		{\rm with}\\
		 {A} = \left[ \begin{matrix} \omega_1 - i \beta_1  & k\\ k & \omega_2 - i \beta_2 \end{matrix} \right].
			\label{thieleequation}
		\end{aligned} 
	\end{small}
\end{equation}
The coupled oscillators combine nonlinear effect and pseudo-Hermitian in  practical spintronic devices which could be a tool for magnetization dynamics control in oscillators.

 Another  platform is cavity magnonics in which   magnon modes couple to  microwave cavity modes.\cite{Zhang2017naturecommun}  Typically, a ferromagnetic sample, often a high-quality YIG sphere known for its low damping, is placed within a microwave cavity. The magnon modes within the YIG couple to the cavity photon modes. PT symmetry can be engineered by actively pumping the cavity mode to introduce gain, balancing it with the magnon loss , or by utilizing dissipative coupling mechanisms where modes interact via a common loss channel. Experiments allow for precise control over parameters like the external magnetic field (which tunes the magnon frequency and thus the detuning), microwave pump power (controlling gain), and the position of the YIG sphere within the cavity (adjusting the magnon-photon coupling strength).

B. Wang et al. designed a PT-symmetric magnon laser in a cavity optomagnonic system by coupling a YIG magnon to two optical whispering-gallery modes, one with gain and one with loss.\cite{PhysRevA.105.053705} When the splitting of the optical supermodes resonantly matches the magnon frequency in the PT-symmetric regime, the magnons experience coherent stimulated emission above a threshold pump power. The PT-balanced optical modes allowed the magnon laser to reach threshold at a lower pump power, and the oscillation frequency was tunable by an external magnetic field, illustrating a frequency-tunable coherent magnon source enabled by PT symmetry. Another mechanism in cavity magnonics is to tune the magnon-photon coupling from Hermitian to dissipative. Yang et al. (2020) engineered an anti-PT symmetric cavity-magnon polariton system by carefully controlling the magnon damping in a microwave cavity.\cite{PhysRevLett.125.147202} By increasing the magnon loss (through an external damping control) relative to the cavity loss, they could reach two EPs in succession, and between these EPs, the system maintained anti-PT symmetry.Besides, microwave transmission and reflection spectroscopy are the standard tools for probing the system's spectral properties and identifying pseudo-Hermitian feature.\cite{Wang18optexpress} 

\subsection{Electromagnetically controlled EPs}

Pseudo-Hermitian magnonics exhibits a range of    phenomena with no direct counterparts in other systems. Especially, EPs with non-Hermitian  degeneracies where not only eigenvalues but also the corresponding eigenvectors coalesce, are the most striking effect. For macrospin model and chain models in Ref. \cite{PhysRevB.91.094416,PhysRevB.94.020408, PhysRevB.97.201411}, by varying gain/loss strength or coupling, the eigenvalues of the coupled magnon modes move together and realize EP. 

\begin{figure}[htbp]
	\includegraphics[width=1.1\textwidth]{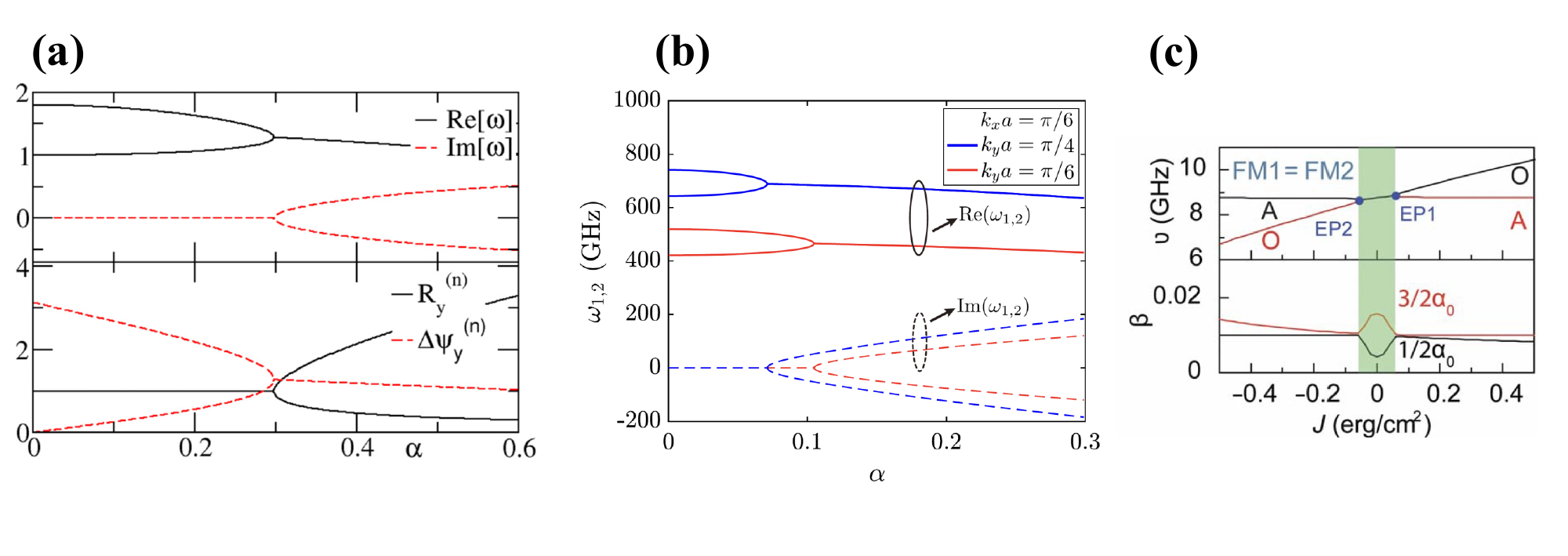}
	\caption{\label{otherep} (a) Eigenfrequencies and eigenvectors of the PT-symmetric models in (Fig. \ref{pseudomodels}(a)) as functions of $\alpha$.\cite{PhysRevB.91.094416} (b) For different magnon modes, the eigenfrequencies dependence of $\alpha$ in the model of Fig. \ref{pseudomodels}(b).\cite{PhysRevLett.121.197201} (c)Eigenfrequencies and damping rates of the Fig. \ref{pseudomodels}(c) model variation with coupling strength $J$ between two magnets.\cite{Haoliangsciadveaax9144} }
\end{figure}

For example, in the above coupled macrospin model with balanced positive and negative damping, the eigenfrequencies are given by,\cite{PhysRevB.91.094416}
\begin{equation}
	\begin{small}
		\begin{aligned} 
			\displaystyle \omega_{1,2} = \frac{1}{1+\alpha^2}(\omega_H + \omega_K \pm \sqrt{\omega_K^2 - \alpha^2 \omega_H(\omega_H + 2\omega_K)}).
			\label{macrospinep}
		\end{aligned} 
	\end{small}
\end{equation}
Increasing the gain and loss parameter $\alpha$ to a critical value $ \alpha_{cr} = \omega_K / \sqrt{\omega_H(\omega_H + 2\omega_K)} $, two eigenfrequencies coalesce to the same value (as demonstrated by Fig. \ref{otherep})(a), indicating the existence of EP. At the EP, there is the degeneracy of the eigenvectors.

In the PT symmetry induced mapped antiferromagnetism model(Fig. \ref{pseudomodels}(b)), two counterclockwise magnon modes are expressed as,\cite{PhysRevLett.121.197201}
\begin{equation}
	\begin{small}
		\begin{aligned} 
			\displaystyle \omega_{1,2}(\vec{k}) = \frac{\gamma J [\lambda + 2\xi(\vec{k})\pm \sqrt{\lambda^2 - 4a^2\xi(\vec{k})[\lambda + \xi(\vec{k})]}]}{(1+\alpha^2)\mu_0 M_s a^3},
			\label{afmhameffectiveEP}
		\end{aligned} 
	\end{small}
\end{equation}
with $$\xi(\vec{k}) = 2 - \cos k_x a - \cos k_y a + (D/J)\sin k_y a.$$
 As the gain and loss parameter ($\alpha$) increases, the EP appears at some critical value, whose value is dependent on the wavevector $\vec{k}$. They found emergent antiferrmagnetism occuring after PT-symmetry is broken, and two opposite precession modes appears beyond the EP when the system is pushed into the regime of nonlinear oscillatory instability. The finding points to a way to alter magnetic order and excitations via gain and loss.

In the passive PT-symmetric system of Fig. \ref{pseudomodels}(c), by tuning the coupling strength via spacer thickness, an EP of the system can be reached at a critical coupling. By setting $\alpha_1 = \delta-\alpha$ and $\alpha_2 = \delta + \alpha$, the eigenfrequencies are expressed as,\cite{Haoliangsciadveaax9144}
\begin{equation}
	\begin{small}
		\begin{aligned} 
			\displaystyle \omega_{1,2} = \frac{(1+i\delta)^2 [\omega_H + \omega_K \pm \sqrt{\omega_K^2 - \alpha^2 \omega_H(\omega_H + 2\omega_K)/(1+i\delta)^2}]}{(1+\alpha^2)(1+i\delta)}.
			\label{passiveEP}
		\end{aligned} 
	\end{small}
\end{equation}
As demonstrated by the results in Fig. \ref{otherep}(c), two coupled modes are distinct below the EP. EP occurs at a critical coupling strength. Beyond the EP, the passive PT-symmetry is broken and one magnon mode become overdamped while the other remain underdamped. The phenomena were testified in experiments, provided direct evidence of PT symmetry breaking in the magnonic device.

The features of EP are discussed in more details for the coupled waveguides with electrically controlled gain and loss via the SOT. For the model of Fig. \ref{pseudomodels}(d), with magnon equation (\ref{coupleequationfmsot}),  the Hamiltonian with finite damping is,\cite{Wang2020nc}
\begin{equation}
	\begin{small}
		\begin{aligned} 
			\displaystyle  H  = (1-i\alpha)\left( \begin{matrix} \omega_H + \omega_{ex} k^2 + q - i \omega_J & -q \\ -q & \omega_H + \omega_{ex} k^2  + q + i \omega_J \end{matrix} \right).
			\label{magnonhamrkkydamp}
		\end{aligned} 
	\end{small}
\end{equation}
The eigenvalues and eigenvectors are generally complex,
 \begin{equation}
	\begin{aligned}  
		\omega_{\pm}  &= (1-i\alpha) (\omega_H + \omega_{ex} k^2 + q \pm \sqrt{q^2 - \omega^2_{\tau}}), \\
		\vec{\psi}_{\pm} &= (\frac{\mp\sqrt{q^2 - \omega^2_{\tau}} + i \omega_J}{q}, 1).
		\label{eprkkyfm}
	\end{aligned}              
\end{equation}
In the limit $ \alpha\to 0 $, the two different eigenvalues are always real and different in the PT-symmetric regime below the gain/loss-balance threshold $ \omega_J/q < 1 $, and the eigenvectors can be given by $ \vec{\psi}_{\pm} = (\mp \exp(\mp i \theta),1) $, where $ \sin(\theta) = \omega_J/q $. At $ \omega_J = q $, the eigenvalues and the eigenvectors of this Hamiltonian coalesce at $ \omega_{\pm} = \omega_H + \omega_{ex} k^2 $ and $ \vec{\psi}_{\pm} = (i,1) $, which is the hallmark of the EP. Above the EP ($ \omega_J > q  $), the eigenvalues turn complex $ \omega_H + \omega_{ex} k^2 \pm  i\sqrt{\omega_J^2 - q^2} $, and the system enters the PT-symmetry broken phase as typical for PT-symmetric systems. The two real parts of $ \omega_{\pm} $ are both $ \omega_H + \omega_{ex} k^2 $, and the two imaginary parts are separated, and the eigenvectors are $ \vec{\psi}_{\pm} = (i\frac{\omega_J \mp \sqrt{\omega_J^{2} - q^2}}{q},1) $. These changes in eigenmodes bring about a non-reciprocal response and power oscillations. Without SOT ($ \omega_J = 0 $), the superposition of the two eigenmodes  $ \vec{\psi}_{\pm} = (\mp 1,1) $ (spin waves modes in the two waveguides are antisymmetric and symmetric) leads to reciprocal wave propagation. The interference between the two modes  leads to a periodic transfer of energy from one waveguide to the other, and the spin waves distributions obey the symmetry of the two waveguides (see Fig. \ref{pseudoep}). The interface pattern is dependent on the wavevector difference of the two modes.  If electric current term $ \omega_J $ increases but is still below the EP, the phase angle changes from the initial value $ \theta = 0 $ reaching eventually   $ \theta = \pi/2$  at the EP. In this range, the superposition of two asymmetric spin wave modes leads to the non-reciprocal wave propagation, where the spin wave distribution at the output end can be entirely different by exchanging the input from one waveguide to the other (Fig. \ref{pseudoep}). 
At EP, the two spin waves modes coalesce to the same mode, and spin waves in the two waveguides travel simultaneously. Surpassing  the EP the magnon always propagates in the waveguide with gain and is quickly damped in the waveguide with loss. The observations resemble  those made for  optical systems which is not surprising because the two systems obey similar equations of motion in the limit under discussion. In the magnonic case,  the EP and the strength of the PT symmetry breaking are reconfigurable by tuning electric current term $ \omega_J $. Note, even in the PT-symmetry preserved case,
i.e.  magnon-density preserving the dynamics is not simply a Hermitian one. This can be inferred from the none-normal  behavior of the modes as quantified by the Petermann factor or the  phase rigidity, shown in Fig. \ref{pseudoep} (e,f). Only at zero charge-current density we have two orthonormal modes an the Petermann factor is unity.

Above the EP, the system becomes unstable when the imaginary part of eigenvalues turn positive, i.e. $ \sqrt{ \omega^2_{\tau} - q^2} - \alpha \omega_H + \omega_{ex} k^2 \ge 0  $. For a very small damping, the critical value of the instability is set by the EP, and a non-vanishing $ \alpha $ shifts the instability range slightly above the EP. Above the instability critical value, the magnon oscillation is soon amplified, which leads to the nonlinear effect and finally reverses the equilibrium magnetization of waveguide with gain. Then, the magnon oscillation is soon damped and can not be sustained.

Allowing for a small damping $ \alpha $ does not alter the modes behavior. In this case, two complex eigenvalues still perfectly merge at the same degenerate EP $ \omega_J = q $. Below the EP ($ \omega_J < q $), the amplitude and separation of imaginary parts of the two eigenvalues are weak due to the smallness of $\alpha$. The separation suddenly becomes obvious above EP ($ \omega_J > q $). Similarly, the real parts of $ \omega_{\pm} $ are obviously distinct below EP, and their difference is very weak above EP. Furthermore, we find the $ \alpha $ dependent  term does not affect the eigen-vectors $ \vec{\psi}_{\pm} $ of the two eigenmodes. 

Near EPs, systems exhibit extreme sensitivity to perturbations. First, we discuss the PT-dependent permeability, which shows high sensitivity enhancement around the EP. Applying 
an external microwave field $ \vec{h}_{p} $ which adds to the effective field in the LLG equation. In frequency space we deduce that 
$ \widetilde{\psi}_p = \sum_{p\prime} \chi_{pp\prime}\,  \gamma \widetilde{h}_{m,p\prime} $ with $ h_{m,p} = h_{x,p} + i h_{z,p} $,  and $ \chi_{pp\prime} $  is the dynamic magnetic susceptibility which has the matrix form,\cite{Wang2020nc}
\begin{equation}
	\begin{small}
		\displaystyle \chi = \frac{1}{(\omega_{k}-i\alpha \omega - \omega) ^2+ \tau^2 - \kappa^2} \left( \begin{matrix} (\omega_k-i\alpha \omega) + (i \tau - \omega) & \kappa \\ \kappa & (\omega_k-i\alpha \omega) - (i \tau - \omega) \end{matrix} \right),
		\label{chi-dynamic}
	\end{small}
\end{equation}
with $ \omega_{k} = \gamma (H_0+ \frac{2 A_{\mathrm{ex}} k_x^2}{\mu_0 M_s} + \frac{J_{\mathrm{RKKY}}}{\mu_0 M_s t_p})  $. Near the EP the system becomes strongly sensitive, for instance  to changes in the charge current term $ \tau $,  the susceptibility is obviously changed. 

Besides, the system near EP is sensitive to weak changes in the magnetic field. This feature is beneficial for sensing variations in the magnetic environment or for increasing the magnetic response in photonic applications. For example, in the above system, by applying a weak perturbation $ \epsilon $ in WG1, where the perturbations $ \epsilon \ll 1 $ negligibly affect the coupling. At the EP $ \omega_J = q $, the obtained eigenfrequencies can be expanded perturbatively  using a Newton-Puiseux series \cite{Hodaei2017nature} that begins with a square-root element, and the first three terms of this series are 
\begin{equation}
	\begin{small}
		\begin{aligned} 
			\displaystyle \omega_{\pm} &\approx q (c_0 \pm c_1 \epsilon^{1/2} + c_2 \epsilon),\\
			c_0 &= (1-i\alpha)(\omega_H + \omega_{ex} k^2)/q,\\
			c_1 &= (1-i\alpha)e^{-i\pi/4}, \mathrm{and}\  c_2 = (1 - i \alpha)/2.
			\label{expand}
		\end{aligned} 
	\end{small}
\end{equation}
These expressions indicate that, the real parts of the eigenfrequencies bifurcate with a square-root dependence on the applied perturbation, i.e. 
$$ \mathrm{Re}[\omega_+ - \omega_-] = (1-\alpha) \sqrt{2\epsilon} \kappa, $$ which quantifies  the sensitivity enhancement.

Above analysis neglect the influence of dipolar interaction, and its influence is weak when the RKKY interaction dominates. To prove its effect, the results with dipolar interaction are discussed in Ref. \cite{Wang2020nc}. In the combination of RKKY coupling and dipolar coupling between two waveguides, still acoustic and optical modes are formed. Due to the nature of dipolar coupling, the gap between two modes is more obvious in the low frequency range with large wave-length. Thus, to merge two separated real parts of eigenfrequencies at EP, the current density becomes larger for a lower $k_x$.

\begin{figure}[htbp]
	\includegraphics[width=0.8\textwidth]{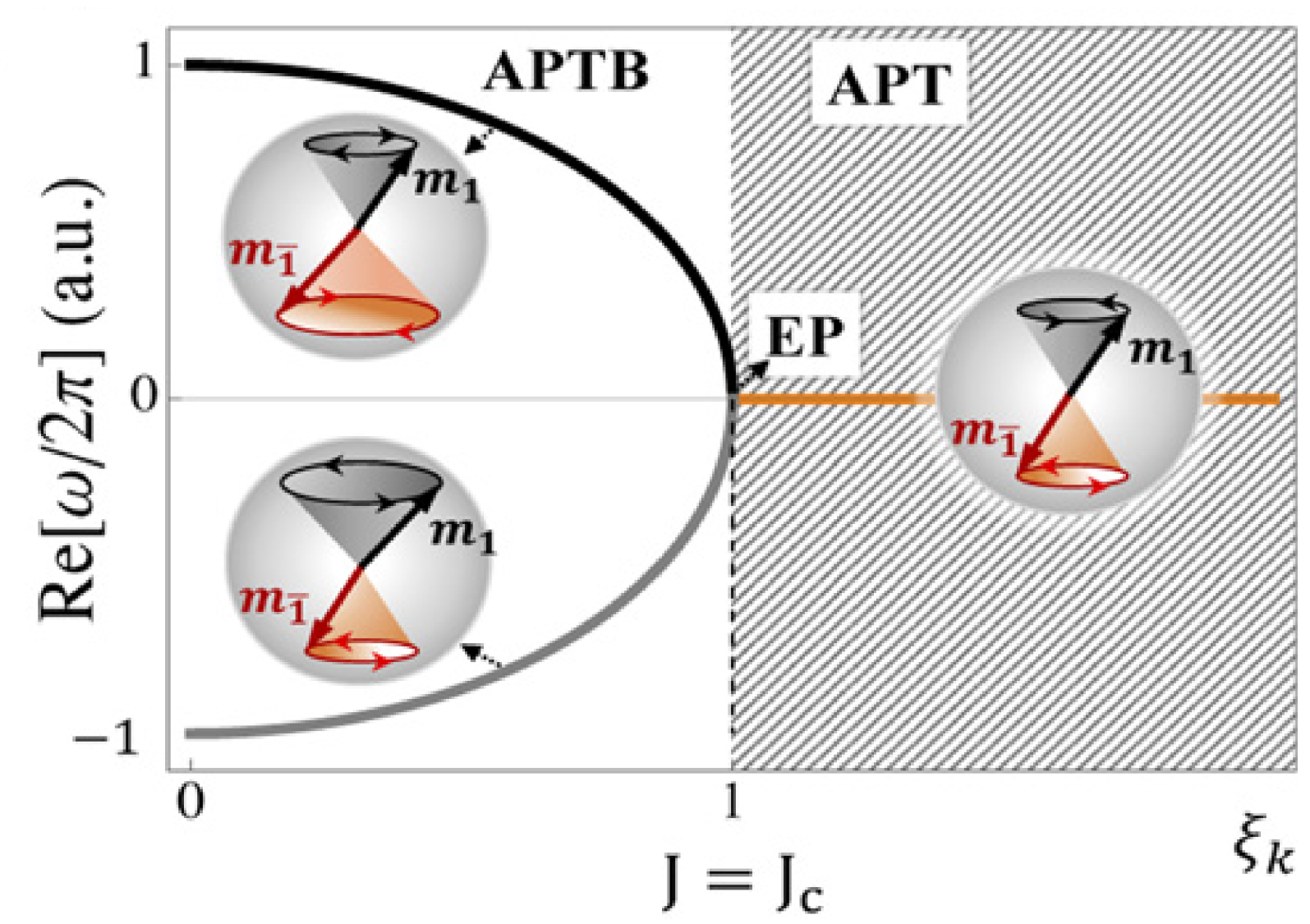}
	\caption{\label{antipt} The schematic of the synthetic antiferromagnets with anti-PT symmetry and EP. Above the EP $J = J_c$, real parts of positive and negative eigenfrequencies merge to 0.\cite{Sui2022newjphys} }
\end{figure}

For SOT driving gain and loss in synthetic antiferromagnets, the eigenvalues are obtained following the above method and in the limit of $ \alpha\to 0 $,\cite{apl50029523}
\begin{equation}
	\begin{small}
		\begin{aligned} 
			\displaystyle \omega_{\pm} = \sqrt{\omega_0^2 - \omega_z^2 - q^2 - \omega_J^2 \pm \sqrt{(\omega_J^2 + q^2)\omega_z^2 - \omega_J^2 \omega_0^2}}.
			\label{eprkkyafm}
		\end{aligned} 
	\end{small}
\end{equation}
For 
$$ \omega_J = \frac{q \omega_z}{\sqrt{\omega_0^2 - \omega_z^2}}, $$ 
the  two magnon modes are separated, and imaginary parts are negligible due to the limit of
 $ \alpha\to 0 $. At
  $$\omega_J = \frac{q \omega_z}{\sqrt{\omega_0^2 - \omega_z^2}}$$, 
  Both $ \omega_{\pm} $
 coalesce at the same value, which is the hallmark of the EP. Above the EP, the real parts of $ \omega_{\pm} $ have the same value, and the two imaginary parts are separated. Different from the FM case, the value of EP is dependent on the wavevector $ k_x $ in $ \omega_0 $, meaning magnons at different frequencies have different EP. Besides, EP is not only determined by the coupling term $q$, the anisotropy  term $ \omega_z $ also affects its value. A finite damping slightly affect the merging at the EP , as demonstrated by. Below EP, the imaginary parts are weak and identical for two modes, and these are still well separated above EP when real parts are almost the same. Without the in-plane anisotropy $ \omega_z $ term, only positive and negative eigenvalues exist. In this case, by varying the interlayer coupling term, the EP for the anti-PT symmetry can be reached, as demonstrated by Fig. \ref{antipt}.\cite{Sui2022newjphys} The authors linearized the magnon equation and derived the Hamiltonian,
\begin{equation}
	\begin{small}
		\displaystyle  H  = \left( \begin{matrix} H_k(1-i\alpha) & J(1-i\alpha) \\ -J(1+i\alpha) & -H_k(1+i\alpha) \end{matrix} \right).
		\label{antiptham}
	\end{small}
\end{equation}
The magnon eigenfrequencies were obtained,
\begin{equation}
	\begin{small}
		\displaystyle \omega_\pm = H_k\!\left(-i\alpha \pm \sqrt{1-\xi_k^{\,2}}\right),
		\label{antipthamep}
	\end{small}
\end{equation}
with $$\xi_k=\dfrac{J\sqrt{1+\alpha^{2}}}{H_k}.$$ At the critical coupling $$J_c=\dfrac{|H_k|}{\sqrt{1+\alpha^{2}}}$$
 the spectrum transits from the anti-PT-symmetric phase to broken anti-PT-symmetric  phase, where unequal cone angles in the two layers generate a finite net magnetization. By introducing interfacial Dzyaloshinskii–Moriya interaction, one can  observe the emergence of the  Friedrich–Wintgen bound states in the continuum and show that they  are protected by anti‑PT symmetry.  Furthermore, the anti-PT-symmetry can be exploited to emulate effects predicted by high-energy field-theory such as KLein-paradox, and  particle-antiparticle pair productions across a potential barrier  but here at meV energy scale and in an experimentally feasible laboratory  setting.
 \cite{Yuan2023}

By varying the  applied magnetic field, one can manipulate the magnetization profile in the magnet, and tune the position of the EP.
 For example, applying a magnetic field $ H_x \vec{x} $ perpendicular to equilibrium magnetization is $\vec{m}_{0,1} = -\vec{m}_{0,2} = (0, 1, 0) $ in the synthetic antiferromagnet, can generate a spin-flop state. Without SOT $ \omega_J = 0 $, $ \vec{m}_{0,1} $ and $ \vec{m}_{0,2} $ exhibit symmetric variations, with a convergence of $x$ and $z$ components. In this case, applying $H_x$ increases the lower magnon mode frequency and decreases the other one. For a certain $H_x$, the two modes cross with equal frequencies. Above the cross point, two modes are separated again. Under a finite SOT, a slight asymmetry is induced in the change of $ \vec{m}_{0,1} $ and $ \vec{m}_{0,2} $. It is found that  $H_x$ induced spin-flop state can significantly decrease the charge current density for EP. Especially, near the cross point of two modes, the EP current density approaches 0, and a tiny $ \omega_J $ is sufficient to attain EP. Thus, the PT-symmetry can be controlled in a non-linear way by a relatively small voltage.\cite{apl50029523}

In the case without RKKY interaction, the SOT driven gain and loss can still lead to the mergence of  PT-symmetric features for dipolarly coupled magnonic modes.\cite{PhysRevApplied.18.024080} For anti-parallel magnetization in Fig. \ref{dipolarep}(a)., the dispersion curves exhibit similar features with backward volume magnon mode, with  negative group velocity in the low wave-vector range. The dipolar coupling between WG1 and WG2 leads to the formation of lower acoustic and higher optic modes. In the larger wave-vector range, the exchange term dominates, and under weaker dipolar interaction the different between two magnon modes becomes very small. Following the same method above for dipolar coupling waveguides and including the coupled gain and loss effect from SOT, it is still possible to achieve EP, where two modes merge at a critical amplitude of $ \omega_J $. As the dipolar coupling strength is weaker for higher wavevector, the value of $ \omega_J $ at EP is smaller. Thus, for a finite amplitude of $ \omega_J $, when lower wavevector magnons are still below the EP, the magnons for higher wave-vector have already surpassed the EP and enter the PT-symmetry broken phase. Still, during the PT-symmetry transition near EP, the electrically reconfigurable non-reciprocal propagation in the coupled wavegudies. 

For parallel configuration in the dipolarly coupled waveguides (Fig. \ref{dipolarep}(b)), there is still splitting between optic and acoustic modes. Compared with the anti-parallel case, the mode splitting becomes slightly smaller under parallel configuration. In such case, the splitting between two modes becomes slightly smaller compared with the above anti-parallel case. Still , one can achieve the electrical current driven PT-symmetry-phase change and EP. Two real parts cross at a certain $k_x$, and the cross modes decrease the required density of EP. For a small $\omega_J$, near the cross point, magnons modes above the EP possess separated imaginary parts and merged real parts. Still, for higher $ k_x $, different between two magnon modes and EP becomes very small, and the small $\omega_J$ drives magnons above the EP.\cite{PhysRevApplied.18.024080}

Exploiting the electrically reconfigurable magnon transfer between two waveguides near EP, potential applications in diode and logic devices are proposed.\cite{PhysRevApplied.18.024080} At a certain frequency, without electric current ($\omega_J = 0$), by switching the magnon injection (input) between two waveguides, two equal magnon outputs are detected. Turing on the electric current, due to the non-reciprocal propagation, the magnon output for input in WG1 becomes twice the out from input of WG2,  which is an important finding in view of realization of an electrically reconfigurable magnon diode. The input in WG1 represents an "on" output with a larger amplitude, and input in WG2 signals an "off" state with an obviously smaller amplitude. Similarly, combing the inject electric current and magnon input in different waveguides, the electrically reconfiguralbe magnon logical operation can be realized around EP. Besides, via the magnon induced heating effect, the thermal information can be embodied in the PT-symmetric magnetic excitations, and controlled by charge current driven magnonic EP, realizing non-reciprocal heat flow, thermal diode and gate operations.

In the PT-symmetric planar dipolarly coupled ferromagnetic waveguides in Fig. \ref{planar}(a), The eigenvalues are expressed in the form of,\cite{PhysRevApplied.18.014003}
\begin{equation}
	\begin{small}
		\begin{aligned} 
			\displaystyle \omega_{1,2} = \omega_0 - i\Gamma_0 \pm \sqrt{\Omega_c^2 - \Gamma_I^2}.
			\label{planarep}
		\end{aligned} 
	\end{small}
\end{equation}
When the electric current term $\Gamma_I = \Omega_c$, two eigenvalues coalesce the the same value $\omega_0 = \omega_0 - i \Gamma_0$, indicating the EP. With the increases in the separation distance between two waveguides, the decrease in the coupling make the EP voltage smaller, as proved by Fig. \ref{planar}.

\begin{figure}[htbp]
	\includegraphics[width=0.9\textwidth]{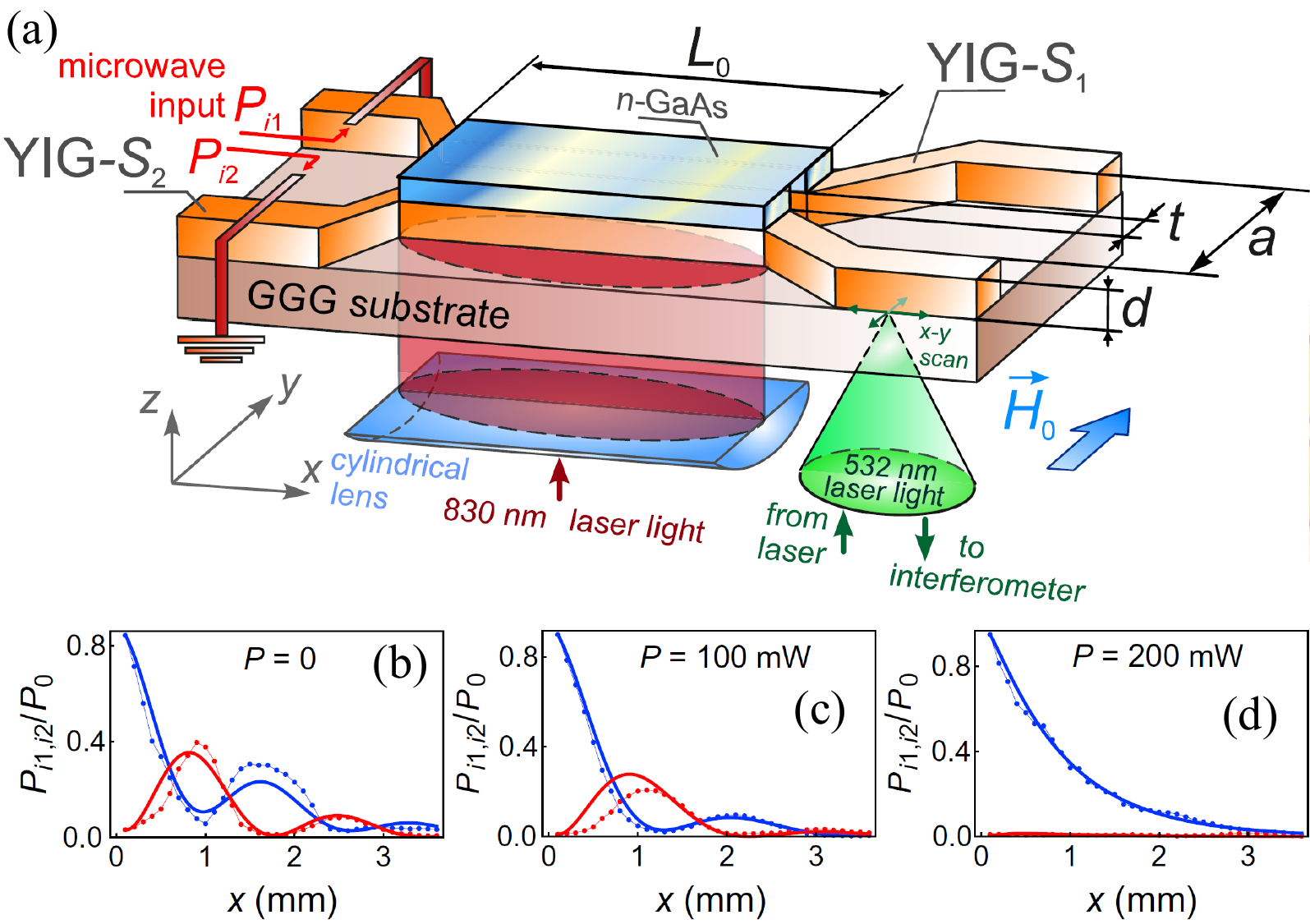}
	\caption{\label{experimentcouple} (a) Schematic of experimental structure with two coupled magnon waveguides. The additional loss in one waveguide is generated by infrared laser irradiation. (b-d) Magnon intensity distribution in two magnonic waveguides under different laser power.\cite{PhysRevApplied.18.024073}}
\end{figure}

A. V. Sadovnikov et al. experimentally identified the magnonic EP and related propagation behavior in the dipolar coupled waveguide (Fig. \ref{experimentcouple}), where the magnonic loss in one waveguide is controlled via the infrared laser irradiation. \cite{PhysRevApplied.18.024073}  By increasing the laser induced loss amplitude, the system gradually approaches EP inducing changes in the eigenfrequencies and magnon modes of coupled waveguides shown in Fig. \ref{experimentcouple}(b-d). The experimental findings  validate the theoretical predication of magnon nonreciprocal propagation around the EP. The non reciprocity in magnonic systems 
can be setup to  active, meaning to be externally controllable  via stimuli. This advantage  can be exploited for applications such as active nonreciprocal cloaking  of magnonic signals.
\cite{PhysRevApplied.22.054046}

In the pseudo-Hermitian coupled spin oscillators of Fig. \ref{oscillator}, the gain in one oscillator and enhanced loss in the other allow for EP physics to emerge in their spectra.\cite{Wittrock2024naturecommun} The eigenfrequencies of Eq. (\ref{thieleequation}) are given by,
\begin{equation}
	\begin{small}
		\begin{aligned} 
			\displaystyle v_{1,2} = \bar{\omega} - i \bar{\beta} \pm \sqrt{k^2 + (\tilde{\omega} - i \tilde{\beta})^2},\\
			\bar{\omega} = (\omega_1 + \omega_2)/2, \bar{\beta}=(\beta_1 + \beta_2)/2,
			\tilde{\omega} = (\omega_1 - \omega_2)/2, \tilde{\beta}=(\beta_1 - \beta_2)/2,
			\label{oscillatorep}
		\end{aligned} 
	\end{small}
\end{equation}
By defining the EP with two eigenfrequencies coalescing  to the same value, the parameter condition of EP is obtained by $k_c - i k_d = \pm(\tilde{\beta} + i\tilde{\omega})$. The EP condition can be achieved by changing one electric current while fixing the other current, as demonstrated by Fig. \ref{oscillator}(b). At the EP, they observed the cessation of oscillations in both oscillators, a dramatic nonlinear effect where the active oscillations quench each other. Slightly away from the EP, they reported other
complex dynamics such as frequency pulling and asymmetric mode hopping. The authors emphasize that the oscillation death at the EP can be harnessed for highly sensitive detectors, and that the setup uses  CMOS-compatible materials making it amenable for integration  in (spin)electronic circuits.\cite{PhysRevB.108.174443,Wittrock2024naturecommun} 

For cavity magnonic systems, tuning the EP often involves adjusting the magnon-photon detuning for coupling strength. For example,  Zhang et al. used a dense ensemble of paramagnetic spins with magnon modes coupled to a coplanar waveguide resonator, reaching an EP by pumping one subset of spins to modify the coupling.\cite{PRXQuantum.2.020307} The observed EP was robust against  varying  spins were pumped, revealing that intrinsic spin–spin cross-relaxation can stabilize the non-Hermitian degeneracy. In more conventional magnet/cavity setups, researchers have observed the polariton mode coalescence by varying the relative damping, and the mechanism can be realized by an extra loss channel to the cavity or a microwave drive that effectively amplifies the magnon.\cite{PhysRevLett.113.156401,Zhang2017naturecommun,PhysRevLett.125.147202,PhysRevLett.121.137203,JIN2022169035} Proposals also exist for achieving higher-order EPs (third-order or beyond) by coupling multiple magnon and photon modes in a cavity. Can and Yan theoretically predicted a cavity magnon-polariton system could be pushed to a third-order EP, at which the transmission spectrum becomes a step-function-like response, accompanied by an ultra narrow mode and dramatically enhanced sensitivity to external perturbations, which can be  for example used for thermal diods \cite{https://doi.org/10.1002/aelm.202300325}. They suggested an experimental scheme involving additional driving to  photons and magnons.\cite{PhysRevB.99.214415}

\subsection{Magnonic metamaterials}

\begin{figure}[htbp]
	\includegraphics[width=0.8\textwidth]{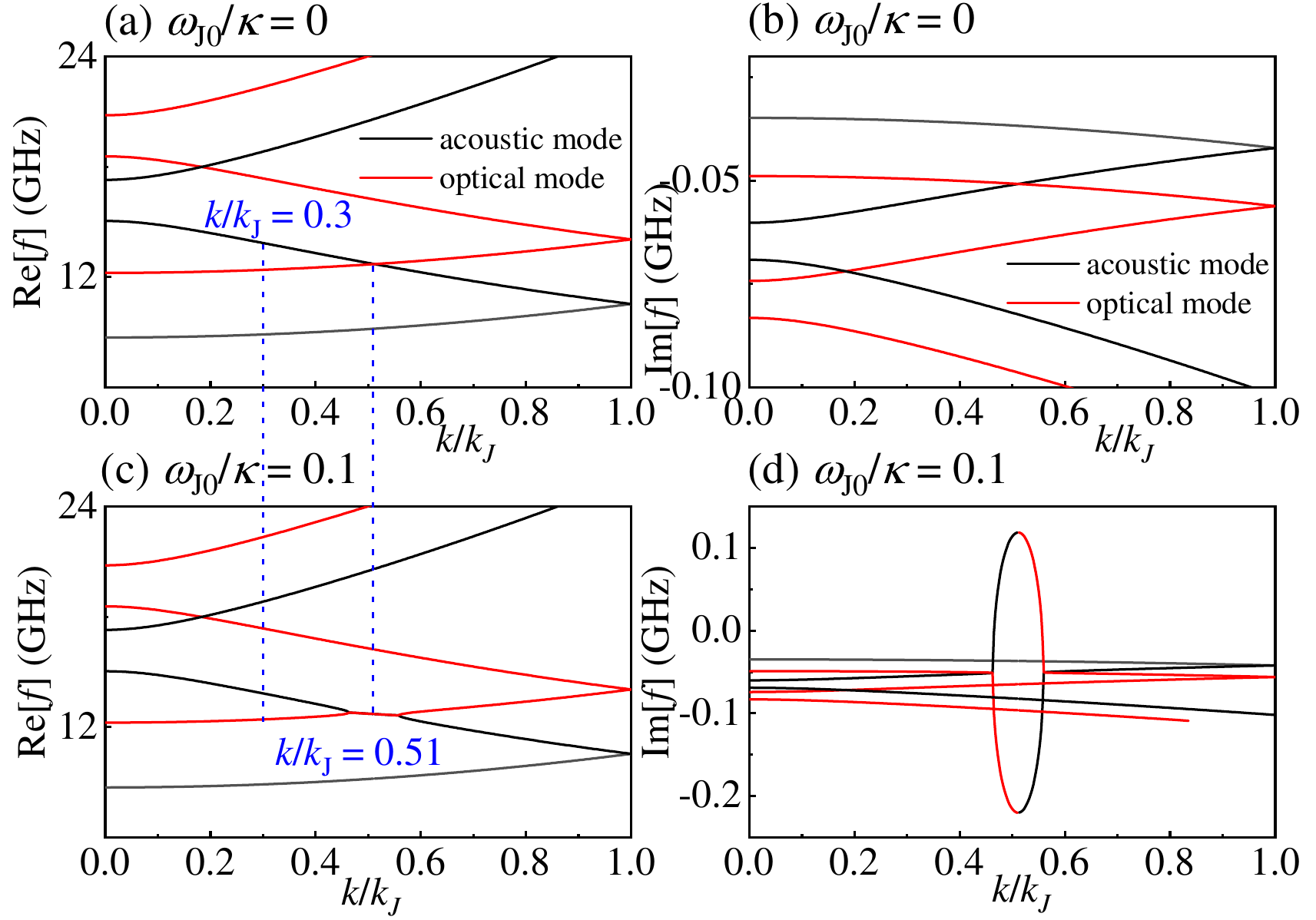}
	\caption{\label{periodic} \color{black} For the periodically coupled waveguides, the real and imaginary components of magnon eigenfrequencies folded in the BZ zone (a-b) without and (c-d) with gain and loss grating.\cite{PhysRevLett.131.186705} }
\end{figure}

{\color{black} The band structure of magnonic metamaterial also can be exploited for EP manipulation. For example, via periodic nanostructuring, adjacent Pt stripes separated with a distance $L$ carry opposite-sign charge current densities  along the $y$-axis. Two waveguides WG1 and WG2 are both magnetized along $+x$ direction with periodic varying coupling strength, forming a magnonic crystal. In such setting, the SOTs are $ \vec{T}_{1(2)} = \gamma c_J(x) \vec{m}_{1(2)} \times (\pm \vec{x}) \times \vec{m}_1(2) $, generating gain/loss grating for the magnonic crystal.\cite{PhysRevLett.131.186705} Using the Bloch theorem, we obtain the eigenvalues for different bands folded into the first BZ, as shown in Fig. \ref{periodic}. Comparing with the cases without and with gain/loss, we find the effect of spatially-periodic   SOT $ \omega_J(x) $ is especially dominant  when two magnon modes approach each other at $ k/k_J = 0.51 $. Here, $ \omega_J(x) $ enlarges the crossing area with collapsed real components, and in the same range the imaginary components are clearly separated. These features indicate the range is above EP. Near the crossing point, the two modes collapse at the EP near $ \omega_{J0} = 0 $. Away from the cross point, the EP value becomes larger. Such feature is also identified in the dipolarly coupled PT-symmetric magnonic crystal \cite{PhysRevApplied.18.024080}. }
\subsection{Hybrid materials}
{\color{black}  X. Li et al. presented a  PT-symmetric gain and loss setup based  on  van der Waals ferromagnetic bilayer. It is shown that multi EPs can appear over extended portions of the first Brillouin zone.\cite{PhysRevB.106.214432}. T. Yu et al. combined non-Hermitian and magnonic crystal. 1D array of ferromagnetic elements on a ferromagnetic film with chiral coupling between each element’s Kittel magnon and unidirectional film spin waves, producing a non-Hermitian skin effect so that modes localize at one edge and yield giant microwave sensitivity.\cite{PhysRevB.105.L180401} B. Flebus et al. adopted an array of spin-torque oscillators connected in series such that each oscillator feeds some of its damping to one neighbor. This could mimic a non-Hermitian Su-Schrieffer-Heeger model exhibiting topologically protected edge states.\cite{PhysRevB.102.180408} }

\subsection{Pseudo-Hermitian magnons in metallic waveguides}

\begin{figure}[htbp]
	\includegraphics[width=0.85\textwidth]{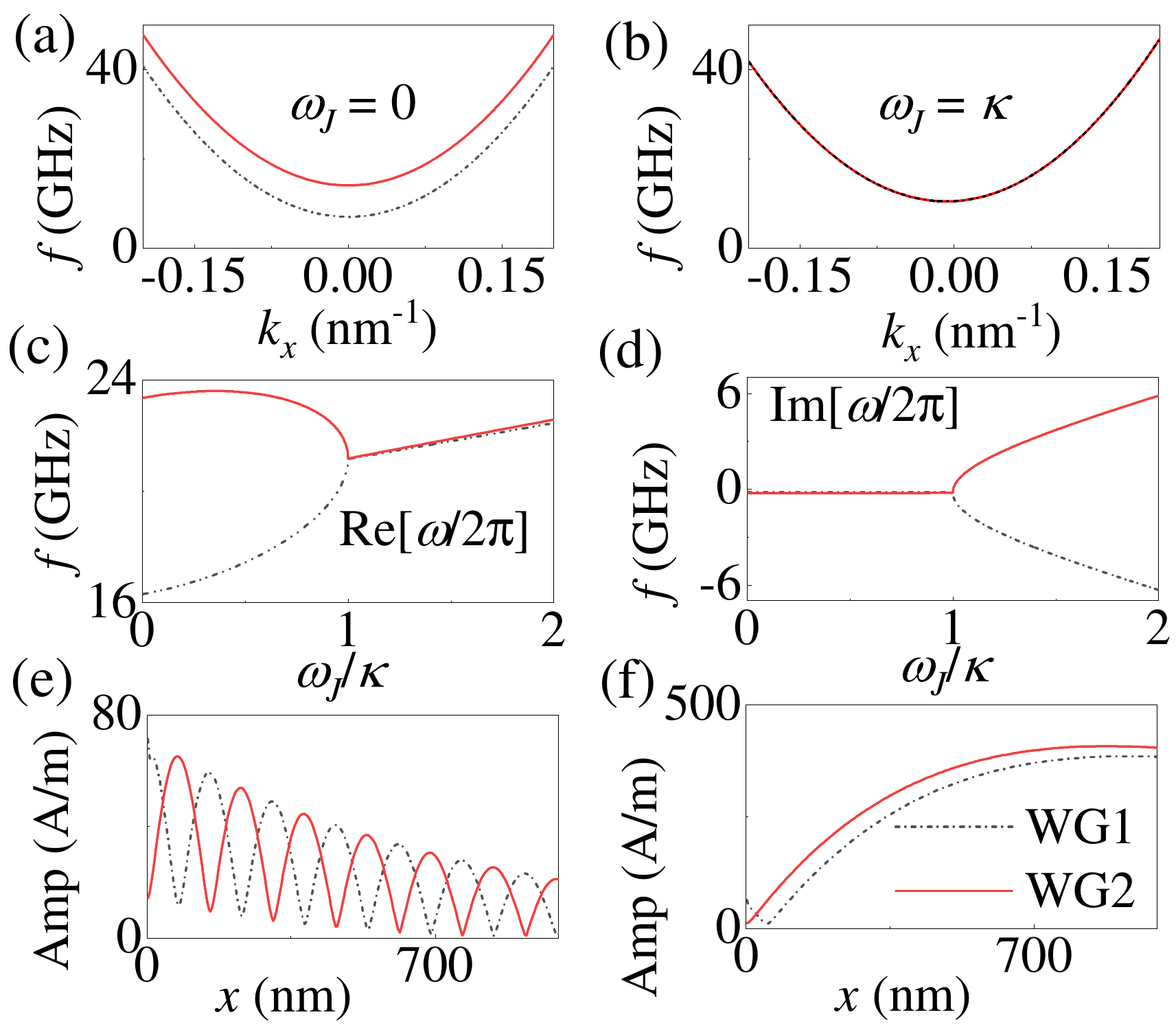}
	\includegraphics[width=0.85\textwidth]{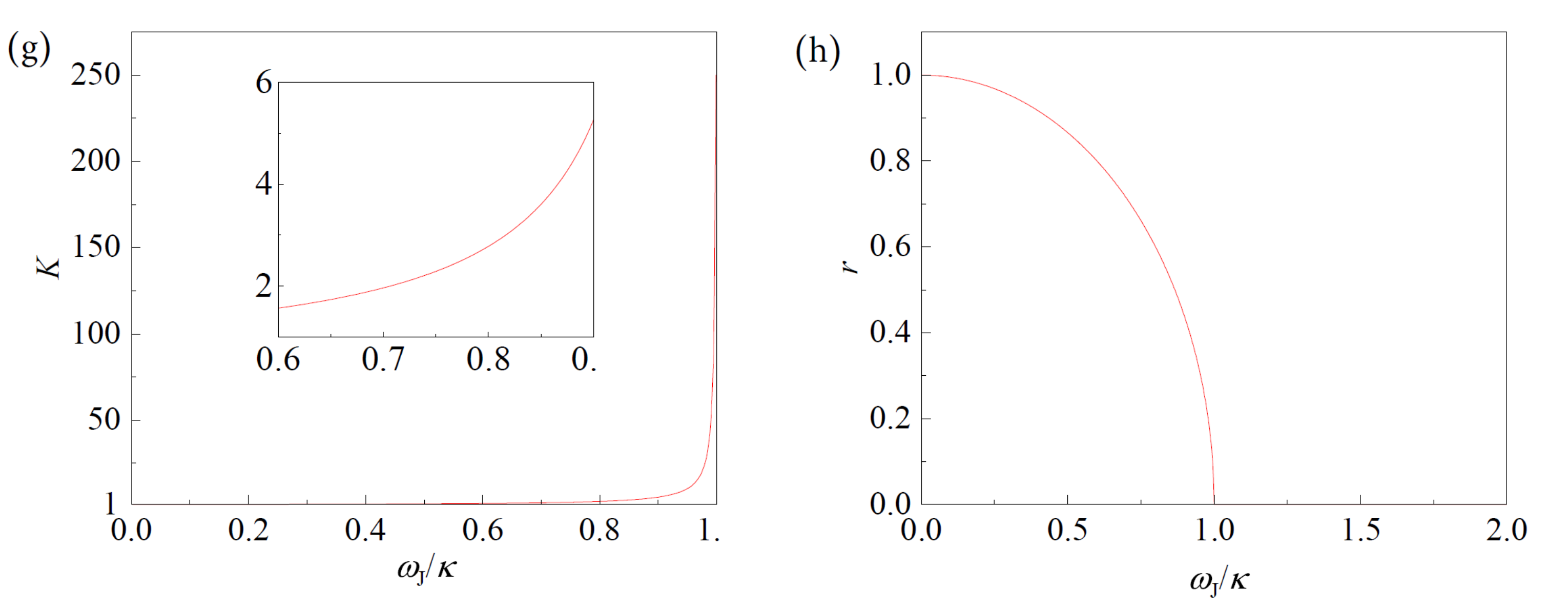}
	\caption{\label{metal}  For two coupled metallic waveguides, (a-b) the two magnon modes dispersion at $ \omega_J = 0 $ and EP $\omega_J = \kappa$. (c-d) Real and imaginary parts of two magnon modes as functions of $\omega_J/\kappa$. (e-f) The spatial distribution of the magnon amplitudes at $ \omega_J = 0 $ and EP $\omega_J = \kappa$.\cite{Wang2020nc} (g) Is the Petermann factor (Eq. (\ref{eq:petermann})) (inset shows small current density regime), and (h) is the phase rigidity (Eq. (\ref{eq:phaserigid})) corresponding to the modes in (c,f).   }
\end{figure}

In the above model of Fig. \ref{pseudomodels}(d) with  electrically controlled EP, the insulating waveguides were adopted. For coupled metallic ferromagnetic waveguides with Pt spacer, the electric current along $x$ axis still apply the same gain and loss torques 
$$ \vec{T}_{1(2)} = +(-) \tau (\vec{m}_{1(2)} \times (\vec{y} \times \vec{m}_{1(2)})). $$
 In addition, an in-plane spin transfer torque (STT) $ b_J \partial_x \vec{m}_p $ is possible in metallic ferromagnetic layer.\cite{RALPH20081190,SLONCZEWSKI1996L1,PhysRevB.54.9353} Here, 
 $$b_J = \frac{g \mu_B P J_e}{2 e M_s},$$
 and   $ g $ is the Land$ \acute{e} $ factor, $ \mu_B $ is the Bohr's magneton, $ P $ is the spin-polarization efficiency, and $ e $ is the electron charge. Including the SOT and additional STT in the LLG equation, the eigenfrequencies for two coupled magnon modes become,
\begin{equation}
	\begin{aligned}  
		\omega_{\pm}  &= (1-i\alpha) (\omega_H + \omega_{ex} k^2 + \omega_{bJ} k_x + q \pm \sqrt{q^2 - \omega^2_{\tau}}),
		\label{eprkkyfmbj}
	\end{aligned}              
\end{equation} 
with $$ \omega_{bJ} = b_J/(1+\alpha^2).$$
 The EP appears  at $\omega_J = \kappa$. The additional STT $b_J$ term leads to a weak asymmetry in two modes dispersion (Fig. \ref{metal}). Above the EP, the increase in STT amplitude $b_J$ can enhance frequency. Also, the magnon propagation features at the EP still hold, as evidenced by  Fig. \ref{metal}(e-f). In particular, the phase rigidity and the difference between the so-called left and right states, signifying $[H^\dagger, H]\neq 0$ indicate 
 that the dynamics is not corresponding to a system with  Hermitian operator, even though the eigenvalues are real (cf. Fig.\ref{metal} (g,h)).

\subsection{Pseudo-Hermitian magnons in ferromagnets with DMI}

In addition to the exchange interaction and dipolar interaction discussed above, there may exist a special type of antisymmetric exchange interaction called Dzyaloshinskii-Moriya interaction (DMI).\cite{DZYALOSHINSKY1958241,PhysRev12091} The physical origin of DMI lies in the relativistic spin-orbit coupling combined with the antisymmetric exchange interaction between neighboring magnetic moments. When inversion symmetry is broken, either at interfaces, surfaces, or in bulk materials lacking inversion centers, the DMI becomes active and competes with the symmetric Heisenberg exchange interaction to determine the magnetic ground state. The DMI favors non-collinear spin configurations and stabilizes chiral magnetic textures such as skyrmions and spin spirals.\cite{Fert2013naturenanotech, RevModPhys89025006} For two neighbouring spins $\vec{S}_i$ and $\vec{S}_j$ , the antisymmetric exchange interaction energy is expressed as $ \vec{D}_{ij} \cdot (\vec{S}_i \times \vec{S}_j) $, where $ \vec{D}_{ij} = -\vec{D}_{ji} $ is the vector characterizing the strength and direction of the interaction.  The vector $ \vec{D} $ depends on the symmetry of the system. In systems with specific crystallographic symmetries, the DMI can be classified into different types. For interfacial DMI of thin magnetic film, the interaction takes on the form $ D \vec{e}_z \cdot (\vec{S}_i \times \vec{S}_j) ,$ with the DMI strength $D$ and $\vec{e}_z$ normal to the film interface. In the continuum limit, the bulk DMI energy density can be written in the form of continuous magnetization $\vec{m}(\vec{r})$, $$ \epsilon_{DMI} = D \vec{m} \cdot (\nabla \times \vec{m}).$$
 The bulk DMI occurs in materials without inversion symmetry such as B20 compounds. For interfacial DMI in thin films, the energy density becomes  \cite{Fert2013naturenanotech,Emori2013naturemater,Ryu2013naturenanotech}
 $$ \epsilon_{DMI} = D \left[ m_z (\nabla \cdot \vec{m}) - (\vec{m} \cdot \nabla) m_z \right] .$$

Besides, the DMI allows for a coupling between magnetization and external electric field $\vec{E}$ and to applied voltage.The coupling energy density is expressed as $\epsilon_{elec} = - \vec{E} \cdot  \vec{P} $, with the spin-driven polarization 
$$ \vec{P} = c_E[(\vec{m} \cdot \nabla) - \vec{m}(\nabla \cdot \vec{m})] .$$\cite{PhysRevLett106247203,Risinggard2016scirep} 
By including the effective field 
$$ \vec{H}_{elec} = -\frac{1}{\mu_0 M_s} \frac{\delta E_{elec}}{\delta \vec{m}} $$
 in the LLG equation (\ref{LLG}), the role of electric field induced effective DMI on the magnon dynamic can be analyzed. Applying electric field $\vec{E} = (0,0,E_z)$ to the single magnet with an uniform magnetic field $\vec{H_a} = H_0 \vec{y} $ applied along $+y$ direction, the magnon dispersion relation under the DMI becomes,
\begin{equation}
	\begin{aligned}  
		\omega = (1-i\alpha)(\omega_H + \omega_{ex} k^2 - \omega_{dmi} k_x),
		\label{singledmi}
	\end{aligned}              
\end{equation}
where $ \omega_{dmi}=  \frac{2 \gamma c_E E_z}{(1 + \alpha^2)\mu_0 M_s}$. The dispersion relation reveals several important features introduced by DMI. Compared to Eq. (\ref{equationswp}) without DMI, the energy gap at $k = 0$ is modified, and the dispersion is asymmetric with respect to  propagation direction along the $x$ axis. This asymmetry leads to non-reciprocal magnon propagation, i.e. $\omega(+k_x) \ne \omega(-k_x)$, where magnons with opposite wave vectors have different frequencies.\cite{PhysRevB.88.184404,CortesOrtuno2013jphyscmatter,PhysRevLett.104.137203} 

\begin{figure}[htbp]
	\includegraphics[width=0.9\textwidth]{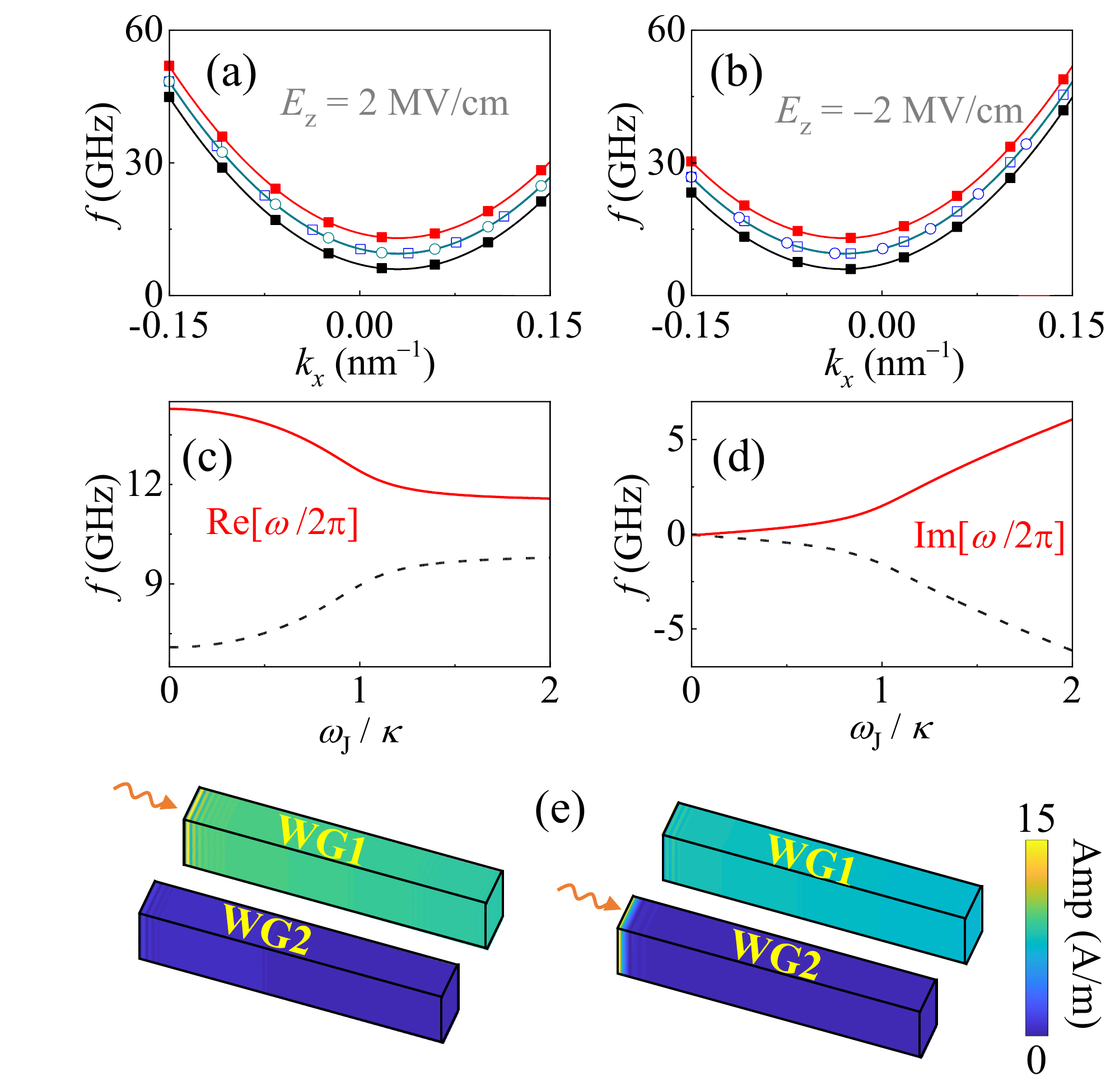}
	\caption{\label{dmieffect}  (a-b) Under an uniform electric $(0, 0, E_z)$ to two coupled waveguides, the magnon dispersion relations with and without electric current at EP. (c-d) For opposite electric fields in two coupled wavegudies, $\vec{E}_1 = (0, 0, E_z)$ in WG1 and $\vec{E}_2 = (0, 0, -E_z)$ in WG2, real and imaginary parts of magnon eigenmodes as function of electric current term $ \omega_J $. (e) Applying the electric field $(0,0,E_z)$ solely in WG1, the spatial profiles of magnons in two coupled waveguides.\cite{Wang2020nc}}
\end{figure}

In the two coupled waveguides (Fig. \ref{pseudomodels}(d)), when the same electric field is applied to the two waveguides, the induced DMI changes the magnon dispersion relations as,\cite{Wang2020nc}
\begin{equation}
	\begin{aligned}  
		\omega_{\pm}  &= (1-i\alpha) (\omega_H + \omega_{ex} k^2 - \omega_{dmi} k_x + q \pm \sqrt{q^2 - \omega^2_{\tau}}).
		\label{eprkkyfmdmi}
	\end{aligned}              
\end{equation}
Compared to Eq. {\ref{eprkkyfm}}, the conditions for PT-symmetry and EP are unaffected. The electric field causes an asymmetry in the magnon dispersion, as shown in Fig. \ref{dmieffect}(a-b). Here, the positive $E_z$ shifts the magnon dispersion to positive $k_x$, and the effect of negative $E_z$ is opposite.

For opposite electric fields in two waveguides, the magnon dispersion relations in WG1 and WG2 is changed to different directions, bringing asymmetric changes in the magnon potential. Thus, the PT symmetry condition is not satisfied, and no EP can be strictly identified in this case, as shown in Fig. \ref{dmieffect}(c-d). Besides, if the electric field is applied only to a single waveguide, it shifts selectively the magnon dispersion relation in this waveguide. As demonstrated in Fig. \ref{dmieffect}(e), the magnon can solely propagate to the waveguide with  electric field, while magnons in the other waveguide are suppressed.\cite{Wang2020nc}

{\color{black} Besides, it was found that combing DMI and spatially nonlocal dissipative term can cause an effective non-Hermitian asymmetry.  Deng et al. considered a 1D chain where each spin has uniform intrinsic damping but is coupled to a common substrate, introducing nonlocal dissipative term.\cite{PhysRevB.105.L180406} With DMI making the coupling directional and the substrate making damping nonlocal, they found that the bulk magnon modes under open boundary conditions all skewed toward one end, i.e. magnonic non-Hermitian skin effect. }

\subsection{Exceptional points in Floquet magnonic systems}

\begin{figure}[htbp]
	\includegraphics[width=0.9\textwidth]{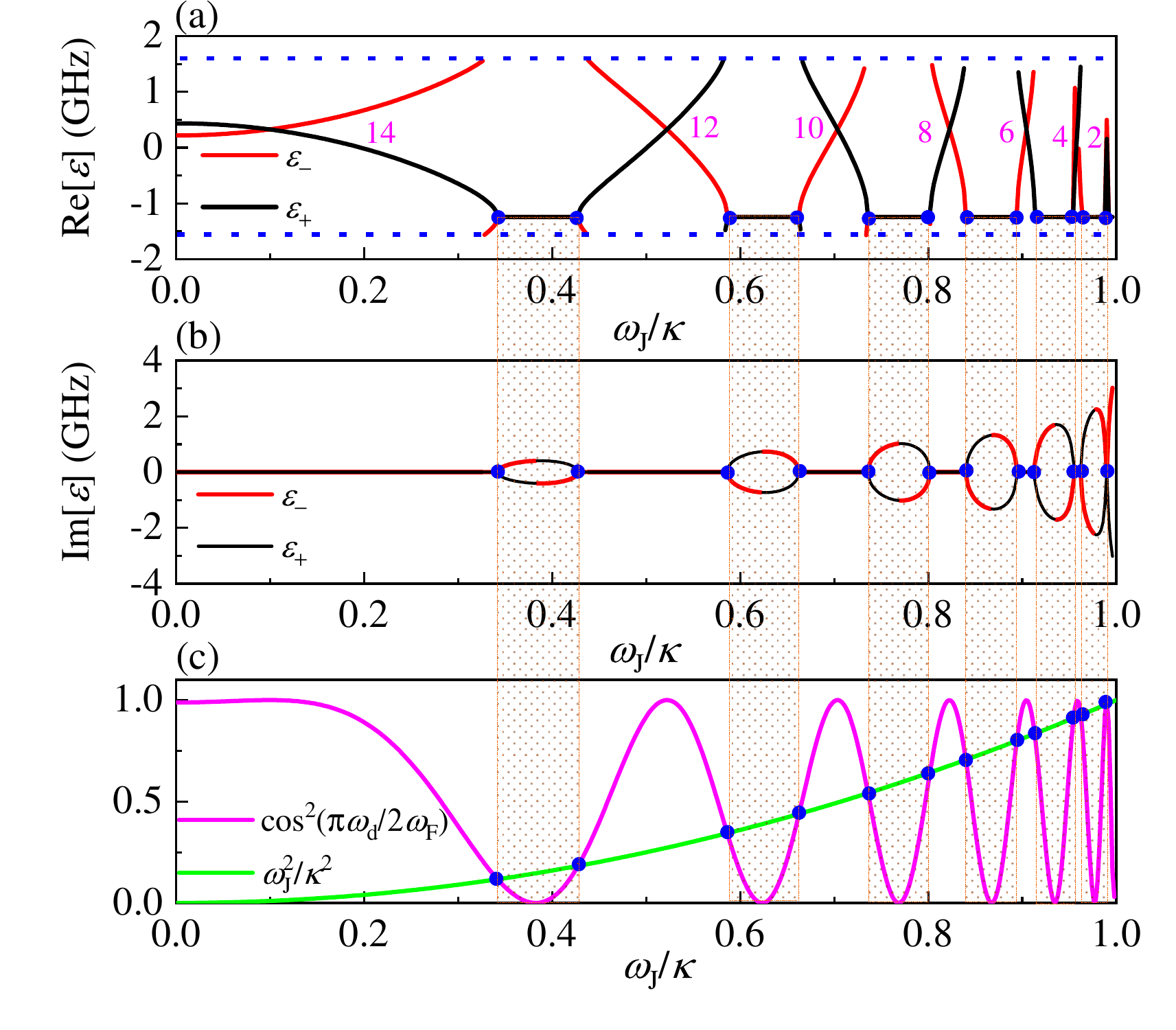}
	\caption{\label{floquetep}  (a) Real and (b) imaginary parts of Floquet quasienergies $ \epsilon_{\pm} $ as functions of the   amplitude   $ \omega_J $ of periodic SOT for $ \alpha \rightarrow 0 $. Real parts of $ \epsilon_{\pm} $ in the zone $ -\omega_F/2 $ and $ \omega_F/2 $ (marked by blue dashed lines) with $T = 2\pi/\omega_F$ being the AC charge current period. The integers (14, 12,...) in (a) are the ratio $ \frac{\omega_d}{\omega_F} $ when $ \epsilon_{\pm} $ cross. (c)  $ \frac{\omega_J^2} {\kappa^2}$ and $ \cos^2(\frac {\pi} {2} \, \frac{\omega_d}{\omega_F}) $. Blue full dots mark Floquet EPs  conditions. Shadowed areas mark the range of  broken PT-symmetry phases.\cite{PhysRevLett.131.186705} }
\end{figure}

In the introduction we discussed for two coupled systems the general expectation for the dynamics of a periodically driven system.
Here, we extend this discussion to PT-symmetric systems.  This can be achieved, for example, by using AC currents to generate time-modulated SOT. Floquet engineering allows for dynamic manipulation of the PT phase transitions, enabling control over the location and number of EPs in the parameter space. This dynamic control is relevant  for  reconfigurable magnonic devices but also on the fundamental side concerning the dynamics of  time-dependent pseudo-Hermitian systems.

To realize the pseudo-Hermitian Floquet magnonic system with time varying gain and loss, a time-dependent electric current is injected in the heavy metal layer of Fig. \ref{pseudomodels}(a).\cite{PhysRevLett.131.186705}  In the case of  time alternating $ \omega_{\tau,t}( 0 \leq t \leq t_s ) = \omega_J  $ and  $ \omega_{\tau,t}( t_s \leq t \leq T ) = -\omega_J $ with period $T = 2 t_s$. Submitting the time-varying function into Eq. (\ref{coupleequationfmsot}). The time evolution of the system state after time $ t_s $ with positive (negative) follows,
\begin{equation}
	\begin{aligned}
		\displaystyle 
		\label{floquetstate} \left( \begin{matrix}
			\psi_1 \\
			\psi_2
		\end{matrix}\right)
		=  {M}_{P(N)}(t_s)
		\left( \begin{matrix}
			\psi_1^0 \\
			\psi_2^0
		\end{matrix}\right).
	\end{aligned}
\end{equation}
 $ (\psi_{1}^0, \psi_{2}^0) $ is the initial state in each evolution, and the propagation matrix $  {M}_{P(N)}(t_s) $ is determined from magnon function Eq. (\ref{coupleequationfmsot}). The combined propagation matrix after one whole period is $  {M}(T) =  {M}_P(t_s)  {M}_N(t_s) $. Under periodic driving, Floquet's theorem states for solutions $ e^{i \epsilon_{\pm} t} \phi_{\pm}(t) $ with the periodic function $ \phi_{\pm}(t) = \phi_{\pm}(t + T) $, and the Floquet's quasi-energy $  \epsilon_{\pm}\in [0,\omega_F]  $ defined  up to multiples of $\omega_F = \pi/ t_s$. For a constant SOT with $ t \rightarrow \infty  $, the Floquet quasienergies $ \epsilon_{\pm} = (1-i\alpha) (\omega_H + \omega_{ex} k^2 \pm \sqrt{q^2 - \omega^2_{\tau}}) $ is exactly the same as for  the above case. 

For the time-periodic SOT,  the  two complex quasienergies are deduced as,\cite{PhysRevLett.131.186705}
\begin{equation}
	\begin{aligned}
		\displaystyle 
		\label{floqueteigen} &\epsilon_{\pm} = -\ln\left\{\frac{e^{-2i w_e}}{2d^2} \left[(e^{-2i d_e} + e^{2i d_e})\kappa^2 - 2\omega_J^2 \right.\right. \\
		&\left. \left. \pm 2\kappa \sinh(i d_e)[2(\kappa^2 + \kappa^2 \cosh(2id_e)-2\omega_J^2)]^{1/2}\right]\right\}\omega_F/(2\pi i).
	\end{aligned}
\end{equation}
Here, we introduced 
$$ d = \sqrt{q^2 - \omega^2_{\tau}} ,$$   $$ w_e = \pi (1 - i\alpha)\omega_0/\omega_F,$$ and  
 $$ d_e = \pi (1 - i\alpha) d/\omega_F .$$
   For small damping, both $\epsilon_{\pm}$ simplify to
\begin{equation}
	\begin{aligned}
		\displaystyle 
		\label{floqueteigendamp} 
		\epsilon_{\pm} = -\ln \left\{\frac{ e^{-2i w_e} \kappa^2}{d^2} [ 
		\cos(\frac {\pi\omega_d}{\omega_F}) -\frac{\omega_J^2}{\kappa^2} 
		%
		\pm 2 i \sin( \frac{\pi\omega_d}{2\omega_F})\,  [\cos^2 (\frac {\pi\omega_d}{2\omega_F})-\frac{\omega_J^2}{\kappa^2}]^{1/2}] \right\} \frac{\omega_F }{2\pi i}.
	\end{aligned}
\end{equation} 
From the above equation one can identify  current tunable cases as follows:\\
a) When  the frequency $\omega_F$ and amplitude  $ \omega_J $ of  the driving current are such that  
$$\cos^2 (\frac\pi 2\frac{ \omega_d} {\omega_F}) > \frac{\omega_J^2}{\kappa^2},$$ the   quasienergies $\epsilon_{\pm}$ are  real and different, indicating a PT-symmetry preserved phase.\\
b) Even in this phase,  Floquet states may become  degenerate when  the level spacing  is a multiple of the driving frequency, i.e., $$\omega_d=2n\omega_F$$ and $n$ is an integer which resembles the standard
multiphoton resonance for weak driving \cite{LARSEN1976254} and is depicted on Fig. \ref{floquetep}.\\
c) At  $$\cos^2 (\frac\pi 2\frac { \omega_d}{\omega_F})=\frac{\omega_J^2}{\kappa^2}$$ the  modes coalesce signaling  EPs in Floquet modes (called henceforth FEPs). 
d) Above the FEP ($\cos^2 (\frac\pi 2\frac{ \omega_d} {\omega_F}) < \frac{\omega_J^2}{\kappa^2}$),   $\epsilon_{\pm}$ turn complex with the two real parts of $\epsilon_{\pm}$ being degenerate, and the two imaginary parts are different which is indicative of the broken PT-symmetry  phase.
Since $\frac{\omega_J^2}{\kappa^2}$ is monotonous but $\cos^2 (\frac\pi 2\frac{ \omega_d} {\omega_F})$ periodic in  $\omega_J$, as we vary the current spacer strength (varying thus $\omega_J$), several broken PT-symmetry  islands appear (shaded areas in Fig. \ref{floquetep} ).  FEPs  mark the boundaries of these islands. Importantly, the values of FEPs occur at much smaller $\omega_J$  (which is proportional to the current density) as the conventional EP (at $ \omega_J = q $).

\begin{figure}[htbp]
	\includegraphics[width=0.9\textwidth]{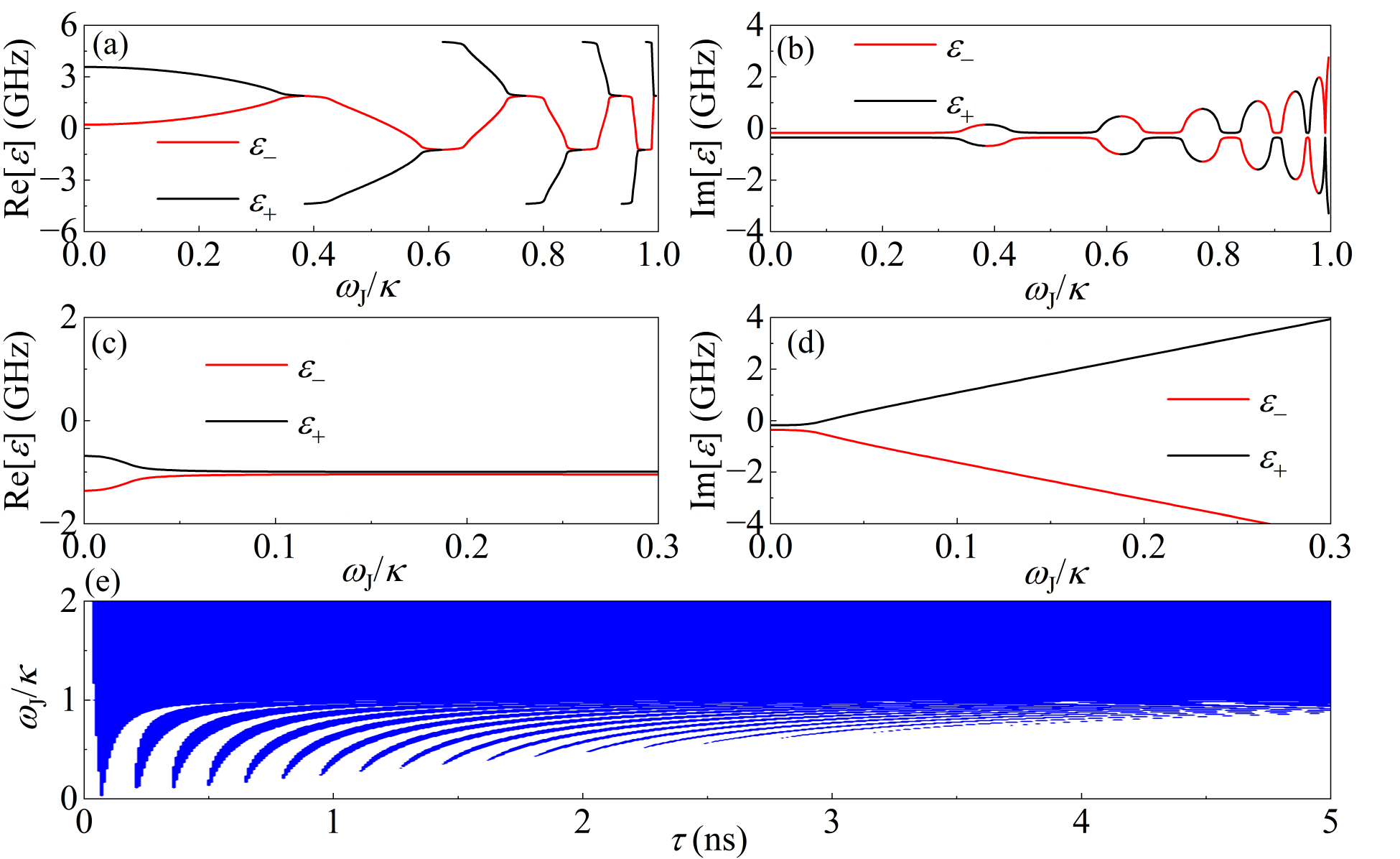}
	\caption{\label{floquetsmall}  For $ \alpha = 0.004 $, real and imaginary parts of quasienergies $ \epsilon_{\pm} $ as functions of the amplitude $\omega_J$ with the period parameters (a-b) $\tau = 1$ ns and (c-d) $\tau = 0.07$ ns. (e) The stability phase diagram on $ \tau $ - $\omega_J$ space. The shaded region corresponds to $ {\rm Im}[\epsilon_{\pm}] > 0 $.}
\end{figure}

For larger damping, in the broken PT-symmetry  phase  small gaps appear between the real parts, and two $\rm{Re}[\epsilon_{\pm}]$ coalesce only at the center, and two imaginary parts $\rm{Im}[\epsilon_{\pm}]$ are more separated there.(Fig. \ref{floquetsmall}(a-b)) Besides, the damping causes  a finite  $\rm{Im}[\epsilon_{\pm}]$ outside the broken PT-symmetry  phases. These larger finite damping induced features make the FEPs not as clear as for small damping,  still one can identify the FEPs regions and broken PT-symmetry  phase from the region with more separated $\rm{Im}[\epsilon_{\pm}]$.

Varying  the charge current strength  (meaning gain/loss strength) the stability behavior can be controlled. Around FEPs, the imaginary parts turn positive, indicating driven instability. In contrast to the constant SOT, the enhanced magnetization oscillation persists long after reaching its maximal. The induced magnon frequencies differ from the frequency of the spacer current, offering so a new method to generate high frequency magnons by an electric current with a very low frequency. The pseudo-Hermiticity induced magnon auto-oscillation is quite different from that in a single waveguide, which always need a tilt angle in the electric polarization with respect to the equilibrium magnetization direction. The above analysis neglect the dipolar interaction. Including dipolar interaction only slightly affects the values of quasienergies and quasi-EPs, and the main features of the Floquet broken PT-symmetry phase and auto-oscillation remain unaltered. 

For a smaller  varying period of the electric current, the current density required for the Floquet EP can be very low, and the number of Floquet EP becomes different. For example, at $\tau = 0.07$ ns, there is only one Floquet EP near $\omega_J= 0.03 \kappa$ (Fig. \ref{floquetsmall}(c-d)), and it is much smaller then the smallest EP at $\tau = 1$ ns. Furthermore, the stability diagram at $\tau$ - $\omega_J$ is provided in Fig. \ref{floquetsmall}(e). Around the Floquet EP, the imaginary parts of $ \epsilon_{\pm} $ can turn positive, and the change in EP current density and EP number with different $\tau$ can be inferred from the stability diagram.

The Floquet pseudo-Hermiticity is realized in the time domain. In a spatially-periodic gain and loss driven by SOT, as the magnon also experiences periodically varying gain and loss during propagation, similar lower EPs and auto-oscillation above EP still exist.\cite{PhysRevLett.131.186705} Also, the system with coupled periodic loss/more loss uncovers similar features FEP, demonstrating the versatility of magnonics when combined with pseudo-Hermiticity. 

 The pseudo-Hermitian Floquet engineering offers a controlling mechanism for high-frequency magnons via a low frequency electric current. The feature is potentially useful for magnon control in antiferromagnets (AFM). In AFM, magnons are usually at Terahertz frequencies and sub-ns lifetimes, and control schemes are required on ps time-scales. Also, in co-linear AFM with vanishing net magnetization, the two magnon modes with opposite circular polarizations are degenerate, and a  control for AFM magnons is needed to affect two modes in opposite ways.\cite{PhysRevB.95.174407,PhysRevLett.111.017204,PhysRevLett.117.217202} The pseudo-Hermitian Floquet engineering can help  controlling ultra high frequency AFM magnons by periodically modulating the gain and loss parameters, and it can works at a much lower thresholds than in a static system. Besides, around the EP, two magnon modes collapse, and thus   tiny variation in gain/loss or coupling can produce large and fast polarization rotations of AFM. These ideas and proposals are of potential use  in   polarization-encoded AFM THz magnonic logic. 

\subsection{Higher order EPs and topological energy transfer}

\begin{figure}[htbp]
	\includegraphics[width=0.9\textwidth]{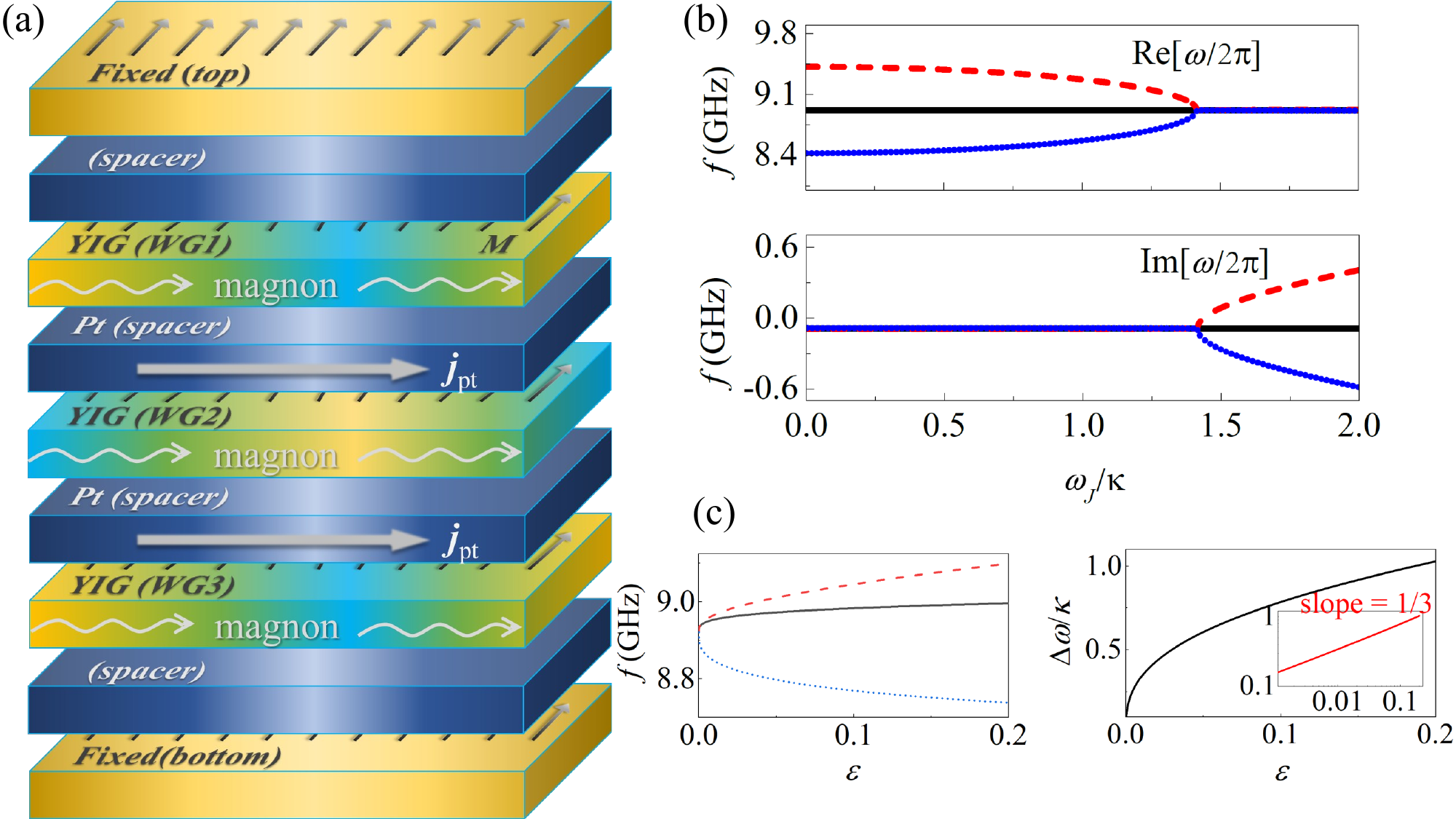}
	\caption{\label{3rdep} (a) Schematics of three coupled magnonic waveguides. Two hard magnetic cap layers are put on top and bottom. Neighboring layers are RKKY coupled. Injecting charge currents in spacers between WG1, WG2 and WG3 leads to SOTs that drive the magnonic loss and gain in WG1 and WG3, while the SOT in WG2 is compensated. (b) Real and imaginary parts of three eigenfrequencies $ \omega_{T1,T2,T3}/(2 \pi) $ as $ \omega_J/\kappa $ varies. (c) At the third order EP, real parts of three eigenfrequencies as functions of the perturbation $ \epsilon $ , $ \Delta\omega/\kappa = {\rm Re}[\omega_{T2} - \omega_{T3}] / \kappa$ and its value on log$_{10}$ scale in the inset.\cite{PhysRevApplied.15.034050} }
\end{figure}

\begin{figure}[htbp]
	\includegraphics[width=0.9\textwidth]{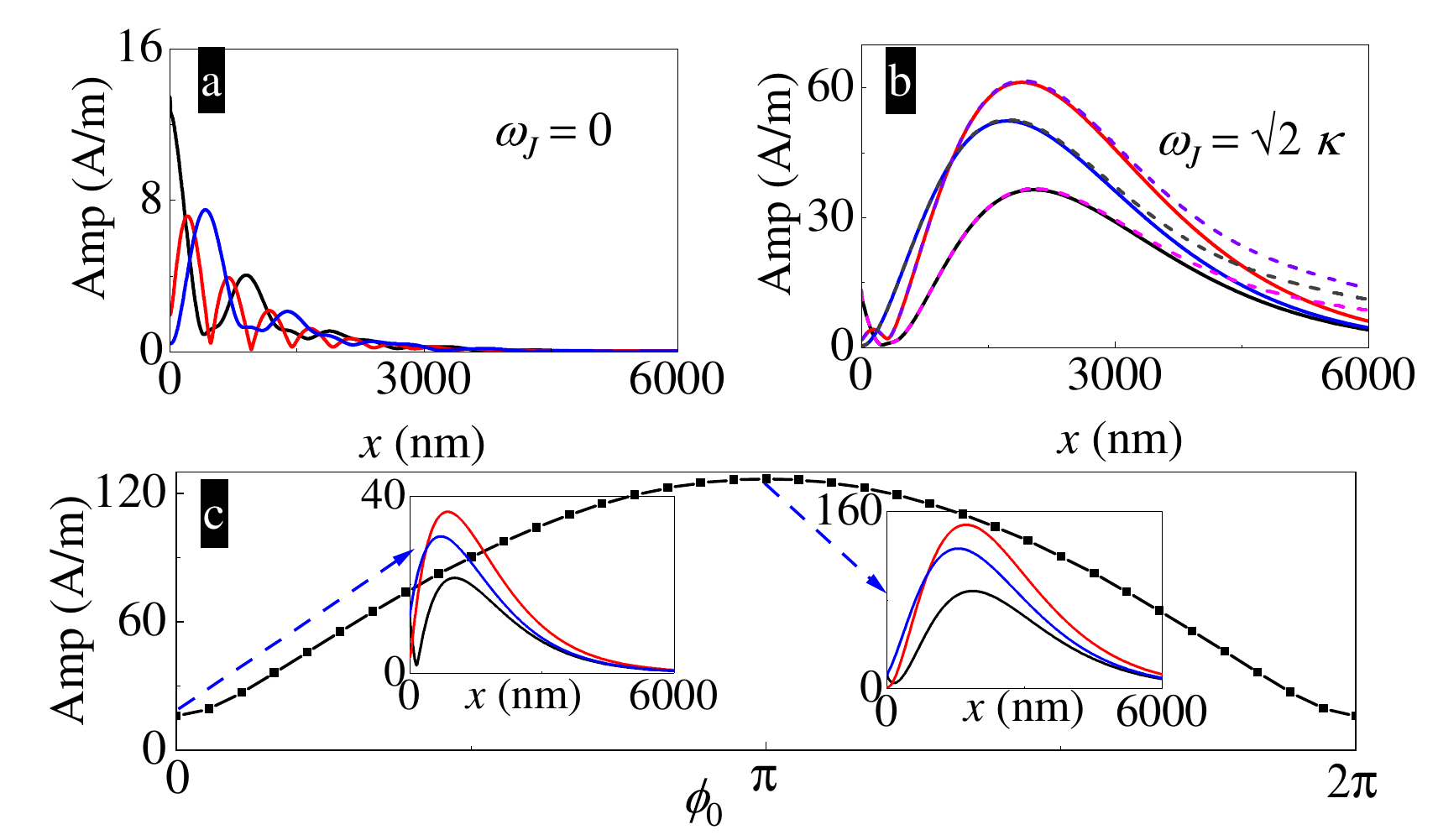}
	\caption{\label{3rpropaga} The spatial profiles of magnon amplitudes at (a) $\omega_J = 0$ and (b) EP. Black (red, blue) solid lines and magenta (violet, gray) dashed lines are magnon amplitudes for perturbation $ \epsilon = 0 $ and 0.01. (c) At EP, the magnon amplitude of WG2 as function of the phase difference between the inputs in WG1 and WG3.\cite{PhysRevApplied.15.034050}}
\end{figure}

\begin{figure}[htbp]
	\includegraphics[width=0.9\textwidth]{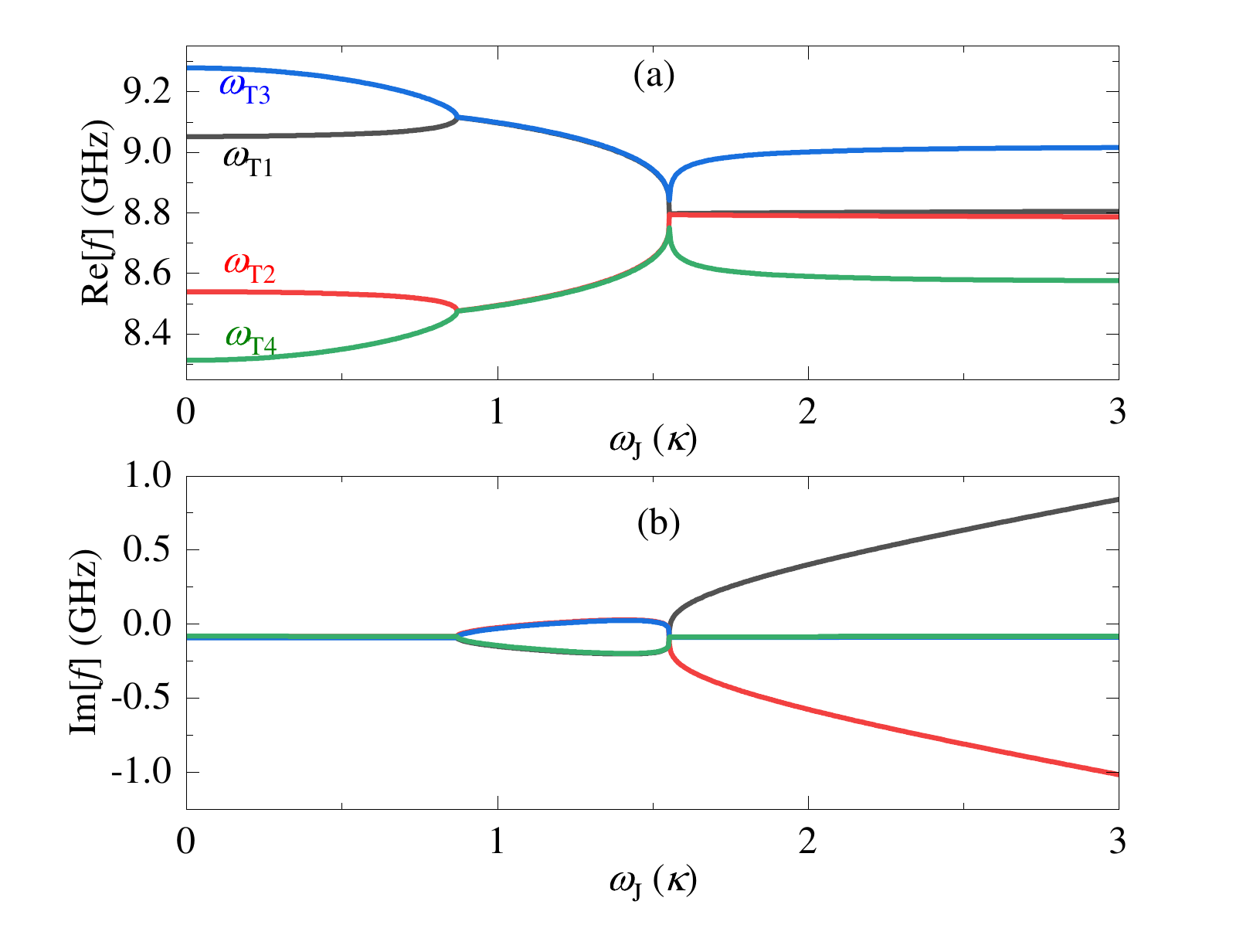}
	\caption{\label{4thep} For four coupled magnonic waveguides, (a) real and (b) imaginary parts of four eigenfrequencies as functions of $ \omega_J / \kappa $.\cite{PhysRevApplied.15.034050}}
\end{figure}

\begin{figure}[htbp]
	\includegraphics[width=0.9\textwidth]{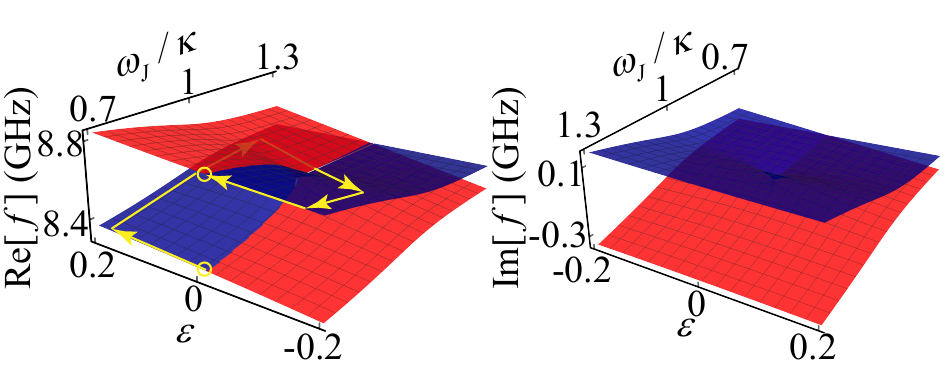}
	\caption{\label{enclosing} For two RKKY coupled magnonic waveguides, real and imaginary parts of eigenfrequencies as a function of $\omega_J$ and $\epsilon$. The loop enclosing the EP is indicated by the yellow loop.\cite{PhysRevApplied.15.034050}}
\end{figure}

From the  above analysis we conclude that  two coupled magnonic waveguides lead  to  second order EP with two merged magnon modes. By carefully engineering the structure and parameters of the system, several evidences also suggest the possibility of realizing higher-order EPs.\cite{PhysRevB.99.214415, PhysRevApplied.15.034050} For example, in cavity magnon-polariton system, Cao and Yan predicted the existence of third order EP by coupling a cavity photon with two magnon modes, where three eigenfrequencies coalesce.\cite{PhysRevB.99.214415} At the EP, they found a “Z-shaped” transmission spectrum featuring two side bands and a nearly flat central band with ultra-narrow line width. As the system approaches the  third order EP, the central mode’s line width vanishes, and exactly at EP the spectrum jumped abruptly. These predictions suggest potentially even greater sensitivity enhancements and rich physics such as nontrivial topology in parameter space. They suggested an experimental scheme involving additional driving to tri-modally couple photons and magnons.

\begin{figure}[htbp]
	\includegraphics[width=0.9\textwidth]{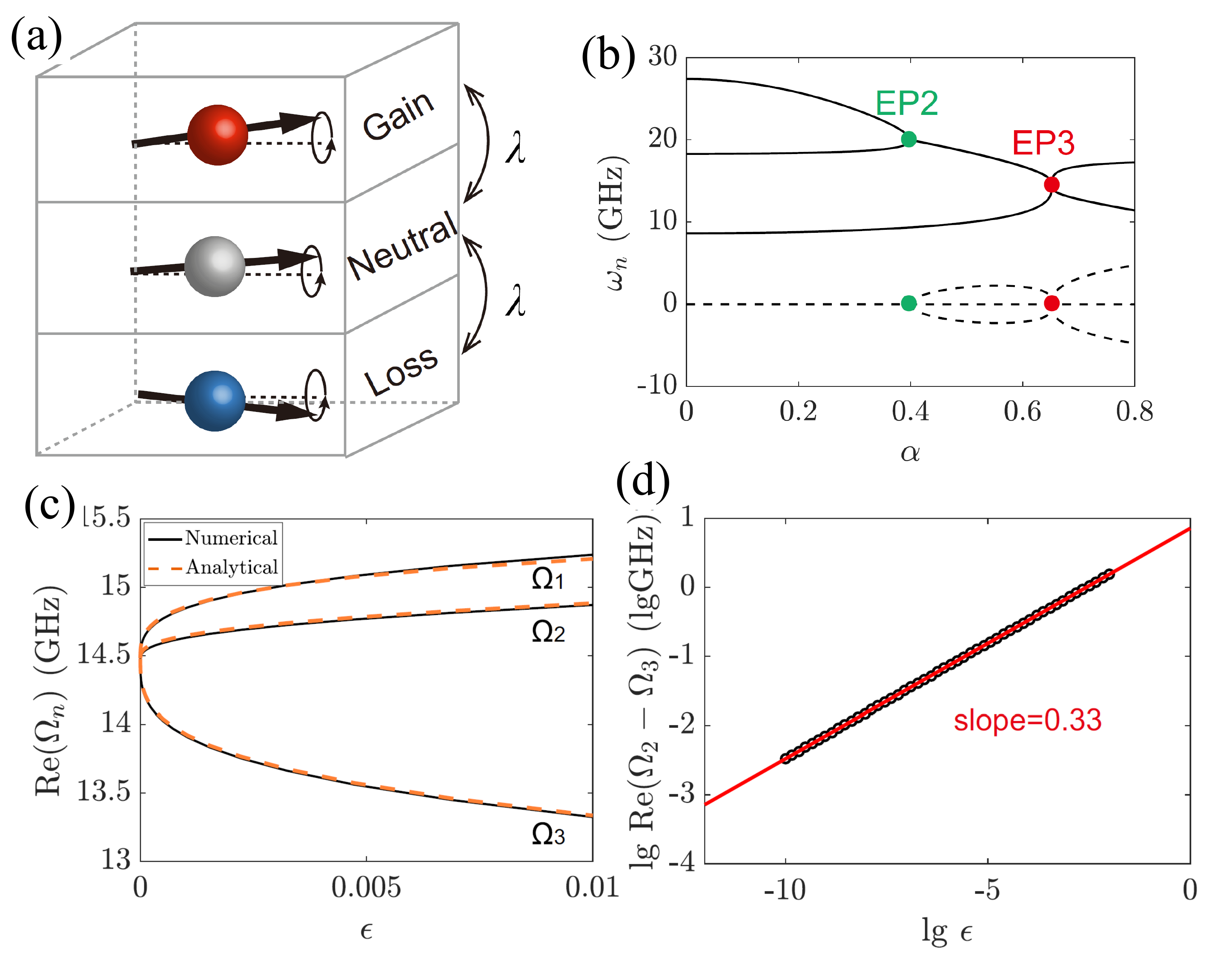}
	\caption{\label{ep3rd2} (a) Schematic of ferromagnetic trilayer heterostructure with third-order EP. (b) The eigenfrequencies of the trilayer system as functions of constant $ \alpha $. (c) Changes in eigenfrequencies with disturbance $\epsilon$. (d) Frequency splitting of ${\rm Re}(\Omega_2 - \Omega_3)$ on a logarithmic scale.\cite{PhysRevB.101.144414}}
\end{figure}

Yu et al. developed a macrospin-based Landau-Lifshitz-Gilbert equations for a gain for negative damping $ \alpha $, neutral for 0 damping, loss for positive $ \alpha $, in ferromagnetic trilayer with third order magnonic EP.\cite{PhysRevB.101.144414} The PT-symmetric $3\times3$ magnonic Humiliation was obtained as,
\begin{equation}
	\begin{small}
		\displaystyle  H  = \left( \begin{matrix} \frac{\omega_{B1}}{1-i\alpha} & -\frac{\omega_{\lambda2}}{1-i\alpha} & 0\\
			-\omega_{\lambda1} & \ \omega_{B2}\ & -\omega_{\lambda1}\\
			0 & -\frac{\omega_{\lambda2}}{1+i\alpha} & \displaystyle \frac{\omega_{B1}}{1+i\alpha} \end{matrix} \right).
		\label{triham}
	\end{small}
\end{equation}
The PT symmetry was achieved  by the exact balance of Gilbert gain ($+\alpha$) and loss ($-\alpha$). By tuning the interlayer exchange constant $\lambda$, the external field $B$ and $\alpha$, the authors derive the third order EP condition,
 \begin{equation}
 	\begin{small}
 		\begin{aligned} 
 		\displaystyle (2\omega_{B1}+\omega_{B2}+\alpha^{2}\omega_{B2})^{2}+3(1+\alpha^{2})\!\left(2\omega_{\lambda1}\omega_{\lambda2}-\omega_{B1}^{2}-2\omega_{B1}\omega_{B2}\right)=0,\\
 		(2\omega_{B1}+\omega_{B2}+\alpha^{2}\omega_{B2})\left(2\omega_{\lambda1}\omega_{\lambda2}-\omega_{B1}^{2}-2\omega_{B1}\omega_{B2}\right) + 9(1+\alpha^{2}) (\omega^2_{B1}\omega_{B2}-2\omega_{B1}\omega_{\lambda1}\omega_{\lambda2})=0.
 		\label{trihamsolve}
 	\end{aligned} 
 	\end{small}
 \end{equation}
The equation showed that third-order and second-order EPs emerge in the same system, see Fig. \ref{ep3rd2}(b). Numerical results reveal the hallmark sub-linear response, where the magnon frequency splitting scale as at the third-order EP (Fig. \ref{ep3rd2}(c-d)), i.e. $\Delta \Omega_{EP3}= c \omega_{\lambda2} \epsilon^{1/3}$ with a tiny perturbing parameter $\epsilon$. Extending the model to real-wave-vector magnons, the work maps out a $k$-dependent phase diagram in which PT-symmetry breaking drives a ferromagnetic-to-antiferromagnetic transition, thus connecting pseudo-Hermitian magnonics with higher-order EP physics.\cite{PhysRevB.101.144414}

Considering the three coupled wavegudies shown  in Fig. \ref{3rdep},  one can achieve that  WG1 and WG3 are  RKKY coupled to the top and bottom hard bias layer. The same charge current in two PT spacer layers results in loss and gains in WG1 and WG3, i.e. $ \vec{T}_{1(3)} = \pm  \tau (\vec{m}_{1(3)} \times (\vec{y} \times \vec{m}_{1(3)})) $. The middle WG2 with compensated SOT is coupled to both WG1 and WG3. For the three coupled waveguides model, the magnon Hamiltonian,\cite{PhysRevApplied.15.034050}
\begin{equation}
	\begin{small}
		\begin{aligned} 
			\displaystyle  H  = (1-i\alpha)\left( \begin{matrix} \omega_3 - i \omega_J & -q & 0 \\ -q & \omega_3 & -q \\ 0 & -q & \omega_3 + i \omega_J \end{matrix} \right),
			\label{magnonhamrkky3}
		\end{aligned} 
	\end{small}
\end{equation}
is non-Hermitian, and we define $ \omega_3 = \omega_H + \omega_{ex} k^2 + 2q $. Three eigenfrequencies $ \omega_{T,p} $ of the Hamiltonian are obtained,
\begin{equation}
\begin{aligned} 
	\displaystyle  \omega_{T1} &= (1-i\alpha)\omega_3,\\
	\omega_{T2} &= (1-i\alpha)(\omega_3 + \sqrt{2q^2-\omega_J^2}),\\
	\omega_{T3} &= (1-i\alpha)(\omega_3 - \sqrt{2q^2-\omega_J^2}).
	\label{eigen3wave}
\end{aligned} 
\end{equation}
Without  SOT, i.e. for  $ \omega_J = 0 $, the eigenvectors   $$ V_{T1} = A_1(-1,0,1) ,$$  $$ V_{T2} = A_2(1,-\sqrt{2},1) ,$$ and 
$$ V_{T3} = A_3(1,\sqrt{2},1) $$
 describe  spatial distributions of guided magnon modes. Increasing the SOT amplitude to the critical value $ \omega_J = \sqrt{2} q $, all three eigenfrequencies coalesce at the same value leading to third-order EP, and three eigenvectors coalesce at $ A_0(-1,i\sqrt{2},1) $.

The third-order EP shows more enhanced sensitivity comparing to the second-order EP.
Applying  a perturbation $ \epsilon $ only the WG1,  at the EP $ \omega_J = \sqrt{2} \kappa $, the three eigenfrequencies are perturbatively  expanded  in a Newton-Puiseux series to yield 
	$$
\begin{aligned} 
	\displaystyle  \omega_{T1} &\approx (1-i\alpha)q (\frac{\omega_3}{q} + e^{i\pi/3}\epsilon^{1/3} - \frac{i\sqrt{2}}{3} e^{-i\pi/3}\epsilon^{2/3}),\\
	\omega_{T2} &\approx (1-i\alpha)q (\frac{\omega_3}{q} + e^{-i\pi/3}\epsilon^{1/3} - \frac{i\sqrt{2}}{3} e^{i\pi/3}\epsilon^{2/3}),\\
	\omega_{T3} &\approx (1-i\alpha)q (\frac{\omega_3}{q} - \epsilon^{1/3} + \frac{i\sqrt{2}}{3}\epsilon^{2/3}).
	\label{ternaryexpand}
\end{aligned} 
$$
Between $ \omega_{T1,T2} $ and  $ \omega_{T3} $, the real parts of the frequency splittings follow the cube-root of $ \epsilon $. For example, 
$$ \mathrm{Re}[\omega_{T2} - \omega_{T3}] \approx (\frac{3-\sqrt{3}\alpha}{2})\epsilon^{1/3} ,$$ while the splitting between the real parts of $ \omega_{T1} $ and $ \omega_{T2} $ is on the order of $ \epsilon^{2/3} $. This cube-root behavior of the perturbation factor is the source of  the sensitivity enhancement. Both the real and the imaginary parts of the eigenfrequencies are affected  resulting in a clear change in the spin-wave wavelength and attenuation. 

One can also test for the case with the perturbation $ \epsilon $ acting on the SOT-neutral WG2. In this case, the resulting bifurcations in the eigenfrequencies read,\cite{PhysRevApplied.15.034050}
$$
\begin{aligned} 
	\displaystyle  \omega_{T1} &\approx (1-i\alpha)q (\frac{\omega_3}{q} - 2^{1/3}e^{i\pi/3}\epsilon^{1/3}),\\
	\omega_{T2} &\approx (1-i\alpha)q (\frac{\omega_3}{q} + 2^{1/3}\epsilon^{1/3}),\\
	\omega_{T3} &\approx (1-i\alpha)q (\frac{\omega_3}{q} - 2^{1/3}e^{-i\pi/3}\epsilon^{1/3}).
	\label{ternaryneutralexpand}
\end{aligned} 
$$
The cube-root frequency splitting occurs  between $ \omega_{T1,T3} $ and  $ \omega_{T2} $. For example, 
$$ \mathrm{Re}[\omega_{T2} - \omega_{T1}]= (\frac{3 + \sqrt{3}\alpha}{2^{2/3}})\epsilon^{1/3}.$$
 The splitting between  $ \omega_{T1} $ and $ \omega_{T3} $ is very small and  determined directly by $ \alpha $ ($ \mathrm{Re}[\omega_{T3} - \omega_{T1}] =  2^{1/3}\sqrt{3}\alpha \epsilon^{1/3} $).

These features also affect the magnon propagation in the waveguides. In the case without SOT $ \omega_J = 0 $, a local microwave field  applied to WG1 excites propagating spinwaves  simultaneously  in three modes. The interaction of the three different magnon modes causes  periodic power transfer between the three WGs. At the EP, magnonic  power in WG2 (with no SOT) is twice as large as  in WG1 and WG3. The magnons travel simultaneously in the three WGs. For inputs  both  in WG1 and WG3,  the excited spin-wave amplitudes become sensitive to the phase differences of the  two inputs. Compared to two inputs with the same phase,  inputs with opposite phases excite much stronger spin waves. The magnon amplitude is sensitive to the phase difference between two inputs. This feature at EP is exploitable for sensitive magnonic  logic operations, as demonstrated by Fig. \ref{3rpropaga}.

Similarly, in a system with four free layers, one can identify a fourth-order EP (four eigenfrequencies coalesce) with a fourth-root enhancement. As an example, considering four coupled free layers (WG1, WG2, WG3 and WG4) with the top guide WG1 is coupled to a fixed layer (via coupling constant $ \kappa_b = \sqrt{\sqrt{2} - 1} q $) and WG2 (via coupling constant $ \kappa_a = q $), WG2 is coupled to WG3 via $ \kappa_b $,  WG3 is coupled to WG4 via $ \kappa_a $, and bottom WG4 is further coupled to a fixed layer via $ \kappa_b $. In this case, the "Hamiltonian" becomes,\cite{PhysRevApplied.15.034050}
\begin{equation}
	\begin{small}
		\begin{aligned} 
			\displaystyle  H _0 = (1-i\alpha) \left( \begin{matrix} (\omega_4 - i\omega_J & -\kappa_a & 0 & 0\\ - \kappa_a & \omega_4 & -\kappa_b & 0 \\0 & -\kappa_b & \omega_4 & -\kappa_a \\ 0 & 0 & -\kappa_a & \omega_4 + i \omega_J \end{matrix} \right).
			\label{fourham}
		\end{aligned} 
	\end{small}
\end{equation}
With
$$ \omega_4 = \omega_H + \omega_{ex} k^2 + \kappa_a + \kappa_b .$$
 Four eigenfrequencies $ \omega_{T,p} $ of the Hamiltonian are obtained as 
\begin{equation}
	\begin{small}
		\begin{aligned} 
			\displaystyle  \omega_{T1} &= (1-i\alpha) (\omega_4 + \sqrt{\kappa_a^2 + \frac{\kappa_b^2 - \omega_J^2}{2} -\sqrt{\kappa_a^2(\kappa_b^2 - \omega_J^2) + \frac{(\kappa_b^2 + \omega_J)^2}{4}}}),\\
			\omega_{T2} &= (1-i\alpha) (\omega_4 - \sqrt{\kappa_a^2 + \frac{\kappa_b^2 - \omega_J^2}{2} -\sqrt{\kappa_a^2(\kappa_b^2 - \omega_J^2) + \frac{(\kappa_b^2 + \omega_J)^2}{4}}}),\\
			\omega_{T3} &= (1-i\alpha) (\omega_4 + \sqrt{\kappa_a^2 + \frac{\kappa_b^2 - \omega_J^2}{2} +\sqrt{\kappa_a^2(\kappa_b^2 - \omega_J^2) + \frac{(\kappa_b^2 + \omega_J)^2}{4}}}),\\
			\omega_{T4} &= (1-i\alpha) (\omega_4 - \sqrt{\kappa_a^2 + \frac{\kappa_b^2 - \omega_J^2}{2} +\sqrt{\kappa_a^2(\kappa_b^2 - \omega_J^2) + \frac{(\kappa_b^2 + \omega_J)^2}{4}}}).
			\label{four-eigen}
		\end{aligned} 
	\end{small}
\end{equation}

Under $$ \kappa_a = q $$ and $$ \kappa_b = \sqrt{\sqrt{2} - 1} q $$
 we find that two of four eigenfrequencies coalesce at the same value when the balanced gain/loss   approaches  the critical value 
 $$  \omega_J = \frac{\sqrt{7} q}{ \sqrt{5 + 3 \sqrt{2}} } = 0.87 q, $$  
 meaning  $ \omega_{T1} $ = $ \omega_{T3} $ and $ \omega_{T2} $ = $ \omega_{T4} $, indicating   the existence of a second-order EP. Increasing the gain/loss ratio to a larger critical value $$  \omega_J = \frac{q}{ \sqrt{\sqrt{2} - 1} } = 1.55 q $$ we find four eigenfrequencies that coalesce simultaneously  at the same value causing so a fourth-order EP (Fig. \ref{4thep}). 

To analyze the enhanced sensitivity at the fourth-order EP, we apply a perturbation $ \epsilon q $ (with $ \epsilon \ll 1 $) in WG. With the increase of $ \epsilon $ the  real parts of the four eigenfrequencies split clearly. By analyzing the logarithmic behavior of the  $ \mathrm{Re}[\omega_{T1} - \omega_{T4}] $, a slope of 1/4 is identified, evidencing  a fourth-root  enhancement at the EP (Fig. \ref{4thep}).

The coalescence of modes at an EP can be exploited for mode management.  Exploiting the topological structure of the dispersion  at   EP,  energy is transferred  between different  modes by encircling  EP.\cite{Hodaei2017nature, Doppler2016nature} E.g., for a bilayer  structure, two eigen-frequencies (optical and acoustic magnon modes) coincide at  EP  with a square-root response to a magnetic perturbation.  The vicinity of  EP exhibits a characteristic structure of a self-intersecting Riemann surface (Fig. \ref{enclosing}). If $ \epsilon $ and $ \omega_J $ are varied as a single closed loop enclosing EP, the  smooth evolution on the eigenvalues  result in a trajectory starting on one sheet and ending on the other, implying an energy transfer between different magnon modes. The fully topological transfer requires large enough $ \epsilon $ and $ \omega_J $ amplitudes. The scanning loop needs to include the EP, and the loop duration should be slow enough to achieve the transition.

\begin{figure}[htbp]
	\includegraphics[width=0.9\textwidth]{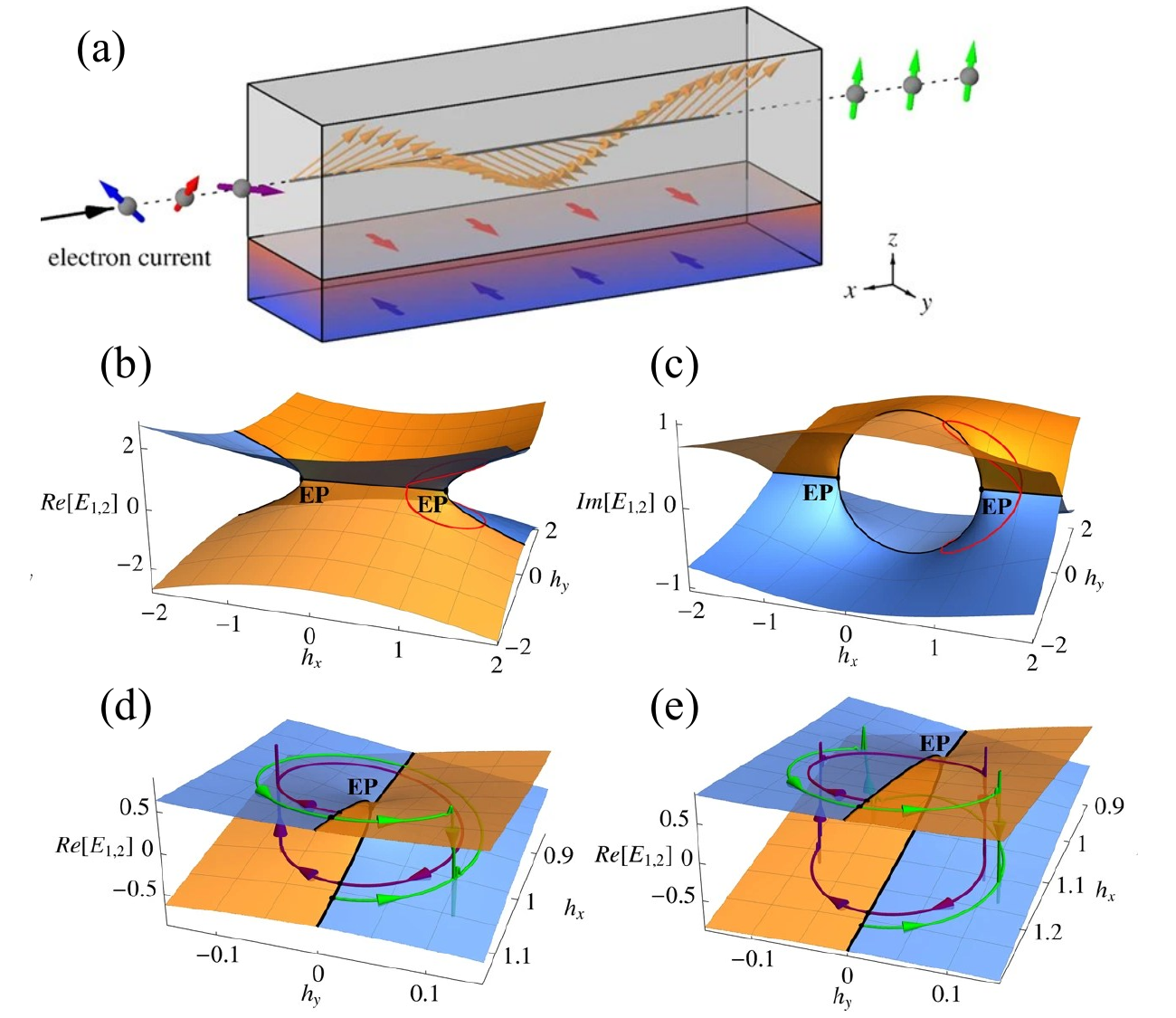}
	\caption{\label{epcircle} (a) Schematic of the magnetic layer with magnons and spin-torque. (b-c) The real and imaginary parts of eigenspectrum. (d-e) Evolution for (d) encircling around the single EP (e) in the vicinity of the EP.\cite{Galda2019scirep}}
\end{figure} 

Glada and Vinokur demonstrated  the EP encircling dynamics in a single classical spin subject (Fig. \ref{epcircle}(a)) to a in-plane filed $(h_x, h_y)$ and a Slonczewski-type spin-torque $\beta$ as the PT-symmetric Hamiltonian,\cite{Galda2019scirep}
\begin{equation}
	\begin{small}
		\begin{aligned} 
			\displaystyle  H _0 = h_x S_x + (h_y + i \beta)S_y,
			\label{singlespin}
		\end{aligned} 
	\end{small}
\end{equation}
and the  spectrum is expressed as for a two-level system,
\begin{equation}
	\begin{small}
		\begin{aligned} 
			\displaystyle E_{1,2} = \pm \sqrt{h_x^2 + (h_y + i \beta)^2}.
			\label{singlespinep}
		\end{aligned} 
	\end{small}
\end{equation}
As shown in Fig. \ref{epcircle}(b-c), two second-order EPs are generated in the spectrum. Using SU(2) coherent-state formalism they derived a spin equation of motion,
\begin{equation}
	\begin{small}
		\begin{aligned} 
			\displaystyle \dot{z} = i \frac{1+\bar{z}z}{2S}\frac{\partial H_0(z,\bar{z})}{\partial \bar{z}}.
			\label{singlespinmotion}
		\end{aligned} 
	\end{small}
\end{equation}
In this way one can numerically track the  spin trajectories while the control parameters trace loops that encircle around the EP, as shown in Fig. \ref{epcircle}(d-e). Dynamical encircling at finite speed forces non-adiabatic state-flip jumps between the Riemann sheets of $E_{1,2}$, producing chiral, direction-dependent final spin orientations. This amounts to  a non-reciprocal magnon/spin current filter whose efficiency is topologically protected by the EPs and tunable via and the loop frequency.

Experimentally, some signs of enhanced sensitivity have been observed. Zhang et al. found that the spin-ensemble/cavity with EP was robust against which spin ensemble was pumped, and they also argued  that the frequency splitting (when slightly off-EP) could be used to detect changes in the system.\cite{PRXQuantum.2.020307} 

\subsection{Non-reciprocity induced by pseudo-Hermiticity}

\begin{figure}[htbp]
	\includegraphics[width=0.9\textwidth]{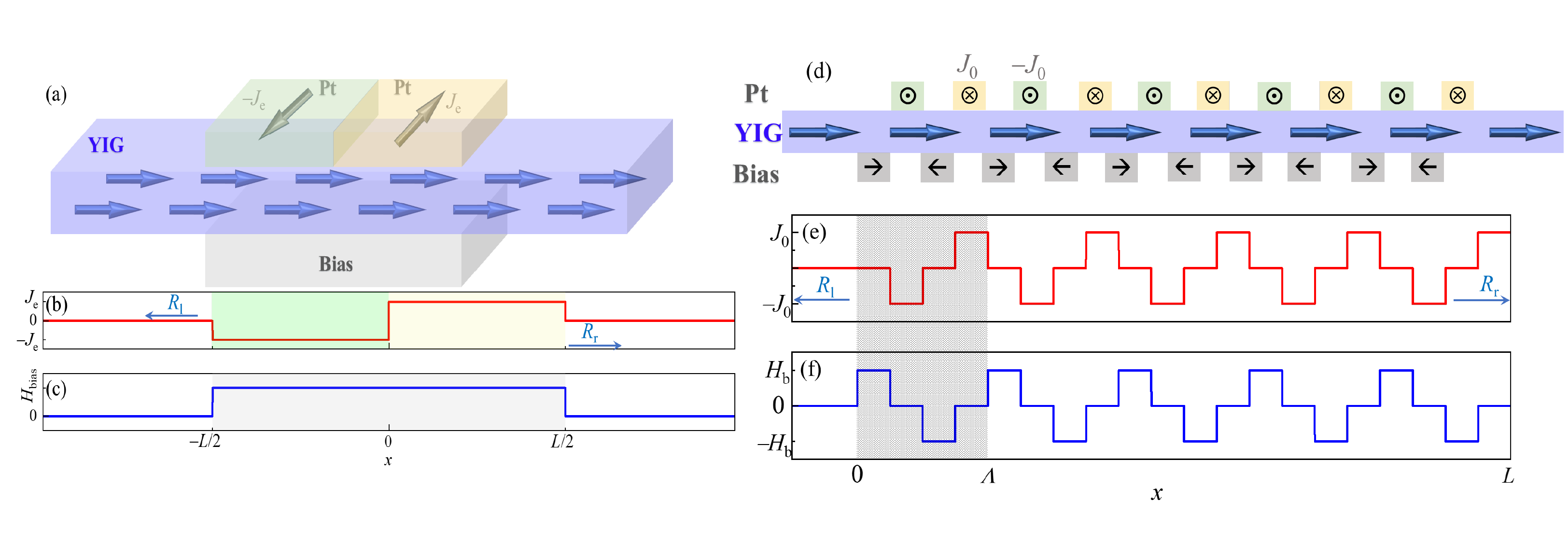}
	\caption{\label{magnonicpotential} Schematics of a magnonic waveguide with (a) single and (d) periodic regions with gain and loss, blue arrows indicate remnant magnetization. Adjacent heavy metal layers with antiparallel charge current resulting in magnonic gain and loss. Spatial dependence of electric current(b,e) and bias field (c, f). Arrow indicates from left (right) reflected wave with reflection coefficient $R_l$ ($R_r$).\cite{PhysRevApplied.22.054046}}
\end{figure}

\begin{figure}[htbp]
	\includegraphics[width=0.9\textwidth]{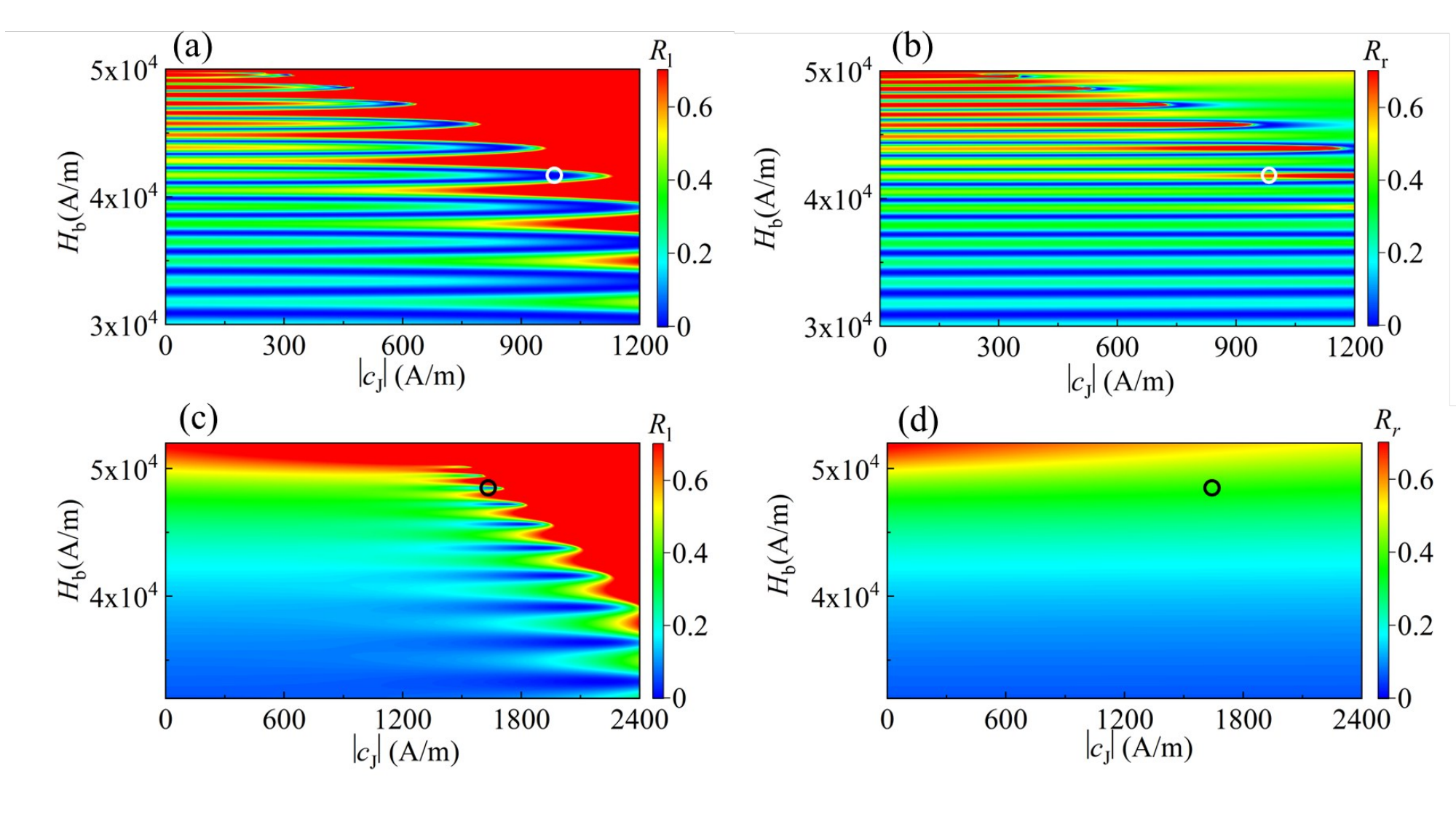}
	\caption{\label{ptreflection1} For the single PT-symmetric potential, $R_l$ and $R_r$ dependence on $c_J$ and bias field $H_b$ with no residual damping $\alpha = 0$ (a-b), or for a finite $\alpha$ (c-d).\cite{PhysRevApplied.22.054046}}
\end{figure}

\begin{figure}[htbp]
	\includegraphics[width=0.9\textwidth]{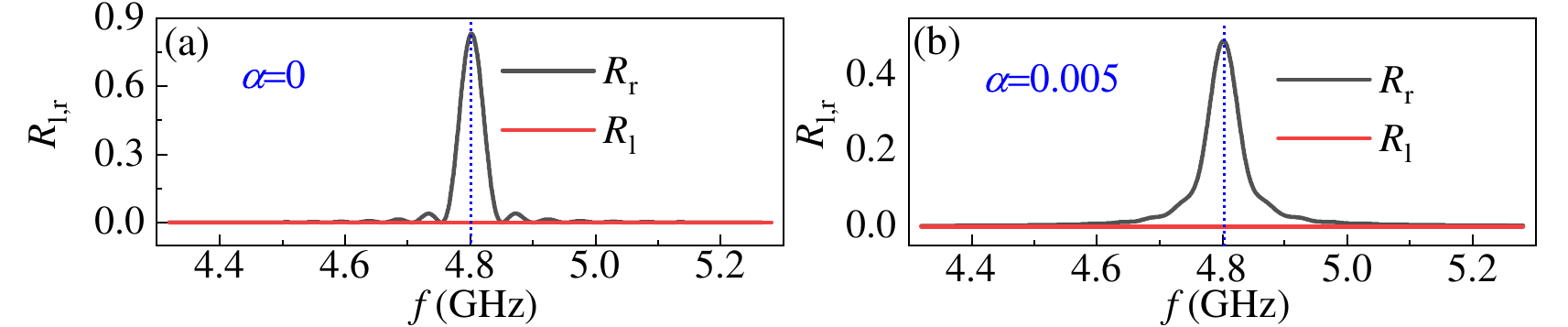}
	\caption{\label{ptreflection2} Reflection coefficients $R_l$ and $R_r$ as functions of magnon frequency $ f $ for (a) $ \alpha = 0 $ and $ \alpha = 0.005 $.\cite{PhysRevApplied.22.054046}}
\end{figure}

The properties of pseudo-Hermitian magnonics   can be exploited for unidirectional invisibility.
 By using SOT and bias fields to create an electrically tunable PT-symmetric potential in a waveguide, the system becomes pseudo-Hermitian. At the EP, which can be actively tuned by the applied electric field/current, the potential can become invisible to magnons incident from one direction while reflecting those from the other.\cite{PhysRevApplied.22.054046}

To demonstrate the electrically reconfigurable unidirectional invisibility magnonic potential, the single magnonic waveguide featuring a spatially varying gain and loss is adopted, as shown in Fig. \ref{magnonicpotential}(a). 
The waveguides initially magnetized along the $+x$ direction are attached to a number of electrically separated heavy metal layers. The charge current density $J_c$  in the metal layer generates the SOT $\vec{T} = \tau \vec{m} \times (\vec{x} \times \vec{m}) $. By spatially varying adjacent heavy metal layers with opposite charge current, the SOT results in the formation of localized magnonic gain (loss) zones. At the same time, attaching bias layers to the waveguide results in bias fields through the  interlayer exchange coupling to the localized potential.

To realize the single PT-symmetric potential,  the electric current's profile ( left region $J(-\frac{L}{2} \le x < 0) = -J_{\mathrm{e}}$, right region $J(0 \le x \le \frac{L}{2}) = J_{\mathrm{e}} $) is used to drive symmetric SOT-induced gain and loss, with the bias field $H_{\mathrm{b}}$ located within the same region of ($-\frac{L}{2} \le x \le \frac{L}{2}$). By defining the wave function $ \psi = \delta m_y - \i \delta m_z $, the LLG equation is changed in the form of Helmholtz equation under the linear assumption,
 \begin{equation}
	\begin{aligned}
		\displaystyle 
		\label{swHelmholtz} \psi''(x)+\left[\frac{\omega-\omega_H-\omega_{\mathrm{b}}(x)}{\omega_k} + i \frac{\omega_J(x)}{\omega_k}\right]\psi(x)=0.
	\end{aligned}
\end{equation}
Here, following notions are introduced: 
$$\omega_k = \frac{2 (1 - i\alpha) \gamma A_{\mathrm{ex}}}{\mu_0 M_{\mathrm{s}} (1+\alpha^2)},$$  
$$\omega_J(x) = \frac{(1- i\alpha) \tau(x)}{1+\alpha^2},$$  $$\omega_H = \frac{(1 -  i\alpha) \gamma H_0}{1+\alpha^2},$$ and  
$$\omega_{\mathrm{b}}(x) = \frac{(1 -  i\alpha) \gamma H_{\mathrm{b}}(x)}{1+\alpha^2}.$$
 With the asymmetric current $c_J(x)$ and symmetric bias field $H_{\mathrm{b}}(x)$, the coefficient of the Helmholtz equation 
 $$O_{\mathrm{H}} = \frac{\omega-\omega_H-\omega_{\mathrm{b}}}{\omega_k} + \i \frac{\omega_J}{\omega_k}$$ satisfies PT symmetry condition $(O^l_{\mathrm{H}})^* = O^r_{\mathrm{H}}$ in the limit of $\alpha \rightarrow 0$, where $ O^l_{\mathrm{H}} $ and $ O^r_{\mathrm{H}} $ represent $ O_{\mathrm{H}} $ in the left and right regions respectively. 

The unidirectional invisibility is defined by reflectionless scattering from left (right), and the condition can be fulfilled by tuning    $|\tau|$ and $ H_{\mathrm{b}} $. When Gilbert damping is neglected ($\alpha = 0$) (pseudo-Hermitian case),  it is possible to precisely determine the unidirectional invisibility condition, as demonstrated by Fig. \ref{ptreflection1}. These unidirectional invisibility conditions mark the collapse of two eigenvalues 
$$ [1 \pm \sqrt{1-M_{11} M_{22}}]/M_{22} $$ of the scattering matrix of the PT-symmetric potential, indicating a spontaneous PT-symmetry breaking point, i.e., EP. A finite $\alpha \ne 0$ slightly affect the  pseudo-Hermiticity, but approximate unidirectional invisibility is still identified, as evidenced  by the results in Fig. \ref{ptreflection1}.

Via a periodic PT-symmetric potential, the unidirectional invisibility can be realized near the Bragg point. As an example, we set a PT-symmetric periodic structure with 
$$\tau(x) = -c_{A} \sin (2 k_J x),$$ and  $$H_{\mathrm{b}}(x) = H_{\mathrm{b}A} \cos(2 k_J x)$$ in the region of $ 0 \le x \le L$. Near the Bragg point $k_x \approx k_J$, the transfer matrix $ {M}$ reads in this case
\begin{equation}
	\label{Matrixperiod} 
	 {M} = \begin{pmatrix}
		\cos(\beta L) + \frac{i \delta_k\sin(\beta L)}{\beta} &
		-i \frac{(\kappa + \xi) \sin(\beta L)}{\beta} \\
		i \frac{(\kappa - \xi) \sin(\beta L)}{\beta} & 
		\cos(\beta L) - \frac{i \delta_k\sin(\beta L)}{\beta}
	\end{pmatrix}.
\end{equation}
With 
$$\kappa =\omega_{bA}/\omega_k,$$ 
$$\omega_{bA} = \frac{(1-i\alpha) \gamma H_{\mathrm{b}A}}{1+\alpha^2},$$
 $$\xi = \omega_{JA} / \omega_k,$$
  $$\omega_{JA} = \frac{(1-i\alpha) \gamma c_{A}}{1+\alpha^2}$$,
   $$ \delta_k = k_0 - k_J, $$ 
   and 
   $$\beta = \sqrt{\delta_k^2 + \xi^2 - \kappa^2}.$$ 
   From $ {M}$ one can determine  the reflection coefficients from the left and right interfaces via 
   $$R_{\mathrm{l}} = |- e^{i k_0 L} M_{21}/M_{22}|^2$$ and 
   $$R_{\mathrm{r}} = | e^{i k_0 L} M_{12}/M_{22}|^2$$.

Without the electric current term ($\xi = 0$), one recovers the standard reflection at a periodic Bragg structure, and the magnon reflection at the periodic potential is reciprocal.  The periodic SOT with $\xi \ne 0$ results in an asymmetry in the reflections in opposite directions. The asymmetry becomes most pronounced at $\kappa = \pm \xi$ causing unidirectional invisibility. With $\kappa = \xi$, $R_{\mathrm{l}} $ is always 0, and $R_{\mathrm{r}} $ reaches its maximum at the Bragg point $\delta_k = 0$ for the Bragg point $ k_0 = k_J $. Near the Bragg point, $R_{\mathrm{r}} $ exhibits several weaker fluctuations, as demonstrated by Fig. \ref{ptreflection2}. Reversing the bias field or electric current ($ H_{\mathrm{b}A} = -c_{A} $), the right incident wave becomes reflectionless ($R_{\mathrm{r}} = 0$ and $R_{\mathrm{l}} \ne 0$). Different from the single PT-symmetric region above, the finite Gilbert damping $ \alpha $ doesn't affect the unidirectional invisibility condition. With $H_{\mathrm{b}A} = +(-) c_{A}$, $\kappa = +(-) \xi$ can be always satisfied independent of $ \alpha $.  Still, the unidirectional invisibility point $\kappa = \pm \xi$ in the periodic PT-symmetric region marks the collapse of two eigenvalues of the scattering matrix, representing EP of the system. By putting the PT-symmetric unidirectional invisibility potential in a ring, one can generate magnon orbital angular momentum (OAM) in ferro- and anitferromagets  \cite{Jia2019nc,Jia2021}.  Orbitaly chiral  magnonic state can be electrically triggered in a magnetic ring; similar effects are expected for magnons in magnetic vortices.\cite{PhysRevLett.122.097202}

\section{Conclusions and future directions}

The purpose has been to give an overview on the current status of  pseudo-Hermitian physics as realized in  magnonic systems which is 
a vibrant and rapidly expanding field of research. From the results achieved   so far it is evident, that  pseudo-Hermitian concepts provide novel  tools for   manipulating magnonic systems via external fields and 
 environmental effects  that can steer the gain and loss in magnon amplitudes. On the fundamental side, magnetic/magnonic platforms 
 offer the possibility of having an intrinsically non-linear system with the nonlinearity being controllable via external fields. For example, a strong magnetic field can drive large amplitude magnetization oscillations beyond the linear regime. Also applying a large voltage to a metallic ferromagent may switch the magnetization. A further feature is the wide-range of frequencies from THz for antiferromagnets to GHz for ferromagnets. Topological PT symmetric magnonic systems are also readily feasible. Magnonic crystal with topological properties is a concept well documented in magnonics and sofar less research have been done in connection with PT and anti PT symmetry in this area. 
The challenges in the field of pseudo-Hermitian magnonics are related to  the development of magnonic system with precisely engineered on-demand spatio-temporally  controllable damping and gain in magnon density. Integrating low-damping materials like YIG with materials providing strong SOT (heavy metals) or other functionalities, while maintaining high interface quality and low intrinsic damping in the magnetic layer, remains a challenge. Exploring novel material platforms, such as 2D van der Waals magnets, holds a significant promise but requires further development for reliable magnon transport and gain-loss engineering. Besides, exploiting the roles of loss in the non-Hermitian magnonics (for example, coupled loss and more loss) may add  new tools  for magnon manipulation via pure loss. 
%
The  nonlinear regime is less studied,  this concerns multiple magnon-magnon scattering in the different PT symmetry regimes. Corresponding studies 
may lead to novel dynamical regimes, such as bistability, self-organization phenomena, chaos, enhanced frequency comb generation, and potentially new paradigms for magnonic neuromorphic computing.
Further  possible future directions are to expand  beyond  1D coupled systems to explore non-Hermitian physics in 2D and 3D magnonic crystals possibly coupled to plasmonic and metamaterials. This will allow for the investigation of higher-order EPs, exceptional surfaces and lines, higher-order skin effects (corner/hinge localization), and magnetophotonic PT-symmetric setups.
 Likewise,  exploration and exploitation of hybrid magnonic platforms, including cavity/circuit magnonics, magnomechanics, magnon-qubit systems, offer further  flexibility in engineering complex Hamiltonians, controlling interactions, tuning parameters, and accessing different physical regimes, thereby facilitating the realization of sophisticated non-Hermitian functionalities.

\section*{Author contributions}

Both authors contributed to this work.

\section*{Acknowledgments}
The work has been supported by the DFG under project nr. 465098690,  National Natural Science Foundation of China (Grants No. T2495212, No. 12274469,  No. 12174452, and No. 12074437), the Natural Science Foundation of Hunan Province of China (Grants No. 2025JJ20005).
We thank   G.-h, Guo ,  C.-L. Jia, H. Jing,  and D. Schulz  for many discussions and collaborations on various aspects related to the topic of this article.

\section*{Financial disclosure}

None reported.

\section*{Conflict of interest}

The authors declare no potential conflict of interests.

\section*{Data Availability Statement}
Further details on the content of this article  are available  from the  authors upon reasonable request.
\bibliography{reffile}

\end{document}